\documentclass[vecphys]{svmult}

\usepackage{makeidx}         
\usepackage{graphicx}        
\usepackage{multicol}        
\usepackage[bottom]{footmisc}
\usepackage{bm}

\makeindex             

\newcommand{\beq}[0]{\begin{equation}}
\newcommand{\eeq}[0]{\end{equation}}
\newcommand{\beqa}[0]{\begin{eqnarray}}
\newcommand{\eeqa}[0]{\end{eqnarray}}
\newcommand{\bequ}[0]{\[}
\newcommand{\eequ}[0]{\]}
\newcommand{\beqau}[0]{\begin{eqnarray*}}
\newcommand{\eeqau}[0]{\end{eqnarray*}}

\newcommand{\be}{\begin{enumerate}}
\newcommand{\ee}{\end{enumerate}}
\newcommand{\bi}{\begin{itemize}}
\newcommand{\ei}{\end{itemize}}

\newcommand{\kf}[0]{k_{\rm F}}

\newcommand{\bfx}[0]{{\bf x}}
\newcommand{\Hhat}{{\widehat H}}
\def\galnab{\stackrel{\leftrightarrow}{\nabla}}
\def\nab{\overrightarrow{\nabla}}
\newcommand{\psibar}{\overline\psi}

\newcommand{\pboxb}[1]{\parbox{.43\textwidth}{%
                             \setlength{\baselineskip}{12pt}%
                             \rule[0pt]{0pt}{12pt}%
                             \raggedright\relax #1 \relax %
                             \rule[-3pt]{0pt}{10pt}%
                         }}

\newcommand{\densx}{\rho({\mathbf x})}

\newcommand{\Vext}{v_{\rm ext}}
\newcommand{\xvec}{{\bf x}}
\newcommand{\yvec}{{\bf y}}

\newcommand{\FHK}{F_{{\rm HK}}}
\newcommand{\Fni}{T_{{\rm KS}}}

\newcommand{\Exc}{E_{{\rm xc}}}

\newcommand{\Lra}{$\Longrightarrow$}

\newcommand{\kt}{\widetilde k}

\newcommand{\Mstar}{M^{\ast}}
\newcommand{\pairj}{j}
\newcommand{\psiup}{\psi_{\uparrow}}
\newcommand{\psidagup}{\psi^\dagger_{\uparrow}}
\newcommand{\psidown}{\psi_{\downarrow}}
\newcommand{\psidagdown}{\psi^\dagger_{\downarrow}}
\newcommand{\grounds}{{\rm gs}}
\newcommand{\GKS}{G_{\ks}}
\newcommand{\FKS}{F_{\ks}}
\newcommand{\ks}{{\rm ks}}

\newcommand{\wt}{\widetilde}

\begin{document}

\title*{EFT for DFT}
\titlerunning{EFT for DFT}
\author{R.J.\ Furnstahl}
\institute{Department of Physics, Ohio State University,
Columbus, OH 43210
\texttt{furnstahl.1@osu.edu}}
%
%
\maketitle

These lectures give an overview of the ongoing application of
effective field theory (EFT) and renormalization group (RG) 
concepts and methods to 
density functional theory (DFT), with special emphasis
on the nuclear many-body problem.
Many of the topics covered are still in their infancy, so rather than
a complete review these lectures aim to provide an introduction
to the developing literature.


\section{EFT, RG, DFT for Fermion Many-body Systems}
\label{sec:1}

\subsection{Overview of Fermion Many-Body Systems}
\label{subsec:1}

There are a wide range of many-body systems featuring
fermion degrees of freedom.
These can be collections of ``fundamental'' fermions   
(electrons, quarks, \ldots)
or of composites each made of \emph{odd} number of fermions (e.g., protons). 
Here are some general categories and examples:

 \begin{enumerate}
  \item Isolated atoms or molecules, which contain 
     electrons interacting via the long-range (screened) Coulomb force.
  \item Bulk solid-state materials, such as 
     metals, insulators, semiconductors, superconductors, etc.
  \item Liquid $^3$He (a superfluid!).
  \item Cold fermionic atoms in (optical) traps
     (note that $^6$Li is a fermion but $^7$Li is a boson).
  \item Atomic nuclei.
  \item Neutron stars, which could mean neutron matter or
     color superconducting quark matter.
 \end{enumerate}
 
\noindent
DFT has been most 
widely applied to systems in categories 1 and 2.  We will
focus in these lectures on categories 4 and 5 (and neutrons stars
are treated in Thomas Sch\"afer's lectures). 

For our purposes, it won't matter whether the fermions are
structureless (as far as we know) such as quarks or electrons, or are
composites such as interacting atoms.  Note that an individual atom,
studied as a many-body system of electrons (with external potential
from the nucleus) is a fermion many-body system, while a collection of
these atoms might be either a boson or 
fermion many-body system \cite{Fermions}.

\begin{figure}[t]
\centering
 \includegraphics*[width=2.8in,angle=0.]{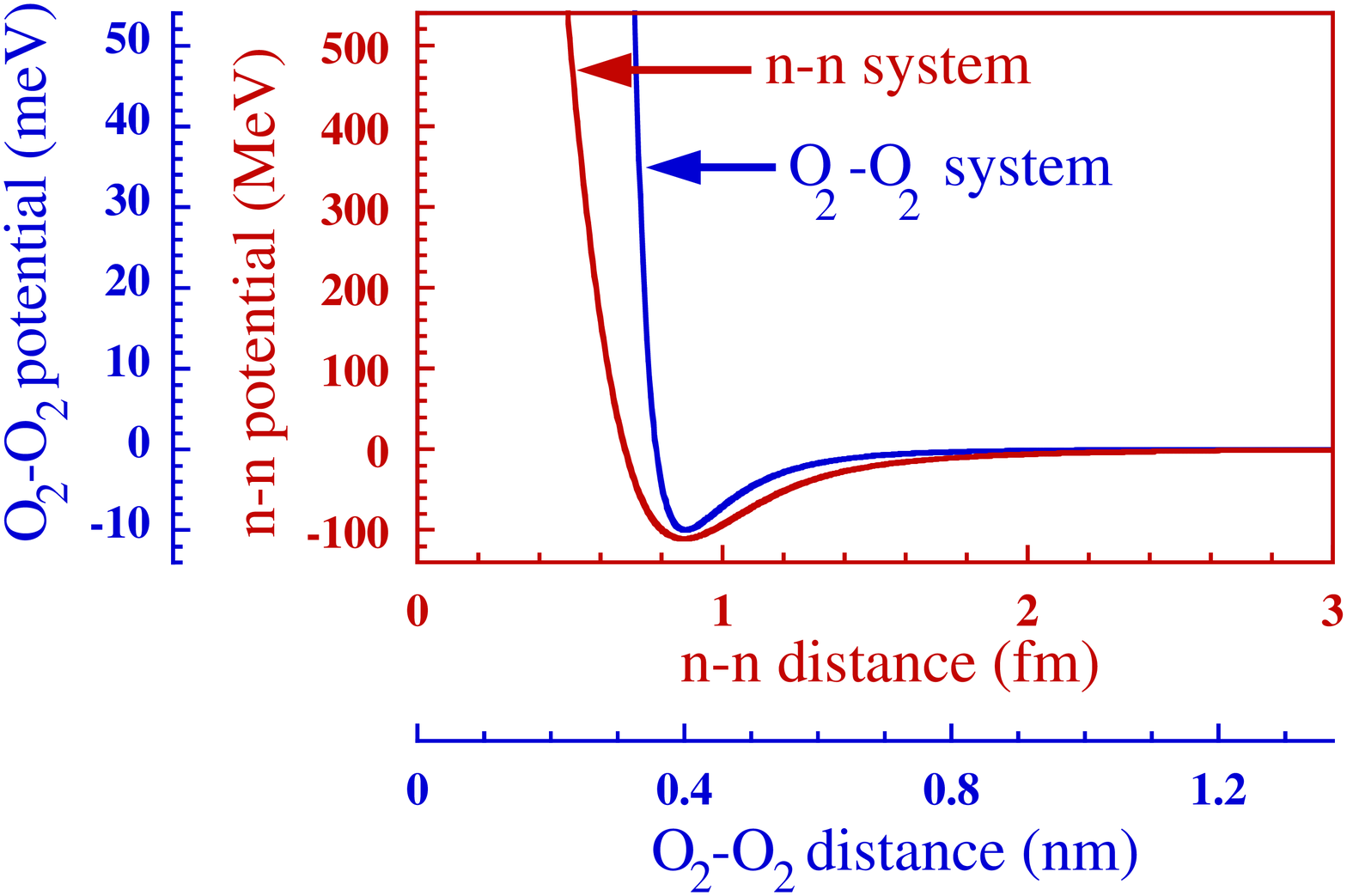}
 \caption{Molecule-molecule and nucleon-nucleon potentials 
 compared~\cite{Jacek}.}
 \label{fig:1}       
\end{figure}

If we label the axes appropriately in Fig.~\ref{fig:1} 
(note the many orders of
magnitude difference!), we see qualitative similarities between 
the central part of a
conventional
nucleon-nucleon 
(NN) potential and potentials between atoms or molecules (a
Lennard-Jones potential is shown). In particular, there is
midrange attraction and
strong short-range repulsion (or a ``hard core''). In the atomic case,
the attraction is van der Waals in nature from induced polarization
(which is why it falls off as $1/r^6$), and the rapid repulsion is from
when the electron clouds overlap. In the nuclear case, the long- and
mid-range attraction is mostly from one- and two-pion exchange.  The
hard core is often described in terms of vector-meson exchange but is
generally  phenomenological. The potential shown 
is the central part of the NN
interaction;  there are also important spin dependences and a non-central
tensor force \cite{Preston75}.

What might one expect qualitatively from a many-body system with such a
potential? Start with the 
equation of state of an ideal gas $PV = nRT$ ($n$ is
number of moles). Hard-core means ``excluded volume'' so $V
\rightarrow (V - nb)$ with $b$ constant. Attraction lowers the pressure on
the container, so we find: $P = \frac{nRT}{V-nb} -
\frac{an^2}{V^2}$.
The end result is a van der Waals equation of state: $(P +
\frac{an^2}{V^2})(V - nb) = nRT$, which has a liquid-gas phase
transition. This is consistent with nuclei! The hard core keeps
particles apart, leading to ``short-range'' correlations in
the wave function.   They make many-body
problems difficult but might seem to be essential in saturating the
nuclear liquid.  Both liquid helium and nuclei can be thought
of as liquid drops, whose radii scale with the number of particles $A$
to the 1/3 power: $R \sim r_0 A^{1/3}$. 
If a hard core of radius $c$ is responsible for saturation, one
can estimate that $0.55c \leq r_0 \leq 2.4c$ \cite{JACKSON92,JACKSON94}.
Liquid $^3$He has
$r_0 \sim 2.4\,\mbox{A} \sim c$, but for nuclei $r_0 \sim
1.1\,\mbox{fm} \sim 2.75c$.  This implies
that nuclear matter is dilute and ``delicately'' bound, which means
that an EFT expansion may be particularly useful.
Other questions we might ask are whether there are common (``universal'')
features of atomic and nuclear systems, and how to we relate 
the many-body physics to
 more fundamental underlying theories?
As we'll see, EFT will also help to address these questions. 

\begin{figure}[t]
\centering
 \includegraphics*[width=3.0in,angle=0.]{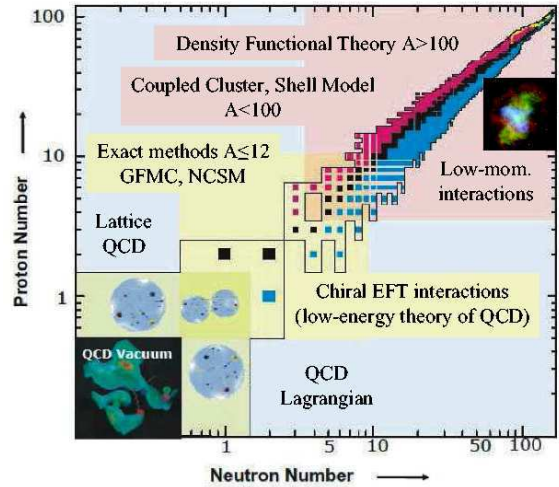}
 \caption{Overview of low-energy many-body nuclear systems \cite{Achim1}.}
 \label{fig:2}       
\end{figure}

In Fig.~\ref{fig:2}, we show the ``big picture'' of low-energy
nuclear physics, which features a specific
scientific goal of predicting properties of unstable nuclei
(the non-black squares), 
and a general scientific
goal, which is connecting the whole picture from quantum chromodynamics
(QCD) to superheavy nuclei in a systematic way with
robust predictions.
In principle, the nuclear part of the problem is simple:  given
internucleon potentials, just solve the many-body 
Schr\"odinger equation.  This turns out to be feasible only
for the smallest nuclei.

Why is this many-body problem so difficult computationally?  
Let's think about the many-body Schr\"odinger
wave function \cite{JoeC}.
How can we represent the wave function for an $A$-body nucleus?
Consider a $^8$Be ($Z=4$ protons, $N=4$ neutrons) wave function
with spin, isospin, and space components:
    \bequ
      |\Psi \rangle = \sum_{\sigma,\tau}
         \chi_\sigma \chi_\tau \phi({\bf R}) \; ,
         \quad \mbox{where ${\bf R}$ are the $3A$ spatial coordinates}
         \; ,
    \eequ
    \bequ
      \chi_\sigma = \downarrow_1 \uparrow_2 \cdots \downarrow_A
      \ \mbox{($2^A$ terms)\ } {= 256 \ \mbox{for\ } A=8} \; ,
    \eequ
    \bequ
      \chi_\tau = n_1 n_2 \cdots p_A
       \ \mbox{($\frac{A!}{N!Z!}$ terms)\ } {= 70\ \mbox{for\ } 
            ^8\mbox{Be}} \; .
    \eequ
So for $^8$Be there are 17,920 complex functions
in $3A-3 = 21$ dimensions!
Suppose for a nucleus of size 10\,fm you represent this with
a mesh spacing of 0.5\,fm.  You would need $10^{27}$ grid points!
Obviously we need to drastically reduce the necessary degrees
of freedom.

An extreme approximation to the full many-body wave function 
is the Hartree-Fock wave function, which is 
the best single Slater determinant in a variational sense:
   \beq
     | \Psi_{\rm HF} \rangle = \det\{\phi_i(\xvec), i=1\cdots A  \}
     \,, \quad \xvec = ({\bf r}, \sigma, \tau)
     \; .
   \eeq
The Hartree-Fock energy in the presence of an external potential
is~\cite{RINGSCHUCK}
  \beqa
   \langle  \Psi_{\rm HF} | \widehat H | \Psi_{\rm HF} \rangle
     &=&
   \nonumber \\ & & \hspace*{-.7in}
     \sum_{i=1}^A \frac{\hbar^2}{2M} \int\!\! d\xvec\,
        \bm{\nabla}\phi^\ast_i\bm{\cdot}\bm{\nabla}\phi_i
   {+ \frac12\sum_{i,j=1}^A
     \int\!\! d\xvec \! \int\!\! d\yvec \, 
      |\phi_i(\xvec)|^2 v(\xvec,\yvec) |\phi_j(\yvec)|^2} 
   \nonumber \\ & & \hspace*{-.7in}
   \null {- \frac12\sum_{i,j=1}^A
     \int\!\! d\xvec \! \int\!\! d\yvec \, 
      \phi^\ast_i(\xvec) \phi_i(\yvec) v(\xvec,\yvec) 
      \phi^\ast_j(\yvec)\phi_j(\xvec) } 
   \nonumber \\ & & \hspace*{-.7in}
    + \sum_{i=1}^A  \int\!\! d\yvec\, \Vext(\yvec) |\phi_j(\yvec)|^2
    \; . 
  \eeqa
We determine the $\phi_i$ by varying with fixed normalization:
      \beq
        \frac{\delta}{\delta \phi_i^\ast({\bf x})} 
	\Bigl(
	    \langle  \Psi_{\rm HF} | \widehat H | \Psi_{\rm HF} \rangle
           - \sum_{j=1}^A \epsilon_j 
	   \int\! d\yvec\, |\phi_j(\yvec)|^2
	\Bigr) = 0 \; .
      \eeq 
We solve this self-consistently, which is non-trivial because
the potential is non-local, but drastically simpler than
solving for the full wave function.
However, while Hartree-Fock is a reasonable starting point for
atoms, it is not for nuclear potentials of the form in
Fig.~\ref{fig:1} (e.g., note that the Argonne $v_{18}$ ``1st order'' curve
in Fig.~\ref{fig:5} is not even bound).

\subsection{Density Functional Theory}
\label{subsec:2}

An alternative to working with the many-body
wave function is density functional theory (DFT)
\cite{DREIZLER90,ARGAMAN00,fi03}, 
which as the 
name implies, has fermion densities as the fundamental ``variables''.
To date, the dominant application of DFT has been to the inhomogeneous
electron gas, which means 
interacting point electrons in the static potentials of 
atomic nuclei.
This has led to
``ab initio'' calculations of atoms, molecules, crystals,
surfaces, and more \cite{caveat1}.
DFT is founded on a theorem of Hohenberg and Kohn (HK):
There \emph{exists} an energy functional   $E_{\Vext}[\rho]$ 
of the density $\rho$ such that
 \beq
   E_{\Vext}[\rho] 
      = F_{\rm HK}[\rho] + \int\!d^3x\, \Vext({\bf x}) \rho({\bf x})
      \; ,
 \eeq
where
 $F_{\rm HK}$ is {\em universal\/}
 (the same for any external potential $\Vext$), the same for $H_2$ to DNA!
This is useful \emph{if} you can approximate the energy functional. 

\begin{figure}[t]
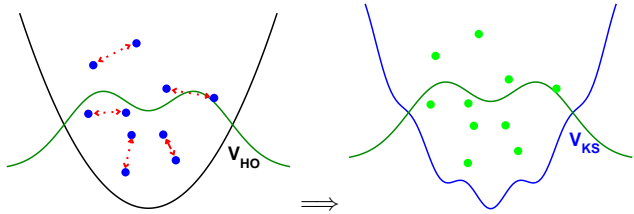

\centering
 \includegraphics*[width=1.5in,angle=0.]{kohn_sham1_new}
 $\Longrightarrow$
 \includegraphics*[width=1.5in,angle=0.]{kohn_sham2_new}
 \caption{Kohn-Sham DFT for a $\Vext = V_{\rm HO}$ harmonic trap.  On
 the left is the interacting system and on the right the Kohn-Sham
 system.  The density profile is the same in each.}
 \label{fig:3}       
\end{figure}

The general procedure is to introduce single-particle 
orbitals and to minimize the energy 
functional to obtain the ground-state energy $E_{gs}$ and density $\rho_{gs}$.
This is called Kohn-Sham DFT, and is illustrated schematically
in Fig.~\ref{fig:3}.
Here the interacting density for $A$ fermions 
in the external potential $V_{\rm HO}$ is equal (by construction)
to the non-interacting density in $V_{\rm KS}$.
Orbitals $\{\psi_i(\bfx)\}$
in the local potential $V_{\rm KS}([\rho],\bfx)$ are solutions to 
 \beq
   [-\bm{\nabla}^2/2m + V_{\rm KS}(\bfx)]\psi_i
   = \varepsilon_i\psi_i
\eeq
and determine the density
\beq    
     \rho(\bfx) = \sum_{i=1}^A |\psi_i(\bfx)|^2
 \eeq
(the sum is over the lowest $A$ states).
The magical Kohn-Sham potential $V_{\rm KS}([\rho],\bfx)$ 
is in turn determined from  $\delta E_{\Vext}[\rho]/\delta \rho(\bfx)$
(see below for an example).
Thus the Kohn-Sham orbitals depend on the potential, which depends
on the density, which depends on the orbitals, so we must 
solve self-consistently (for example, by iterating until convergence).

DFT for solid-state or molecular systems starts with the
HK free energy for an inhomogeneous electron gas~\cite{ARGAMAN00}:
  \beq
     \FHK[\densx] = \Fni[\densx] +  \frac{e^2}{2} \!
          \int\! d^3x\, d^3x' 
            \frac{\densx \rho({\mathbf x'})}{|{\bf x}-{\bf x'}|}
      + \Exc[\densx] \; .
  \eeq
Then $V_{\rm KS} = \Vext -e\phi + v_{\rm xc}$
with $v_{\rm xc}({\mathbf x}) = \delta E_{\rm xc}/\delta\densx$.
To calculate the Kohn-Sham kinetic energy ${\Fni[\densx]}$, find 
the normalized $\{\psi_i,\epsilon_i\}$ from
 \beq
   \Bigl( -\frac{\hbar^2}{2m}\nabla^2 + V_{\rm KS}({\mathbf x}) 
   \Bigr) \psi_i({\mathbf x}) = \epsilon_i \psi_i({\mathbf x})
 \eeq
and, with $\densx = \sum_{i=1}^A |\psi_i({\mathbf x})|^2$,
  \beq
    \Fni[\densx] = \sum_{i=1}^A \langle \psi_i | 
    -\frac{\hbar^2}{2m}\nabla_i^2 | \psi_i \rangle
    = \sum_{i=1}^A \epsilon_i - \int\!d^3x\, \densx V_{\rm
    KS}({\mathbf x}) \; .
  \eeq

In practice, the DFT is usually based on the
local density approximation (LDA):
$\Exc[\densx] \approx \int\! d^3x \,  {\cal E}_{{\rm xc}}(\densx)$ 
with  ${\cal E}_{{\rm xc}}(\rho)$ 
fit to a Monte Carlo calculation of the uniform electron gas.
For example, one  parametric formula for the energy density
is~\cite{ARGAMAN00} 
 \beq
   {\cal E}_{\rm xc}(\rho)/\rho =
     -0.458/r_s - 0.0666G(r_s/11.4) \; ,
 \eeq 
with
\beq
   G(x) = \frac12\left\{(1+x)^3 \log(1+x^{-1}) - x^2 + \frac12x -
      \frac13 \right\}  \; . 
\eeq 
This is just like a simple Hartree approach with the additional
potential:
    \beq v_{\rm xc}({\bf x}) = \left.\frac{d[{\cal E}_{\rm
    xc}(\rho)]}{d\rho}\right|_{\rho=\densx} \; .
    \eeq
The LDA is improved with the Generalized Gradient Approximation (GGA),
such as the van Leeuwen--Baerends GGA~\cite{ARGAMAN00}, 
  \beq
    v_{\rm xc}({\bf r})  = -\beta \rho^{1/3}({\bf r})
      \frac{x^2({\bf r})}{1 + 3\beta x({\bf r})
        \sinh^{-1}(x({\bf r}))}
  \eeq
with $x = |\nabla \rho| / \rho^{4/3}$.
For these Coulomb systems, Hartree-Fock is generally a good starting point, 
 DFT/LDA is better, and DFT/GGA is better still.
 
There are some concerns, however, about DFT. 
Here are some quotes from the DFT literature that help motivate
the application of EFT to DFT:

  \begin{quote}From \emph{A Chemist's Guide to DFT} 
      \cite{KOCH2000}:
    ``To many, the success of DFT appeared somewhat miraculous,
    and maybe even unjust and unjustified.  Unjust in view of the easy
    achievement of accuracy that was so hard to come by in the wave
    function based methods.  And unjustified it appeared to those who
    doubted the soundness of the theoretical foundations. '' 
  \end{quote}

  \begin{quote}From \emph{Density Functional Theory} \cite{ARGAMAN00}:
    ``It is important to stress that all practical
    applications of DFT rest on essentially uncontrolled
    approximations, such as the LDA \ldots''
  \end{quote}

   \begin{quote}{From \emph{Meta-Generalized Gradient Approximation}
       \cite{PERDEW99}}
     ``Some say that `there is no systematic way to construct
     density functional approximations.'  But there are more or
     less systematic ways, and the approach taken \ldots here is
     one of the former.''
  \end{quote}
  
\noindent
Thus, a microscopic, controlled, and systematic approach to DFT
would be welcome.  

We end this section with a preview of DFT as an effective
action~\cite{Polonyi2001}.
Recall ordinary thermodynamics with $N$ particles at $T=0$.
The thermodynamic potential is related to the partition
function, with the chemical potential $\mu$ acting as a  
source to change $N = \langle \widehat N\rangle$,
 \beq
    \Omega(\mu) = -kT \ln Z(\mu)
    \qquad
    \mbox{and}
    \qquad
    N = -\left(\frac{\partial\Omega}{\partial\mu}
	 \right)_{TV} \; .
 \eeq
If we \emph{invert} to find $\mu(N)$ and apply a Legendre transform,
we obtain
 \beq
   F(N) = \Omega(\mu(N)) + \mu(N) N \; .
 \eeq
This is our (free) energy function of the particle number, which is 
analogous to the DFT energy functional of the density.
Indeed, if we
generalize to a spatially dependent chemical potential $J(\xvec)$, then
  \beq
    Z(\mu) \longrightarrow Z[J(\xvec)]
    \qquad
    \mbox{and}
    \qquad
    \mu N = \mu\int\psi^\dagger\psi
      \longrightarrow
      \int J(\xvec)\psi^\dagger\psi(\xvec) \; .
  \eeq
Now Legendre transform from $\ln Z[J(\xvec)]$ to 
$\Gamma[\rho(\xvec)]$, where $\rho = \langle\psi^\dagger\psi\rangle_J$,
and we have DFT with $\Gamma$ simply proportional to 
the energy functional! 


\subsection{DFT for Nuclei: EFT and RG Approaches}
\label{subsec:3}

\begin{figure}[t]
\centering
 \includegraphics*[width=2.6in,angle=-90.]{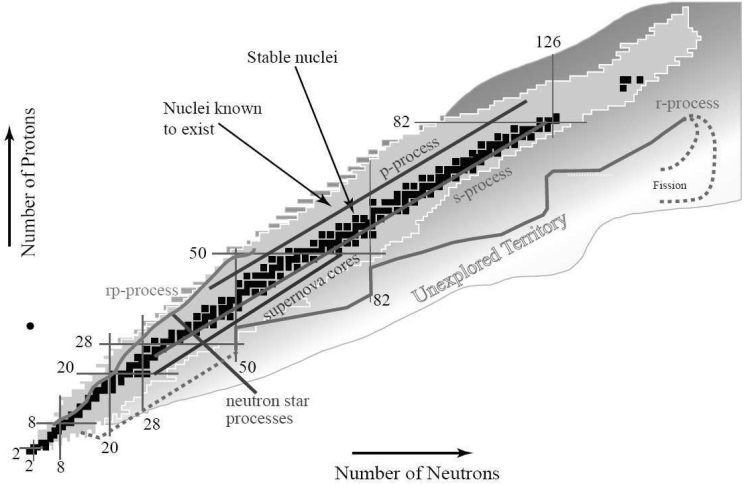}
 \caption{Table of the nuclides.}
 \label{fig:4}       
\end{figure}
Figure~\ref{fig:4} shows the table of the nuclides, labeled by the
total numbers of protons and neutrons.  For example, $^{208}$Pb is
found at the intersection of the horizontal line labeled ``82''
(protons) and the vertical line labeled ``126'' (neutrons).
Stable nuclei are in black, known in pink.  
Here are some basic
nuclear physics questions (which \emph{everyone} should know
how to answer):
   \bi
    \item  Why is the slope of the black region less than a 45 degree
     angle once it is past $Z = N = 20$ or so?
    \item  How do the binding energies of the nuclei in black compare?
     (E.g., do they vary over a wide range?  Do they have a regular
     pattern?)
    \item  What happens to the binding energy as you move perpendicular
     to the black line?
   \item   What is the difference between being unstable and unbound?
     What are the drip lines?
    \ei
(See \cite{Preston75,RINGSCHUCK} for nuclear physics background.)    
The nucleosynthesis r-process lives largely in ``Unexplored Territory."
Radioactive beam facilities (existing and proposed) 
address this physics, as well as
exotic nuclei such as halo nuclei and many other phenomena.
As one gets further from stability,
the importance of pairing grows, which highlights
a difference between nuclear many-body physics
and the physics of Coulomb systems, like atoms and molecules.

\begin{figure}[t]
\centering
 \includegraphics*[width=2.8in,angle=0.]{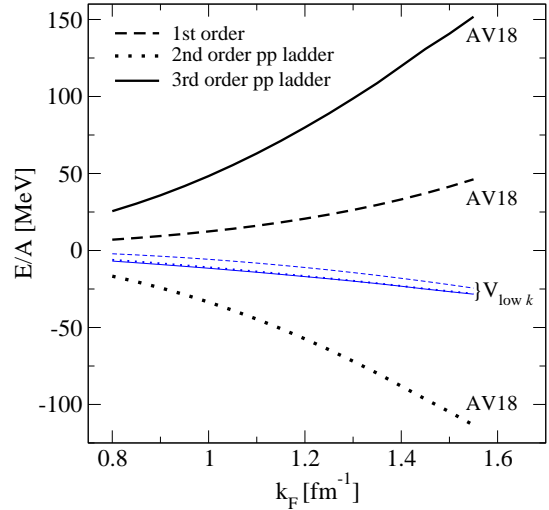}
 \caption{Nuclear matter in perturbation theory for a conventional NN potential
 (Argonne $v_{18}$) and a low-momentum potential \cite{Bogner_nucmatt}.}
 \label{fig:5}       
\end{figure}

Let's try solving 
nuclear matter in low-order perturbation theory.
One expects a minimum in the energy/particle ($E/A$) versus
density (here the Fermi momentum $\kf \propto \rho^{1/3}$)
at about $\kf \sim 1.35\,\mbox{fm}^{-1}$ and $E/A \sim -16\,$MeV.
The standard Argonne $v_{18}$ potential \cite{AV18} is used
in Fig.~\ref{fig:5}, with 
``Brueckner ladder'' contributions shown order-by-order.
First order is Hartree-Fock and it is not even bound!
The repulsive core means that the series diverges badly.

But there is an energy density functional approach that seems
to be based on Hartree-Fock (HF), which works quite well throughout
Fig.~\ref{fig:4}.
This is the phenomenological Skyrme HF
approach \cite{RINGSCHUCK,VB72,Dobaczewski:2001ed,BENDER2003}.
Recall from our earlier Hartree-Fock discussion that solving 
self-consistently is somewhat tricky because the potential
is non-local.
It would be much simpler if 
$v(\xvec,\yvec) \propto \delta(\xvec-\yvec)$.
This is the case with the Skyrme interaction 
$V_2^{\rm Skyrme} + V_3^{\rm Skyrme}$, where 
$\langle {\bf k} | V_2^{\rm Skyrme} | {\bf k'} \rangle
 = t_0 + \frac{1}{2}t_1({\bf k}^2+{\bf k'}^2) + t_2 {\bf k}\cdot{\bf k'}
 + iW_0(\mathbf{\sigma}_1+\mathbf{\sigma}_2)\cdot{\bf k}\times{\bf k'}$.
This motivates the Skyrme energy density functional (for
$N=Z$)~\cite{RINGSCHUCK}:
  \beqa
    {\cal E}[{\rho},{\tau},{\bf J}] 
      &\!\!=\!\!& {1\over 2M}{\tau} + {3\over 8} t_0 {\rho^2}
    + {1\over 16} t_3 {\rho^{2+\alpha}}
   + {1\over 16}(3 t_1 + 5 t_2) {\rho} {\tau}  \nonumber
    \\ & & \hspace*{-.1in}\null
    + {1\over 64} (9t_1 - 5t_2) {(\nabla \rho)^2}  
    - {3\over 4} W_0 {\rho} \nabla\cdot{\bf J}
    + {1\over 32}(t_1-t_2) {\bf J}^2 \; ,  
  \eeqa
where ${\rho(\xvec) = \sum_i |\phi_i(\xvec)|^2}$
and ${\tau(\xvec) = \sum_i |\nabla\phi_i(\xvec)|^2}$ (see 
\cite{RINGSCHUCK} for the ${\bf J}(\xvec)$ formula).
We minimize $E = \int\!d{\bf x}\,{\cal E}[\rho,\tau,{\bf J}] $
by varying the (normalized) $\phi_i$'s,
  \beq
   \Bigl( - \mathbf{\nabla} {\frac{1}{2M^*({\bf x})}} \mathbf{\nabla} 
     + {U({\bf x})} + 
     \frac{3}{4}W_0\mathbf{\nabla} \rho\cdot \frac{1}{i}
     \mathbf{\nabla} \times \bf{\sigma} \Bigr)\,  
     \phi_{i}({\bf x}) =
   \epsilon_{i}\,\phi_{i}({\bf x}) \; ,
  \eeq
with
 $U = \frac{3}{4}t_0\rho 
      + (\frac{3}{16}t_1 + \frac{5}{16}t_2)\tau
    + \cdots$ and  
   $ \frac{1}{2M^{*}({\bf x})}
     =\frac{1}{2M} + ({3\over {16}}t_1
     +{5\over {16}}t_2)\rho$.
One iterates until the $\phi_i$'s and $\epsilon_i$'s
are self-consistent.
    
While phenomenologically successful,
there are many questions and possible criticisms of Skyrme HF.    
Typical [e.g., SkIII] model parameters are:
     $t_0=-1129$,  $t_1=395$, $t_2=-95$, $t_3=14000$, $W_0=120$
(in units of MeV-fm$^n$). 
These seem large; is there an expansion parameter?
Where does $\rho^{2+\alpha}$ come from?
Why not $\rho^{2+\beta}$?
Is this just parameter fitting?  A famous quote from 
von Neumann (via Fermi via Dyson) says: 
         \emph{ ``With four parameters I can fit an elephant, 
         and with five I can make him wiggle his trunk.''}   
One might also say that Skyrme HF is only a mean-field model,
which is too simple to accommodate NN correlations.
How do we improve the approach?  Is pairing treated
correctly?
How does Skyrme HF relate to NN (and NNN) forces?
And so on.
There is also the observation that Skyrme functionals
works well where there is already data, but elsewhere fails to
give consistent predictions (and the theoretical error bar is
unknown).

Rather than focus on the Skyrme \emph{interaction}, we consider the
Skyrme energy functional as an approximate DFT functional.
(Note: this is the viewpoint of modern practitioners.)
Our master plan is to use EFT and renormalization group (RG)
methods to elevate something close to Skyrme HF to a full DFT treatment.
We want to use EFT and RG to provide a systematic input potential, 
including three-body potentials,
and to generate systematically improved energy functionals.
At the same time, we want to be able to estimate theoretical errors,
so that extrapolation is under control.

\begin{table} 
 \caption{(Nuclear) Many-Body Physics: {``Old'' Approach} }
 \renewcommand{\tabcolsep}{10pt}
 \centering 
 \begin{tabular}{l|l}
   \hline
   {\pboxb{One Hamiltonian for all problems and
   energy/length scales (not QCD!)}} 
   & 
   {\pboxb{For nuclear structure, protons and
    neutrons with a \emph{local} potential~\cite{Table1} fit to NN  data }}
   \\[16pt]\hline
   {\pboxb{Find the ``best'' potential}}
   &
   {\pboxb{NN potential with
     $\chi^2/\mbox{dof} \approx 1$ 
     up to $\sim 300\,$MeV energy~\cite{Table2}}}
   \\[12pt]\hline
   {\pboxb{Two-body data may be sufficient;
     many-body forces as last resort}}
   &
   {\pboxb{Let phenomenology dictate
    whether three-body forces are needed (answer: yes!~\cite{Table3})}}
   \\[16pt]\hline
   {\pboxb{Avoid (hide) divergences}}
   &
   {\pboxb{Add ``form factor'' to suppress high-energy
       intermediate states; don't consider cutoff dependence~\cite{Table4}}}
   \\[16pt]\hline
   {\pboxb{Choose approximations by ``art''}}
   &
   {\pboxb{Use physical arguments (often handwaving)
      to justify the subset of diagrams used~\cite{Table5}}}
   \\[16pt] \hline
 \end{tabular}
 \label{tab:1}
\end{table}

This is a relatively new and different approach.  
In Table~\ref{tab:1} we 
summarize aspects of the ``traditional''
approach to the (nuclear) many-body problem that are being challenged by
the EFT approach.
There are many continuing successes of
conventional many-body approaches.  The idea is not to prove standard
methods wrong but to highlight where new
insight can be provided.  
For each ``old'' item in this table (see endnotes for further
explanations), we'll have a ``new''
perspective from EFT (see Table~\ref{tab:2}).

\subsection{EFT Analogies}
\label{subsec:4}

From an effective field theory we desire
systematic calculations with error estimates.  We want them to
be robust and model independent, which will
enable reliable extrapolation.  To help understand
how this is accomplished, we can explore useful
analogies between EFT and sophisticated numerical analysis.
 \begin{itemize}
   \item Naive error analysis: 
     pick a method and reduce the mesh size
     (e.g., increase grid points) 
   until the error is ``acceptable''. 
   Sophisticated error analysis: understand scaling and sources
   of error  (e.g., algorithm vs.\ round-off errors).
   \emph{Does it work as well as it should?}
   \item 
   Representation dependence means that not all 
   ways of calculating are equally effective!
   \item Reliable extrapolation requires 
   completeness of an expansion basis.  An EFT lagrangian
   provides the analog of a complete basis.
 \end{itemize}
Note: quantum mechanics makes EFT trickier than ``classical''
numerical analysis (see Lepage's lectures)!

\begin{figure}[t]
\centering
 \includegraphics*[width=2.8in,angle=0.]{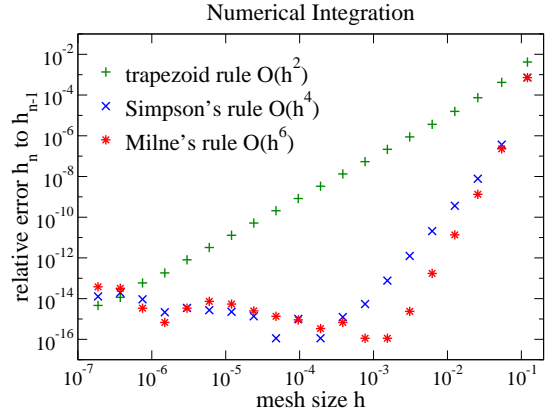}
 \caption{Error plots in numerical integration.}
 \label{fig:5a}       
\end{figure}

\begin{figure}[t]
\centering
 \includegraphics*[width=2.8in,angle=0.]{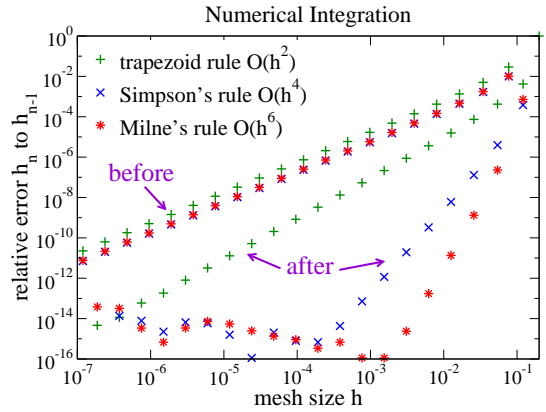}
 \caption{The representation can make a difference:
 errors for (\ref{eq:int1}) vs.\ (\ref{eq:int2}).}
 \label{fig:6}       
\end{figure}
    
Consider the numerical calculation of an integral using equal-spaced
integration rules of increasing sophistication.     
How do the \emph{numerical} errors behave?
Log-log plots of the relative error against a parameter such as the
mesh size are very helpful;  a straight line indicates a power law
 and the exponent is read off from the slope.
An example is shown in Fig.~\ref{fig:5a}.  Similar plots can be made
for order-by-order EFT calculations, as described in Peter Lepage's lectures. 
Just like computer math is not equivalent
to ordinary math, EFT is not the same
as a theory applicable at all energies.  It breaks down at high
resolution, as the numerical calculations break down and degrade
because of round-off errors at small mesh sizes.  
(Note: Don't carry this analogy to extremes!)
Next consider an elliptic integral:
 \beq
   \int_0^1\! \sqrt{(1-x^2)(2-x)} \, dx
   \label{eq:int1}
 \eeq
with errors plotted in Fig.~\ref{fig:6}.  Something is wrong; the
errors do not behave as expected.
However, after a simple transformation:
     \beq        
       \int_0^{\pi/2}\! \sin^2 y \, \sqrt{2-\cos y}\,  dy \; .
       \label{eq:int2}
     \eeq
As seen in Fig.~\ref{fig:6}, the transformation makes a big difference!
If you have freedom to change the representation, you can make a big
difference in the ability to calculate accurately and easily.  This is
a moral we will apply when using EFT and RG methods to many-body problems.
     
\subsection{Principles of Effective Low-Energy Theories}   

 There are some basic physics principles underlying \emph{any} low-energy
effective model or theory.
A high-energy, short-wavelength probe sees detail down to scales comparable
to the wavelength.  Thus, high-energy
electron scattering at Jefferson Lab resolves the quark substructure
of protons and neutrons in a nucleus. 
But at lower energies, details are not resolved,
and one can replace short-distance structure with something simpler, as in
a multipole expansion of a complicated charge or current distribution.
So it is not necessary to do full QCD to do strong interaction physics
at low energies, we can replace quarks and gluons by neutrons
and protons (and pions).
Chiral effective field theory is a systematic approach to carrying out
this program using a local lagrangian framework.  

It is not obvious that working at low resolution 
will work in quantum mechanics as
it does for pixels or point dots or the classical multipole expansion,
because \emph{virtual} states can have high energies that are not, in
reality, simple.  But renormalization theory says it can be done!  (See
Lepage's lectures and \cite{LEPAGEren,Lepage}.)  Note that
this doesn't mean that we are \emph{insensitive} to all short-distance
details, only that their effects at low energies
can be accounted for in a simple way.
    
We can use the possibility of changing the resolution scale to
change the ``perturbativeness'' of nuclei.    
There are several sources of nonperturbative physics for nucleon-nucleon
interactions: 
\begin{enumerate}    
 \item
  Strong short-range repulsion (``hard core'').
 \item Iterated tensor ($S_{12}$) interactions (e.g., from pion
 exchange).
 \item
 Near zero-energy bound states.
\end{enumerate}
In Coulomb DFT, Hartree-Fock gives the dominate contribution to the
energy,
and correlations are small corrections.  This may be why
DFT works.  In contrast, for NN interactions,
correlations $\gg$ HF; does this mean DFT fails??
However, the first two sources depend on 
the resolution (i.e., the cutoff of high-energy physics), and
the third one is affected by Pauli blocking.
Thus we might use the freedom of low-energy theories to simplify
calculations.

\begin{figure}[t]
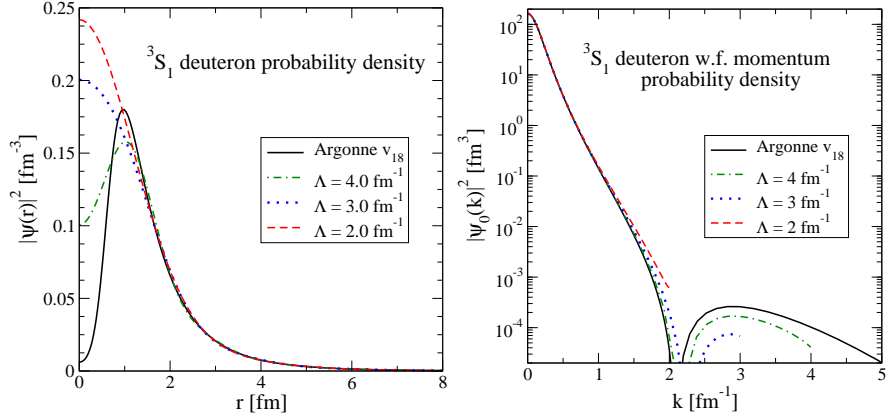

\centering
       \includegraphics*[width=2.3in]{3s1rspacewfsq_rev2}
       \includegraphics*[width=2.25in]{3s1kspacewfsq_rev2}
 \caption{The deuteron probability density
 at different resolutions (as indicated
  by the sharp momentum cutoff $\Lambda$).}
 \label{fig:7}       
\end{figure}

\begin{figure}[t]
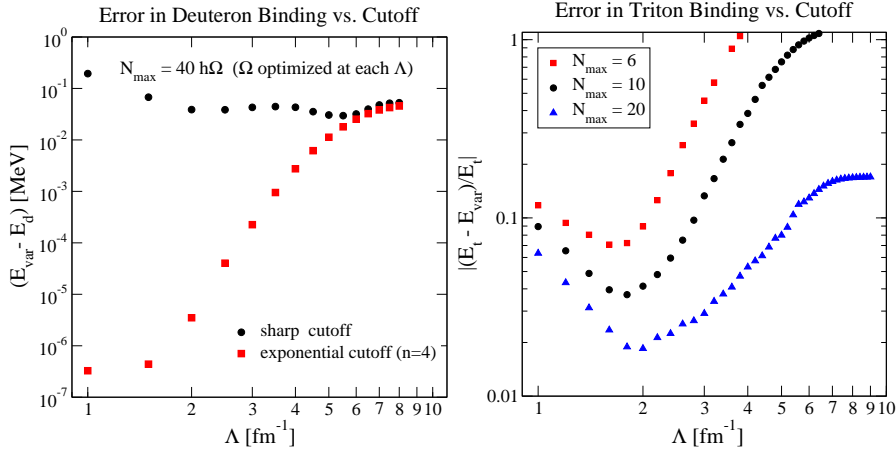

\centering
      \includegraphics*[width=2.3in]{deuteron_var_ho}
      \includegraphics*[width=2.3in]{triton_talk}
 \caption{Variational calculations at different resolutions
 \cite{Bogner:2005fn,Bogner:2006a}.}
 \label{fig:8}       
\end{figure}
     
We can see the impact of different resolutions on the deuteron
wave function in Fig.~\ref{fig:7}.     
The repulsive core leads to short-distance suppression 
and important high-momentum (small $\lambda$) components.
This makes the many-body problem complicated.       
In contrast, potentials evolved by renormalization group (RG) methods to
low momentum, generically called 
$V_{{\rm low\,}k}$ \cite{Vlowk1,Vlowk2,VlowkRG},
have much simpler wave functions!
(See Andreas Nogga's lectures for more detail on such potentials.)
We note that
chiral EFT potentials are naturally low-momentum potentials, but
lowering their cutoffs further is generally advantageous.
The consequence of lower resolution
for variational calculations is illustrated
in Fig.~\ref{fig:8} \cite{Bogner:2005fn,Bogner:2006a}.  
Note that the improvement 
for the deuteron comes with smooth (exponential)
cutoffs \cite{Bogner:2006tw},
which is another example of how the representation can make a difference.

\begin{figure}[t]
\centering
      \includegraphics*[width=2.3in]{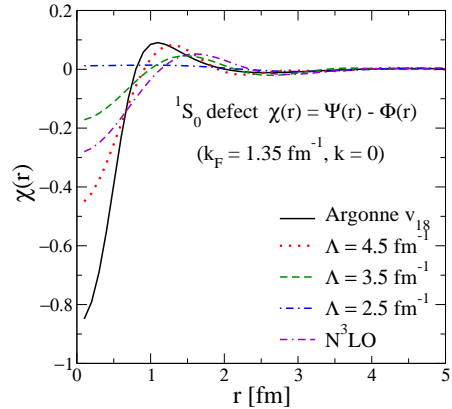}
 \caption{Two-body correlations in nuclear matter.}
 \label{fig:9}       
\end{figure}

These observations carry over to nuclear matter as well, as seen in
Fig.~\ref{fig:9} (although the effect with lowered cutoff in the 
important $^3$S$_1$ channel is less dramatic). 
In medium,
the \emph{phase space} in the pp-channel
is strongly suppressed:
        \beq \int_{\kf}^{{\infty}} q^2\,dq
        \frac{V_{NN}(k',q)V_{NN}(q,k)}
             {k^2-q^2}
        \eeq
compared to
        \beq \int_{\kf}^{{\Lambda}} q^2\,dq
        \frac{V_{{\rm low\,}k}(k',q)V_{{\rm low\,}k}(q,k)}
             {k^2-q^2}
	     \; .
        \eeq
(In addition, the potential itself gets smaller in magnitude
in the integration region.) 
This tames the hard core, tensor force, \emph{and} the bound state.
If we return to Fig.~\ref{fig:5}, we see the consequence, which
is very rapid convergence (by 2nd order) of the in-medium T-matrix
for $V_{{\rm low\,}k}$ \cite{Bogner_nucmatt}.
But there is no saturation in sight!

There were active attempts to transform away hard cores and soften
the tensor interaction in the late 
sixties and early seventies~\cite{KERMAN66,KERMAN67,KERMAN73}.
But the requiem for soft potentials was given by Bethe (1971):
\begin{quote}
  ``Very soft potentials must be excluded because they do not give
  saturation; they give too much binding and too high density.  In
  particular, a substantial tensor force is required.''
\end{quote}     
The next 30+ years were spent
trying to solve accurately with ``hard'' potentials.
But the story is not complete: three-nucleon forces (3NF)!
When they are added consistently, we have the advantages of
soft potentials while answering Bethe's criticism. 

\begin{figure}[t]
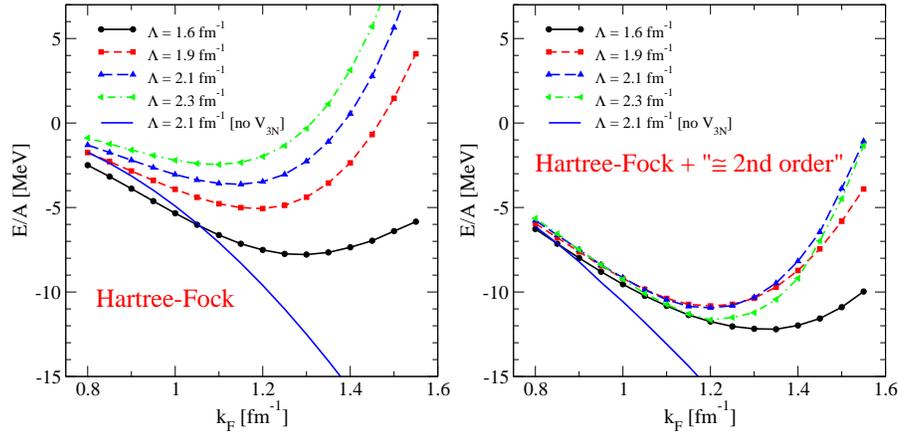

\centering
      \includegraphics*[width=2.3in]{paper_hf_2+3_rev4}
      \includegraphics*[width=2.3in]{paper_2ndorder_mstar_2+3_fullP_rev3}
 \caption{Nuclear matter energy per particle \cite{Bogner_nucmatt}.}
 \label{fig:10}       
\end{figure}

Ideally we would 
start with chiral NN + 3NF EFT interactions and then
evolve downward in $\Lambda$.
What is possible now is
to run the NN and \emph{fit} 3NF EFT at each $\Lambda$ \cite{Vlowk3N}.
The consequence is shown in Fig.~\ref{fig:10}.
There is saturation even at the Hartree-Fock level,
which is now driven by the three-body force.  
At second order, the cutoff dependence is greatly reduced and the
minimum moves closer to the empirical point.
One might worry
that the three-body force is unnaturally large, but it is consistent
with EFT power counting.
Excellent results are also found for neutron matter.
While there is much to be done, these results are very encouraging
and motivate a microscopic DFT for nuclei.
     
\subsection{Summary}

In summary, there is a new attitude for many-body physics
inspired by effective field
theory.  Bethe wrote about the nuclear case:
\begin{quote}
``The theory must be such that it can deal with any
nucleon-nucleon (NN) force, including hard or `soft' core, tensor
forces, and other complications.  It ought not to be necessary to tailor
the NN force for the sake of making the computation of nuclear matter
(or finite nuclei) easier, but the force should be chosen on the basis
of NN experiments (and possibly subsidiary experimental evidence, like
the binding energy of H$^3$).''
\end{quote}
The new attitude is to instead  seek to make the problem easier.
It's like the old vaudeville joke about a doctor and his patient: \\
{{\bf Patient:} Doctor, doctor, it hurts when I do this!} \\
{{\bf Doctor:} Then don't do that.} \\
We also follow Weinberg's Third Law of Progress in Theoretical Physics: 
   \emph{``You may use any degrees of freedom you like to
        describe a physical system,  but
          if you use the wrong ones, you'll be sorry!''}
We conclude with a new table (Table~\ref{tab:2}) 
of many-body physics, contrasting the
old and the new approaches.

\begin{table} 
 \caption{(Nuclear) Many-Body Physics: {``Old''} vs.\ {``New''}}
 \renewcommand{\tabcolsep}{10pt}
 \centering 
 \begin{tabular}{l|l}
   \hline
   {\pboxb{One Hamiltonian for all problems and
   energy/length scales}} 
   & 
   {\pboxb{\emph{Infinite} \# of low-energy potentials;
   different resolutions \Lra different dof's and Hamiltonians }}
   \\[10pt]\hline
   {\pboxb{Find the ``best'' potential}}
   &
   {\pboxb{There \emph{is} no best potential\\ \Lra
      use a convenient one! }}
   \\[10pt]\hline
   {\pboxb{Two-body data may be sufficient;
     many-body forces as last resort}}
   &
   {\pboxb{Many-body data needed and
      many-body forces \emph{inevitable}}}
   \\[10pt]\hline
   {\pboxb{Avoid divergences}}
   &
   {\pboxb{Exploit divergences (cutoff dependence as tool)}}
   \\[10pt]\hline
   {\pboxb{Choose diagrams by ``art''}}
   &
   {\pboxb{Power counting determines diagrams and 
      truncation error}}
   \\ \hline
 \end{tabular}
 \label{tab:2}
\end{table}
     

\section{EFT/DFT for Dilute Fermi Systems}
\label{sec:2}

\subsection{Thermodynamics Interpretation of DFT}

As an analogy,
consider a system of spins $S_i$ on a lattice
with interaction strength $g$~\cite{NEGELE88}.
The partition function has all the information about the energy and 
magnetization of the system:
 \beq
   {\cal Z} = {\rm Tr\,} e^{-\beta g \sum_{\{i,j\}} S_i S_j} 
   \; .
 \eeq
The magnetization $M$ is
 \beqa
   M &\!\!=\!\!& \Bigl\langle \sum_i S_i \Bigr\rangle
   \\ &\!\!=\!\!&
    \frac{1}{\cal Z} \, 
   {\rm Tr\,}\biggl[ 
     \Bigl(\sum_i S_i \Bigr)
     e^{-\beta g \sum_{\{i,j\}} S_i S_j}
    \biggr] \;.
 \eeqa
Now add an external magnetic probe source $H$.
The source adjusts the spin configurations near the ground state,
 \beq
     {\cal Z}[H] = 
     e^{-\beta F[H]} = {\rm Tr\,} 
       e^{-\beta (g \sum_{\{i,j\}} S_i S_j {- H\sum_i S_i})} \;.
 \eeq
Variations of the source yield the magnetization
 \beq
   M = \Bigl\langle \sum_i S_i \Bigr\rangle_H 
       = -\frac{\partial F[H]}{\partial H}
       \; ,
 \eeq
where $F[H]$ is the Helmholtz free energy.
For the ground state,
we set $H=0$ (or equal to a real external source) at the end.

Now if we find $H[M]$ by inverting $M[H]$,
  \beq
    M = \Bigl\langle \sum_i S_i \Bigr\rangle_H 
      = -\frac{\partial F[H]}{\partial H}
      \; ,
  \eeq
we can Legendre transform to the {Gibbs} free energy
  \beq
    \Gamma[M] = F[H] + H\, M \; .
  \eeq 
Then the ground-state magnetization $M_{\rm gs}$ follows by minimizing
$\Gamma[M]$:
 \beq
   H = \frac{\partial\Gamma[M]}{\partial M}
   \longrightarrow
     \left.\frac{\partial\Gamma[M]}{\partial M}\right|_{M_{\rm gs}}
     = 0 \; .     
 \eeq
Thus, we have a function of $M$
(or functional if $H$ is inhomogeneous) that
is minimized at the ground-state free energy and magnetization. 

DFT has an analogous structure as an effective action~\cite{Polonyi2001}.
An effective action is generically the
Legendre transform of a generating functional
with an external source (or sources).  For DFT, we use a source to adjust
the density instead of the magnetization.
The 
partition function in the presence of $J(x)$ coupled to the density is
(we'll use a schematic notation here):
 \beq
     {\cal Z}[J] = 
     e^{-W[J]} \sim {\rm Tr\,} 
       e^{-\beta (\Hhat + J\,\widehat \rho) }
     \longrightarrow \int\!{\cal D}[\psi^\dagger]{\cal D}[\psi]
     \,e^{-\int [{\cal L} + J\,\psi^\dagger\psi]}
     \; . 
 \eeq
The density $\rho(x)$ in the presence of $J(x)$ is
(keep in mind that we want $J=0$ eventually), 
 \beq
   \rho(x) \equiv \langle \widehat \rho(x) \rangle_{J}
    = \frac{\delta W[J]}{\delta J(x)}
    \; .
 \eeq  
After inverting 
to find $J[\rho]$, we Legendre transform from $J$ to
$\rho$:
 \beq
    \Gamma[\rho] = W[J] - \int\! J\, \rho
    \quad
    \mbox{and}
    \quad
    J(x) = -\frac{\delta \Gamma[\rho]}{\delta \rho(x)}
    \; .
 \eeq 

Now consider the partition function in the zero-temperature limit of
a Hamiltonian with time-independent source $J({\bf x})$
\cite{ZINNJUSTIN,Minkowski}:
 \beq
   { \Hhat}(J) = {\Hhat}  + \int\!  J\, \psi^\dagger\psi \; .
 \eeq
\emph{If} the ground state is isolated (and bounded from below),
 \beq
   e^{-\beta \Hhat} = e^{-\beta E_0}
     \left[
       | 0 \rangle \langle 0 |
	+ {\cal O}\bigl(e^{-\beta (E_1 - E_0)}\bigr)
     \right]
     \; .
 \eeq
 As  $\beta \rightarrow \infty$, ${\cal Z}[J]$ yields the
ground state of ${\Hhat}(J)$ with energy $E_0(J)$:
  \beq     
  { E_0(J)} = \lim_{\beta\rightarrow \infty} -\frac{1}{\beta} \log 
  {\cal Z}[J]
    = \frac{1}{\beta}W[J] \; .
 \eeq
Substitute and separate out the pieces:
 \beq
 E_0(J) = \langle {\Hhat }(J) \rangle_J 
      = \langle \Hhat \rangle_J 
	 + \int\! J \langle \psi^\dagger\psi \rangle_J
      = \langle {\Hhat} \rangle_J + \int\! J\, \rho(J)
      \; .
 \eeq
Rearranging,
the expectation value of ${\Hhat}$ in the ground state
generated by $J[\rho]$ is 
 \beq
   { \langle {\Hhat} \rangle_J} =  E_0(J) - \int J\, \rho
    = \frac{1}{\beta}\Gamma[\rho]
    \; .
 \eeq
Let's put it all together:
   \beq
      \frac{1}{\beta}\Gamma[\rho] = \langle {\Hhat} \rangle_J 
     \stackrel{J\rightarrow 0}{\longrightarrow}
       E_0
     \quad \mbox{and} \quad
    J(x) = -\frac{\delta \Gamma[\rho]}{\delta \rho(x)}
     \stackrel{J\rightarrow 0}{\longrightarrow}
    {
    \left.\frac{\delta \Gamma[\rho]}{\delta \rho(x)}
            \right|_{\rho_{\rm gs}(\bfx)} =0 }
	    \; .
   \eeq
So for static $\rho(\bfx)$, $\Gamma[\rho]$ is proportional to 
the DFT energy functional $F_{\rm HK}$!  
Furthermore, the true ground state (with $J=0$) is a  variational
minimum,
so additional sources should be better than just
one source coupling to the density (as we'll consider below).
The simple, universal dependence on external potential follows
directly in this formalism:
    \beq
      \Gamma_{v}[\rho] = W_{v}[J] - \int\! J\,\rho
       = W_{v=0}[J+v] - \int\! [(J+v)-v]\, \rho
       = \Gamma_{v=0}[\rho] + \int\! v\,\rho
       \; .
    \eeq
[Note: the    
functionals will change with resolution or field redefinitions; 
only stationary points are observables.]

There are a number of paths to the DFT effective action:
  \be
    \item Follow the usual Coulomb Kohn-Sham DFT by 
      calculating the uniform system as function of density,
      which yields an    
      LDA (``local density approximation'') 
      functional and a standard Kohn-Sham procedure. 
      Improve the functional with a semi-empirical gradient expansion.
    \item Derive the functional with an RG approach~\cite{Schwenk:2004hm}. 
    \item Use the auxiliary field method \cite{NAGAOSA,STONE}. 
      Couple $\psi^\dagger\psi$ to an auxiliary field $\varphi$,
      and eliminate all or part of $(\psi^\dagger\psi)^2$.
      Add a source term
       $J\varphi$ and perform a 
       loop expansion about the expectation value $\phi
        =\langle\varphi\rangle$.
      A Kohn-Sham version uses 
	the freedom of the expansion to require the density 
	be unchanged at each order.
    \item The inversion method~\cite{Fukuda:im,VALIEV97,VALIEV,RASAMNY98}
      yields a systematic Kohn-Sham DFT, based on
      an order-by-order expansion.  For example, we can
      apply the EFT power
         counting for a dilute system.
  \ee

\noindent
We'll expand here on the last path.

\subsection{EFT for Dilute Fermi Systems}

\begin{figure}[t]
\centering
      \includegraphics*[width=2.5in]{hard_sphere_ps}
      \hspace*{.3in}
      \includegraphics*[width=1.2in]{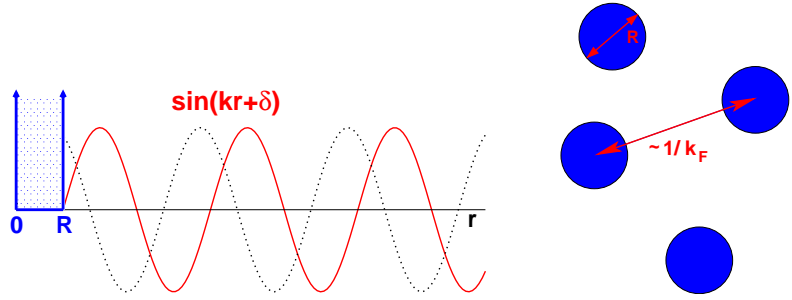}
 \caption{Hard-sphere phase shifts and scales at finite density.}
 \label{fig:11}       
\end{figure}

 We consider first one of the simplest many-body systems: a
collection of ``hard spheres,'' which means that the potential is
infinitely repulsive at a separation $R$ of the fermions. Since the
potential is zero outside of $R$ and the wave function must vanish in
the interior of the potential (so that the energy is finite), we can
trivially write down the $S$-wave scattering solution for momentum
$k$:  it is just a sine function shifted by $kR$ from the origin
(see Fig.~\ref{fig:11}). Our
problem will be to find the energy per particle (and other
observables) of a system of particles interacting with this potential
at $T=0$, given the density.

Let's do a quick review of scattering.
(More details on scattering at this level can be found in
practically any first-year graduate quantum mechanics text.  For a
more specialized but very readable account of nonrelativistic
scattering, check out ``Scattering Theory'' by Taylor.)
Consider relative motion with total momentum $P=0$:
 \beq
   \psi(r) \stackrel{r\rightarrow\infty}{\longrightarrow}
        e^{i{\bf k \cdot r}} + { f(k,\theta)} \frac{e^{ikr}}{r}
	\; , 
 \eeq	
where $k^2={k'}^2 = ME_k$ and $\cos\theta = \hat
k\cdot \hat k'$.
The differential cross section is $d\sigma/d\Omega = |f(k,\theta)|^2$.
For a central potential, we use partial waves:
\beq
   f(k,\theta) = \sum_l (2l+1)f_l(k) P_l(\cos\theta) \; ,
\eeq   
where~\cite{Extra}
 \beq
   f_l(k) = \frac{e^{i\delta_l(k)}\sin\delta_l(k)}{k}
     = \frac{1}{k\cot\delta_l(k) - i k}
 \eeq
and the $S$-wave phase shift is defined by
 \beq
   u_0(r) \stackrel{r\rightarrow\infty}{\longrightarrow}
      \sin[kr + \delta_0(k)]
      \quad\Longrightarrow\quad \delta_0(k) = -kR
      \ \mbox{for hard sphere} \; .
 \eeq
Note: we can do a partial wave expansion even if the potential is not
central (as in the nuclear case!); it merely means that different
$l$'s will mix.  The more important question is how many total $l$'s
do we need to include to ensure convergence.

As first shown by Schwinger, $k^{l+1}\cot\delta_l(k)$
has a power series expansion.  For $l=0$,
 \beq
   k\cot\delta_0 = -\frac{1}{ a_0} + \frac12 { r_0} k^2
      - Pr_0^3 k^4 + \cdots \; ,
 \eeq
which defines the \emph{scattering length} $a_0$ and the \emph{effective range}
$r_0$. While $r_0 \sim R$, the range of the potential, $a_0$ can be anything;
if $a_0 \sim R$, it is called ``natural''. 
The other case $|a_0| \gg R$
(unnatural) is particularly interesting; it is the case for
nucleon-nucleon interactions and can be studied in detail
with cold atoms. The
effective range expansion for hard sphere scattering is:
 \beq
  k \cot(-kR) = -\frac{1}{R} + \frac13 R k^2 + \cdots
      \quad \Longrightarrow \quad
      a_0 = R\, \quad r_0 = 2R/3
      \; ,
 \eeq
so the low-energy effective theory is natural.
is
Schwinger first derived the effective range expansion back in
the 1940's and then Bethe showed an easy way to derive (and
understand) it.  
 The implicit assumption here is that the potential is
short-ranged; that is, it falls off sufficiently rapidly with
distance.  This is certainly satisfied by any potential that actually
vanishes beyond a certain distance.  Long-range potentials like the
Coulomb potential must be treated differently (but a Yukawa potential
with finite range is ok).

So now consider the EFT for a natural, short-ranged interaction~\cite{HAMMER00}.
A simple, general interaction is a sum of delta functions and
derivatives of delta functions. In momentum space,
 \beq
   \langle {\bf k} | V_{\rm eft} | {\bf k'}\rangle
    = C_0 + \frac12 C_2 ({\bf k}^2 + {\bf k'}^2)
      + C'_2\, {\bf k\cdot k'} + \cdots
 \eeq
Or, we construct the effective lagrangian ${\cal L}_{\rm eft}$ from
the most general local (contact) interactions: 
\beqa
   {\cal L}_{\rm eft}   &=& 
       \psi^\dagger \Bigl[i\frac{\partial}{\partial t} 
               + \frac{\nab^{\,2}}{2M}\Bigr]
                 \psi - \frac{{ C_0}}{2}(\psi^\dagger \psi)^2
            + \frac{{ C_2}}{16}\bigl[ (\psi\psi)^\dagger 
                                  (\psi\!\galnab\!{}^2\psi)+\mbox{ h.c.} 
                             \bigr]   
  \nonumber \\[5pt]
   & & \null +
         \frac{{ C_2'}}{8}(\psi\! \galnab\! \psi)^\dagger \cdot
              (\psi\!\galnab\!\psi)
   - \frac{{ D_0}}{6}(\psi^\dagger \psi)^3 +  \ldots
\eeqa
Dimensional analysis (with a bit of additional insight to
give us the $4\pi$'s) implies
\beq
  C_{2i} \sim \frac{\textstyle 4\pi}{\textstyle M } R^{2i+1}\; ,
  \quad
   D_{2i} \sim 
  \frac{\textstyle 4\pi}{\textstyle M} R^{2i+4}
  \; , 
\eeq   
which will enable us to make quantitative power-counting estimates.

The ingredients for an effective field theory are nicely
summarized in the ``Crossing the Border'' review~\cite{CROSSING}:
  \begin{enumerate}
    \item \emph{Use the most general ${\cal L}$ with low-energy 
     degrees-of-freedom consistent
      with  global and local symmetries of underlying theory.} Here, 
   \beq
      {\cal L}_{\rm eft} = 
         \psi^\dagger \bigl[i\frac{\partial}{\partial t} 
         + \frac{\nabla^{\,2}}{2M}\bigr]
           \psi - \frac{{ C_0}}{2}(\psi^\dagger \psi)^2
              - \frac{{D_0}}{6}(\psi^\dagger \psi)^3 +  \ldots
   \eeq

   \item  \emph{Declare a regularization and renormalization scheme.}
       For a natural $a_0$, using
       dimensional regularization and minimal subtraction
       is particularly convenient and efficient.	   

   \item \emph{Establish a well-defined power counting}, which means
     identifying small expansion parameters, typically
     using the separation of scales.
     Here, $\kf/\Lambda$ 
     with $\Lambda \sim 1/R$, which implies $\kf a_0$, $\kf r_0$, etc.\
     are expansion parameters.  In the end, this will be manifest in the
     energy density:
       \beq
        {\cal E} =
           { \rho} \frac{{ \kf^2}}{2M}
            \biggl[\frac{3}{5} + {\frac{2}{3\pi}}{ (\kf a_0)}
            +
              { \frac{4}{35\pi^2}(11-2\ln 2)}{ (\kf a_0)^2}
              + \cdots
            \biggr] \;.  
         \eeq      
  \end{enumerate}

\begin{figure}[t]
\centering
      \includegraphics*[width=3.8in]{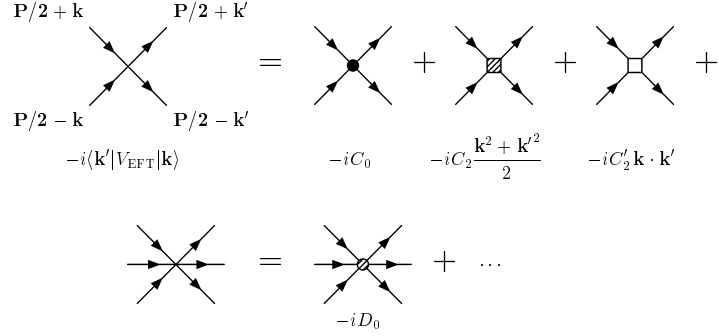}
 \caption{Feynman rules in free space \cite{HAMMER00}.}
 \label{fig:12}       
\end{figure}

The Feynman rules for the EFT lagrangian
are summarized in Fig.~\ref{fig:12} \cite{HAMMER00}.  
We need to    
reproduce $f_0(k)$ in perturbation theory (the Born series):
 \beq 
    {f_0(k)} 
     \propto a_0 - ia_0^2 { k} - (a_0^3 - a_0^2 r_0/2)
         { k^2} + {\cal O}({ k^3} a_0^4) \; . 
 \eeq
The leading potential $V^{(0)}_{\rm EFT}(\bfx) = C_0 \delta(\bfx)$ or
 \beq
   \langle {\bf k} | V^{(0)}_{\rm eft} | {\bf k'}\rangle
    \ \Longrightarrow\
    \raisebox{-.2in}{\includegraphics*[width=.4in,angle=0]{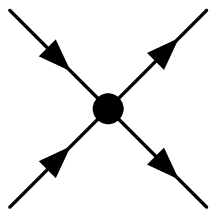}}
   \ \Longrightarrow\ C_0 \; .
  \eeq
Thus, choosing $C_0 \propto a_0$ gets the first term.  Next is 
$\langle {\bf k} | VG_0V | {\bf k'}\rangle$:
 \beq
  \raisebox{-.2in}{\includegraphics*[width=0.7in,angle=0]{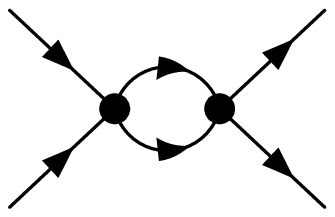}}
   \ \Longrightarrow\ 
  C_0 M \int\! \frac{d^3q}{(2\pi)^3} \frac{1}{k^2-q^2 + i\epsilon}\, C_0
     \longrightarrow \infty!
 \eeq
This is a linear divergence.
If the integral is cutoff at $\Lambda_c$, we can absorb the linear
dependence on $\Lambda_c$
into $C_0$, but we'll have all powers of $k^2$:
 \beq 
   \int^{\Lambda_c}\! \frac{d^3q}{(2\pi)^3} \frac{1}{k^2-q^2 + i\epsilon}
     \longrightarrow \frac{\Lambda_c}{2\pi^2}
       - \frac{ik}{4\pi} + {\cal O}(\frac{k^2}{\Lambda_c})
       \; .
 \eeq
A more efficient scheme is 
dimensional regularization with minimal subtraction (DR/MS), which implies 
only one power of $k$ survives:
 \beq
   \int\! \frac{d^Dq}{(2\pi)^3} \frac{1}{k^2-q^2 + i\epsilon}
     \stackrel{D\rightarrow 3}{\longrightarrow} 
       - \frac{ik}{4\pi} \; . 
 \eeq

The diagrammatic power counting with DR/MS is very simple, with each
loop adding a power of $k$:

\begin{center}
  \includegraphics*[width=3.5in,angle=0]{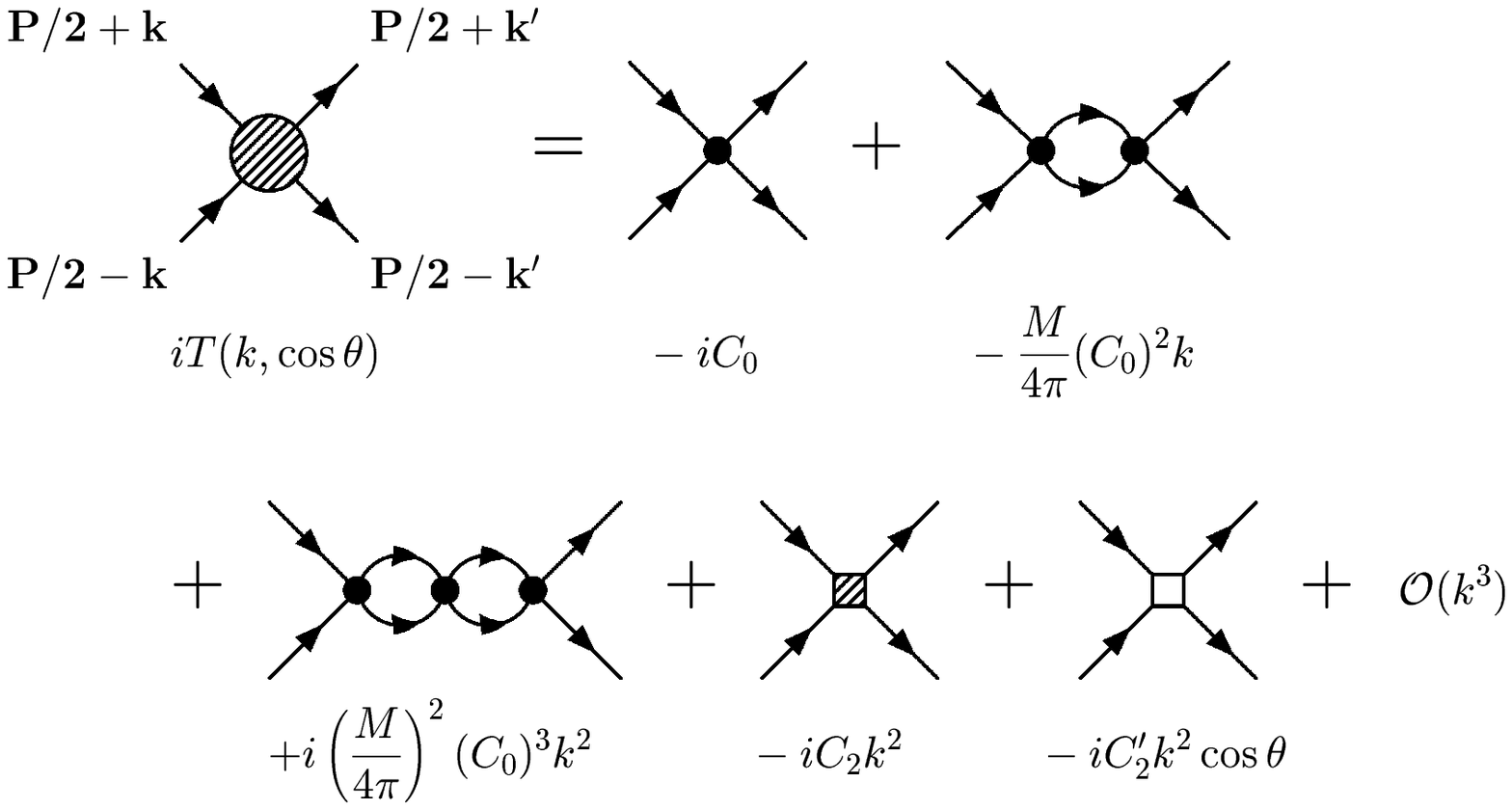}
\end{center}

\noindent After matching to the scattering amplitude,
 \beq
     C_0 = \frac{\textstyle 4\pi}{\textstyle M}a_0
        = \frac{\textstyle 4\pi}{\textstyle M}R\; ,
     \quad C_2= \frac{\textstyle 4\pi}{\textstyle M} 
          \frac{\textstyle  a_0^2 r_0}{\textstyle  2}
          = \frac{\textstyle 4\pi}{\textstyle M} 
           \frac{\textstyle R^3}{\textstyle 3}
          \; , 
              \quad
    C_2' = \frac{\textstyle 4\pi}{\textstyle M} a_p^3
          \; ,    \quad
              \cdots  
 \eeq
recovers the effective range expansion order-by-order with diagrams:
 \beq
   \frac{4\pi}{M}\left(
      a_0  -i a_0^2 p - a_0^3 p^2 + a_0^2 r_0 p^2 + \cdots
     \right) \; ,
 \eeq
with one power of $k$ per diagram and \emph{natural} coefficients,
so we can
estimate truncation errors from simple dimensional analysis.

    
\subsection{Apply at Finite Density}

Consider a noninteracting Fermi sea of particles at $T=0$.    
Put the system in a large box ($V=L^3$) with periodic boundary
conditions
and spin-isospin degeneracy $\nu$ (e.g., for nuclei, $\nu=4$).
Fill momentum states up to Fermi momentum $\kf$, so that
 \beq
   N = \nu\sum_{\bf k}^{\kf} 1 \; , \qquad
   E = \nu\sum_{\bf k}^{\kf} \frac{\hbar^2k^2}{2M}
   \; .
 \eeq
We can evaluate the sums using 
 \beq
   \int F(k)\,dk \approx \sum_i F(k_i)\Delta k_i
     = \sum_i F(k_i)\frac{2\pi}{L}\Delta n_i
     = \frac{2\pi}{L} \sum_i F(k_i) \; .
 \eeq
In one dimension (try finding the 1-D analogs of the following
results!),
 \beq
   N = \nu \frac{L}{2\pi}\int_{-\kf}^{+\kf}\!dk =
     \frac{\nu\kf}{\pi}L
    \ \Longrightarrow \ \rho =
    \frac{N}{L}=\frac{\nu\kf}{\pi};
    \quad
    \frac{E}{L} = \frac{1}{3}\frac{\hbar^2\kf^2}{2M} \rho
    \; ,        
 \eeq
while in three dimensions:   
 \beq
   N = \nu \frac{V}{(2\pi)^3}\int^{\kf}\!d^3k =
     \frac{\nu\kf^3}{6\pi^2}V
    \ \Longrightarrow \ \rho =
    \frac{N}{V}=\frac{\nu\kf^3}{6\pi^2};
    \quad
    \frac{E}{V} = \frac{3}{5}\frac{\hbar^2\kf^2}{2M} \rho 
    \; .       
 \eeq
The volume/particle $V/N = 1/\rho \sim 1/\kf^3$,
so the spacing $\sim 1/\kf$, as implied by Fig.~\ref{fig:11}.
     
We find the energy density by summing over the Fermi sea.
In leading order, we found $V^{(0)}_{\rm EFT}(\bfx) = C_0 \delta(\bfx)$,
so that $V^{(0)}_{\rm EFT}({\bf k},{\bf k'}) = C_0$, and
 \beq
   \raisebox{-.19in}{\includegraphics*[width=.4in,angle=0]{fig_C0}}
    {\quad \Longrightarrow \quad}
   \raisebox{-.08in}{%
  \includegraphics*[width=0.7in,angle=0]{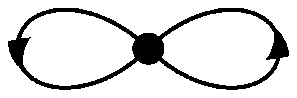}}
  \qquad {\cal E}_{\rm LO} 
     = \frac{C_0}{2}\nu(\nu-1)\left(\sum_{\bf k}^{\kf} 1\right)^2
  \ { \propto a_0 \kf^6} \; .
 \eeq
At the next order, we get a linear divergence again:
 \beq
  \raisebox{-.18in}{\includegraphics*[width=0.7in,angle=0]{fig_C0sq}}
 {\quad \Longrightarrow \quad}
  \raisebox{-.2in}{%
  \includegraphics*[width=.4in,angle=0]{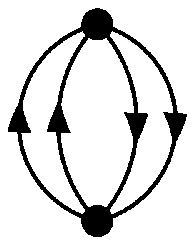}}
  \qquad {\cal E}_{\rm NLO} \propto
         \int_{\kf}^\infty\! \frac{d^3q}{(2\pi)^3} \frac{C_0^2}{k^2-q^2}
	 \; .
 \eeq
The \emph{same} renormalization fixes it!  
 \beq
         {\int_{\kf}^\infty\! \frac{1}{k^2-q^2}}
     =     
       { \int_{0}^\infty\!  \frac{1}{k^2-q^2}}
      { - \int_{0}^{\kf}\!  \frac{1}{k^2-q^2}}
    \stackrel{ D\rightarrow 3}{\longrightarrow}
     { - \int_{0}^{\kf}\!  \frac{1}{k^2-q^2}}
  \propto a_0^2 \kf^7 \; .
 \eeq
We also note that particles $\longrightarrow$
holes through the renormalization.

The Feynman rules for the energy density ${\cal E}$ at $T=0$,
which is the sum of \emph{Hugenholtz} diagrams~\cite{NEGELE88}
(closed, connected Feynman diagrams with symmetry factors) with
the same vertices as free space (and the same renormalization!), are:
\begin{enumerate}
   \item Each line is
   assigned conserved $\kt \equiv (k_0,\vec{k})$  and
   [$\omega_{\vec{k}} \equiv k^2/2M$],
  \beq
      iG_0 (\kt)_{\alpha\beta}=i\delta_{\alpha\beta}
      \left( \frac{\theta(k-\kf)}{k_0-\omega_{\vec{k}}+i\epsilon}
        +\frac{\theta(\kf-k)}{k_0-\omega_{\vec{k}}-i\epsilon}\right)
	\; . 
  \eeq
  \vspace*{-.1in}
 \item  
 \raisebox{0ex}{\parbox{.5in}{%
    \includegraphics*[width=0.4in,angle=0.0]{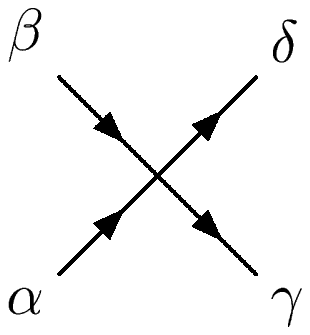}}}
 \raisebox{-0ex}{\parbox{3in}{%
   $\longrightarrow (\delta_{\alpha\gamma}\delta_{\beta\delta}
   +\delta_{\alpha\delta}\delta_{\beta\gamma})%
    \quad \mbox{(if spin-independent).}$
 }}
 \item After spin summations, $\delta_{\alpha\alpha}\rightarrow -\nu$ 
       in every closed fermion loop.
 \item Integrate $\int\! d^4 k /(2\pi)^4$ 
        with $e^{ik_0 0^+}$  for tadpoles
  \item The symmetry factor $i/(S \prod_{l=2}^{l_{\rm max}} (l!)^k)$  
    counts vertex permutations  and
   equivalent  $l$--tuples of lines (see \cite{HAMMER00} for examples).
 \end{enumerate}

\noindent
These Feynman rules in turn lead to power counting rules: 
\begin{enumerate}
   \item for every propagator (line): $M/{ \kf^{\!2}}$
   \item for every loop integration: ${ \kf^{\!5}}/M $
   \item for every $n$--body vertex with $2i$  derivatives: 
                 ${ \kf^{\!2i}}/M\Lambda^{2i+3n-5}   $
\end{enumerate}
Thus, a diagram with $V^n_{2i}$ $n$--body vertices 
scales as ${ (\kf)^\beta}$ \emph{only}:
 \beq
  \beta=5+\sum_{n=2}^\infty \sum_{i=0}^\infty (3n+2i-5) V_{2i}^n\; .
 \eeq
For example, at leading order,
 \beq
  \raisebox{0ex}{\parbox{0.7in}{%
   \includegraphics*[width=0.7in,angle=0.0]{fig_hugenholtz6}}}%
    \  \Longrightarrow      
       V_0^2=1 \Longrightarrow 
         \beta = 5 + (3\cdot 2 + 2\cdot 0 - 5)\cdot 1  = 6 \Longrightarrow 
         {\cal O}({ \kf^{6}}) \; ,
  \eeq	  
and at next-to-leading order, 
 \beq
  \raisebox{0ex}{\parbox{0.4in}{%
   \includegraphics*[width=0.4in,angle=0.0]{fig_hugenholtz7}}}%
    \  \Longrightarrow      
       V_0^2=2 \Longrightarrow 
     \beta = 5 + (3\cdot 2 + 2\cdot 0 - 5)\cdot 2  = 7 \Longrightarrow 
         {\cal O}({ \kf^{7}}) \; .
 \eeq

We emphasize that
Pauli blocking doesn't change the free-space ultraviolet (short distance)
renormalization, since the density is a long-distance effect.
As noted before, particles become holes:
  \beq
       \int_{\kf}^\infty = \int_0^\infty - \int_0^{\kf} \longrightarrow
           -\int_0^{\kf}
	   \; . 
  \eeq	   
The power counting is exceptionally clean, with a
separation of vertex factors $\propto a_0,r_0,\ldots$ and
a dimensionless geometric integral times $\kf^n$,
with each diagram contributing to exactly one order in the
expansion.
This is a systematic expansion:
the ratio of successive terms is $\sim \kf R$, so you can 
\emph{estimate} excluded contributions.

The full result for the energy density
through ${\cal O}(\kf^8)$ is~\cite{HAMMER00}: 
  \beqa
    \frac{E}{V}  &=&   { \rho} \frac{{ \kf^2}}{2M}
    \biggl[ \frac{3}{5} 
    {\null + (\nu-1)\frac{2}{3\pi}{ (\kf a_0)}}
    {\null +(\nu-1)
      \frac{4}{35\pi^2}(11-2\ln 2){ (\kf a_0)^2}}
     \nonumber \\[12pt] & &   \null
      {\null + (\nu-1)\bigl(0.076 
      + 0.057(\nu-3)\bigr){ (\kf a_0)^3}} 
      {\null 
      + (\nu-1)\frac{1}{10\pi}{ (\kf r_0)(\kf a_0)^2}} 
     \nonumber \\[7pt] & &  \null
      {\null 
      + (\nu+1)\frac{1}{5\pi}{ (\kf a_p)^3} + \cdots}
    \biggr] 
    \; . 
  \eeqa
This looks like a power series in $\kf$, but it's not!
There are new \emph{logarithmic} divergences in 3--3 scattering,
in these diagrams:
 \beq
  \raisebox{-.2in}{\includegraphics*[width=2.3in,angle=0.0]{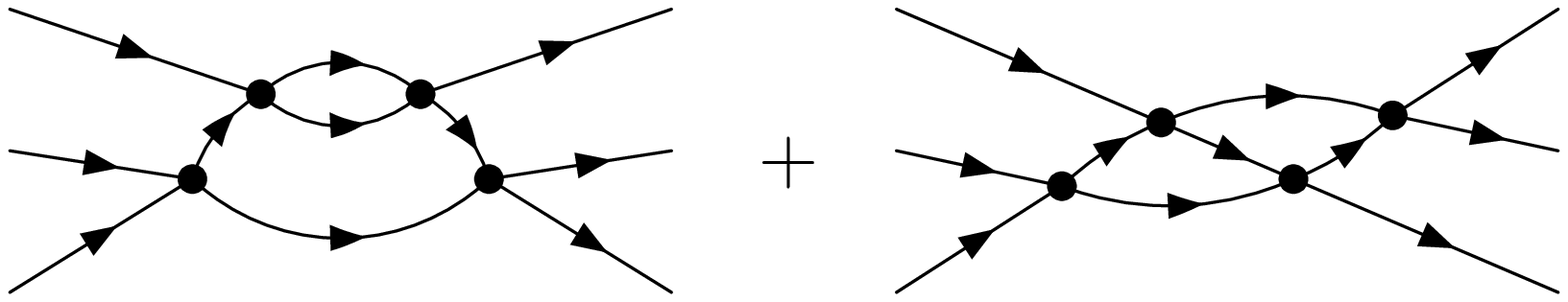}}      
       \quad   \propto { (C_0)^4 \ln(k/\Lambda_c)}
       \; .
 \eeq
Changes in $\Lambda_c$ must be absorbed by the \emph{3-body} coupling
$D_0(\Lambda_c)$, so~\cite{BRAATEN97}
 \beq	  
    D_0(\Lambda_c) \propto { (C_0)^4 \ln (a_0\Lambda_c)}
      + \mbox{const.}
 \eeq    
Then requiring the sum to be independent of $\Lambda_c$,    
 \beq
   \frac{d}{d\Lambda_c}\biggl[
   \raisebox{-.2in}{%
      \includegraphics*[width=2.4in,angle=0.0]{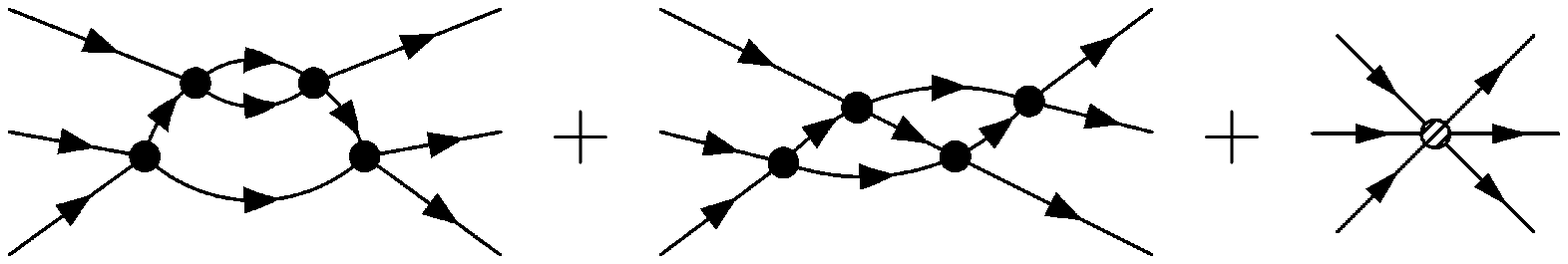}}
   \biggr] = 0
 \eeq
fixes the coefficient!
This implies for the energy density,
  \beq 
    \raisebox{-.28in}{\includegraphics*[width=2.5in]{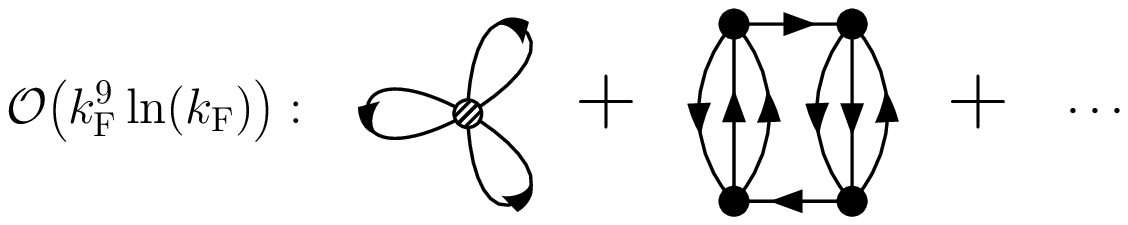}}       
    \propto (\nu-2)(\nu-1)\,(\kf a_0)^{4} \, {\ln(\kf a_0)}
  \eeq
without actually carrying out the calculation.
Similar analyses can identify the higher logarithmic terms in the
expansion of the energy density~\cite{BRAATEN97,HAMMER00}.  
   
Divergences indicate sensitivity to short-distance behavior.
The cutoff $\Lambda_c$ here serves as a resolution scale;  as we
increase $\Lambda_c$, we see more of the short-distance details.
Observables (such as scattering amplitudes) must not change
when $\Lambda_c$ changes, so they must be absorbed in a coupling.
But it can't be a coupling from 2--2 scattering, because
we already took care of all the divergences there.
So there must be a point-like three-body force included,
whose coupling $D_0$ can absorb the changes.
    
Let's summarize the dilute Fermi system with natural $a_0$.
We find that
the many-body energy density is perturbative (but not analytic) 
in $\kf a_0$, and
is efficiently reproduced by the EFT approach.
Power counting gives us error estimates from omitted diagrams.
Three-body forces are \emph{inevitable} in the low-energy
effective theory
and not unique;  they depend on the two-body potential.
The case of a natural scattering length is under control 
for a uniform system, but 
what if the scattering length is not natural?
We'll come back to that situation in the last lecture.
First we consider a non-uniform system:
a finite number of fermions in a trap,
which takes us back to DFT.


\subsection{DFT via EFT
\cite{Puglia:2002vk,Bhattacharyya:2004qm,Bhattacharyya:2004aw}}
  
Return to the thermodynamic version of DFT through
the effective action and ask:
What can EFT do for DFT?  
We can construct the effective action as a path integral
by finding $W[J]$ order-by-order in an EFT expansion.
For a dilute short-range
system, this means the same diagrams as before,
but now the
propagators (lines) are in the background field $J({\bf x})$:
  \beq
    G^0_{J}({\bf x},{\bf x'}; \omega)
    = \sum_\alpha \psi_\alpha({\bf x})\psi^\ast_\alpha({\bf x'})
    \left[
     \frac{\theta(\epsilon_\alpha - \epsilon_{\rm F})}
          {\omega - \epsilon_\alpha + i\eta}
	+
     \frac{\theta(\epsilon_{\rm F} - \epsilon_{\alpha})}
          {\omega - \epsilon_\alpha - i\eta}
    \right]
    \; ,
  \eeq
where $\psi_\alpha(\bfx)$ satisfies:
\beq
       \bigl[ -\frac{{\nabla}^2}{2M}  +  \Vext({\bf x}) - J({\bf x})
       \bigr]\, \psi_\alpha(\bfx) = \epsilon_\alpha \psi_\alpha(\bfx)
       \; .
       \label{eq:psis}
\eeq
Applying this to the leading-order (LO) contribution $W_1[J]$, 
which is Hartree-Fock,
yields
 \beqa
   W_1[J] &=& \frac{1}{2} \nu(\nu-1) C_0
     \int\! d^3{\bf x}\, 
     \int_{-\infty}^{\infty}\!\frac{d\omega}{2\pi}
     \int_{-\infty}^{\infty}\!\frac{d\omega'}{2\pi}\
       G^0_{J} ({\bf x},{\bf x}; \omega)
       G^0_{J} ({\bf x},{\bf x}; \omega')
     \nonumber \\
     &=&
     -\frac{1}{2}\frac{(\nu-1)}{\nu} C_0
     \int\! d^3{\bf x}\, [\rho_J({\bf x})]^2 \; ,
 \eeqa
where $\rho_J({\bf x}) \equiv \nu\sum_\alpha^{\epsilon_{\rm F}}
	  |\psi_\alpha({\bf x})|^2$.
Expressions for the other $W_i$'s proceed directly from the Feynman
rules using the new propagator.	  

Given $W[J]$ as an EFT expansion, how do we perform the Legendre
transformation, 
 \beq
   \Gamma[\rho] = W[J] - \int\! J\rho \; ,
 \eeq
in a systematic way?   
The EFT power counting gives us a means to invert to find
$J[\rho]$.
In particular,
the ``inversion method'' provides an order-by-order inversion from
$W[J]$ to $\Gamma[\rho]$~\cite{Fukuda:im,VALIEV97,VALIEV}.  It proceeds by
decomposing $J(x) = J_0(x) + J_{\rm LO}(x) + J_{\rm NLO}(x) + \ldots $
with two conditions on $J_0$:
  \beq
    \rho(x) = \frac{\delta W_0[J_0]}{\delta J_0(x)}
    \quad\mbox{and}\quad
    \left.J_0(x)\right|_{\rho = \rho_{\rm gs}} = 
      \left.\frac{\delta\Gamma_{\rm interacting}[\rho]}
        {\delta \rho(x)}\right|_{\rho = \rho_{\rm gs}}
	\; .
  \eeq 
We are using the freedom to split $J$ into $J_0$ and the rest,
in the same way that one adds and subtracts a single-particle
potential $U$ to a Hamiltonian: $H = T + V = (T + U) + (V - U)$ and
then uses the freedom to choose $U$ to improve many-body convergence.
In our case, we choose $J_0$ so that there are no corrections to
the zeroth order density at each order in the expansion.
The interpretation is that $J_0$ is the external potential that yields 
for a noninteracting system the exact density.
This is the Kohn-Sham potential!
The two conditions involving $J_0$ imply a self-consistent
procedure.

The inversion method  for effective action DFT~\cite{Fukuda:im,VALIEV97,VALIEV} is an
order-by-order matching in a counting
parameter $\lambda$  (e.g., an EFT expansion):
  \beqa
   \mbox{ diagrams}&\Longrightarrow&
   W[J,\lambda] = W_0[J] + \lambda W_1[J] + \lambda^2 W_2[J] + \cdots \\
   \mbox{ assume}&\Longrightarrow&
   J[\rho,\lambda] = J_0[\rho] + \lambda J_1[\rho] 
       + \lambda^2 J_2[\rho] + \cdots \\
   \mbox{ derive}&\Longrightarrow&
   \Gamma[\rho,\lambda] = \Gamma_0[\rho] 
            + \lambda \Gamma_1[\rho] + \lambda^2 \Gamma_2[\rho] + \cdots 
  \eeqa
We start with the exact expressions for $\Gamma$ and $\rho$
[note: $\beta\ \mbox{and}\ T=1$ here],
  \beq
    \Gamma[\rho] = W[J] - \int\!d^4x\, J(x)\rho(x) \; , 
  \eeq 
  \beq
    \rho(x)
       = \frac{\delta W[J]}{\delta J(x)} \; ,
    \quad
     J(x) = -\frac{\delta \Gamma[J]}{\delta \rho(x)} \; .    
  \eeq 
Then plug in each of the expansions, with $\rho$ treated as order unity.
Zeroth order is the noninteracting system with potential $J_0(x)$, 
 \beq
   \Gamma_0[\rho] = W_0[J_0] - \int\!d^4x\, J_0(x)\rho(x) 
   \quad \Longrightarrow \quad \rho(x)
      = \frac{\delta W_0[J_0]}{\delta J_0(x)}  \; ,  
 \eeq     
which is the Kohn-Sham system with the exact density!
(Note: $J_0 \equiv V_{\rm KS}$ here.)
To evaluate $W_0[J_0]$, we introduce the orbitals from (\ref{eq:psis}),
which diagonalize $W_0$, so that it yields a sum of $\varepsilon_i$'s
for the occupied states.
Then we find $J_0$ for the ground state via a  self-consistency loop:
\beq
  { J_0} \rightarrow W_1 \rightarrow \Gamma_1 \rightarrow J_1
   \rightarrow W_2 \rightarrow \Gamma_2 \rightarrow \cdots
   \Longrightarrow
     { J_0(x) = \sum_{i>0} 
         \frac{\delta\Gamma_{i}[\rho]}{\delta\rho(x)}}
	 \; ,    
\eeq
which is the second of our two conditions on $J_0$.

We note that the Kohn-Sham potential is local: 
 \beq
   J_0({\bf x}) = \frac{\delta \Gamma_{\rm int}[\rho]}{\delta
      \rho({\bf x})} 
      \; ,
      \label{eq:J0d}
 \eeq
in stark contrast to the non-local and state-dependent
self-energy  $\Sigma^\ast({\bf x},{\bf x}';\omega)$.
Evaluating the functional derivatives is easiest
if $\Gamma$ is approximated so that the dependence on the density
is explicit, as with the LDA or DME (see below).
Otherwise we need to use a 
chain rule with the ``inverse density-density correlator''
\cite{VALIEV97}
\begin{center}
    \includegraphics*[angle=0.0,width=4.3in]{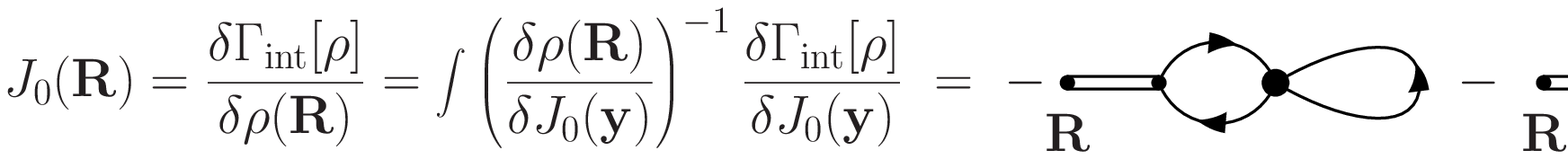}
\end{center}
There are new Feynman rules for $\Gamma_{\rm int}$ for evaluating
such diagrams~\cite{VALIEV97}.
(A related
approach is the OEP method~\cite{fi03,Ba05,Go05,Ba05b}.)

In constructing the diagrams for $W[J]$ and new diagrams
for $\Gamma[\rho]$ order by order in the expansion
(e.g., EFT power counting),
the source $J_0({\bf x})$ is now the background field (rather
than the full $J({\bf x})$).
Propagators (lines) in the background field $J_0({\bf x})$ are
 \beq
   G^0_{\rm KS}({\bf x},{\bf x'}; \omega)
   = \sum_\alpha \psi_\alpha({\bf x})\psi^\ast_\alpha({\bf x'})
   \left[
    \frac{\theta(\epsilon_\alpha - \epsilon_{\rm F})}
         {\omega - \epsilon_\alpha + i\eta}
       +
    \frac{\theta(\epsilon_{\rm F} - \epsilon_{\alpha})}
         {\omega - \epsilon_\alpha - i\eta}
   \right]
   \; ,
 \eeq
where $\psi_\alpha(\bfx)$ satisfies:
  \beq
   \bigl[ -\frac{{\nabla}^2}{2M}  +  v({\bf x}) - J_0({\bf x})
   \bigr]\, \psi_\alpha(\bfx) = \epsilon_\alpha \psi_\alpha(\bfx)
   \; .
  \eeq
For example, if we apply this prescription
to the short-range LO contribution (i.e., Hartree-Fock),
we obtain
    \beqa
      W_1[J_0] &=& \frac{1}{2} \nu(\nu-1) C_0
        \int\! d^3{\bf x}\, 
	\int_{-\infty}^{\infty}\!\frac{d\omega}{2\pi}
	\int_{-\infty}^{\infty}\!\frac{d\omega'}{2\pi}\
	  G^0_{\rm KS} ({\bf x},{\bf x}; \omega)
	  G^0_{\rm KS} ({\bf x},{\bf x}; \omega')
	\nonumber \\
	&=&
	-\frac{1}{2}\frac{(\nu-1)}{\nu} C_0
	\int\! d^3{\bf x}\, [\rho_{J_0}({\bf x})]^2
	\; ,
    \eeqa
where
\beq
  \rho_{J_0}({\bf x}) \equiv \nu\sum_\alpha^{\epsilon_{\rm F}}
	  |\psi_\alpha({\bf x})|^2 \; .
\eeq

Let us construct the $T=0$ local density approximation (LDA).
In a uniform system, each line is a non-interacting propagator.
The energy density in the uniform system evaluates to:
 \beqa
   \frac{E}{V}  &=&   { \rho} \frac{{ \kf^2}}{2M}
   \biggl[ \frac{3}{5} 
   {\null + (\nu-1)\frac{2}{3\pi}{ (\kf a_0)}}
   {\null +(\nu-1)
     \frac{4}{35\pi^2}(11-2\ln 2){ (\kf a_0)^2}}
    \nonumber \\[12pt] & &   \null
     {\null + (\nu-1)\bigl(0.076 
     + 0.057(\nu-3)\bigr){ (\kf a_0)^3}} 
     {\null 
     + (\nu-1)\frac{1}{10\pi}{ (\kf r_0)(\kf a_0)^2}} 
    \nonumber \\[7pt] & &   \null
     {\null 
     + (\nu+1)\frac{1}{5\pi}{ (\kf a_p)^3} + \cdots}
   \biggr]
   \; .  
 \eeqa
with $\kf = (6\pi^2\rho/\nu)^{1/3}$.  Using this relation to
replace $\kf$ everywhere by $\rho(\xvec)$,
we directly obtain the LDA expression for $\Gamma[\rho]$,
 \beqa
   \Gamma[\rho]  &=&  \int\!d^3 x\,\biggl[ T_{\rm KS}({\bfx})
   {+\frac12\frac{(\nu-1)}{\nu}
      \frac{4\pi a_0}{M} [\rho(\bfx)]^2} 
   {\null + d_1
         \frac{a_0^2}{2M}[\rho(\bfx)]^{7/3}}
    \nonumber \\[12pt] & &   \qquad\qquad\null
     {\null 
     + d_2\,  a_0^3 [\rho(\bfx)]^{8/3}}
     {\null 
     + d_3\, a_0^2\, r_0[\rho(\bfx)]^{8/3}} 
    \nonumber \\[7pt] & &   \qquad\qquad\null
    {\null 
     + d_4\, a_p^3[\rho(\bfx)]^{8/3} + \cdots \biggr]}
     \; .
     \label{eq:Gammalda}
 \eeqa
The Kohn-Sham $J_0$ according to the EFT expansion       
follows immediately in the LDA from (\ref{eq:J0d}):
      \beqa
        J_0(\bfx)  &=&   \biggl[ 
        -\frac{(\nu-1)}{\nu}
           \frac{4\pi a_0}{M} \rho(\bfx) 
         - c_1 \frac{a_0^2}{2M}[\rho(\bfx)]^{4/3}
          - c_2\,  a_0^3 [\rho(\bfx)]^{5/3}
         \nonumber \\[12pt] & &  \quad \null
          - c_3\, a_0^2\, r_0[\rho(\bfx)]^{5/3} 
          - c_4\, a_p^3[\rho(\bfx)]^{5/3} + \cdots \biggr]
  \; .
    \label{eq:J0lda} 
      \eeqa
(Finding the $\{d_i\}$'s and $\{c_i\}$'s is left as an
exercise for the reader.)      
 
\begin{figure}[t]
\centering
      \includegraphics*[width=3.2in]{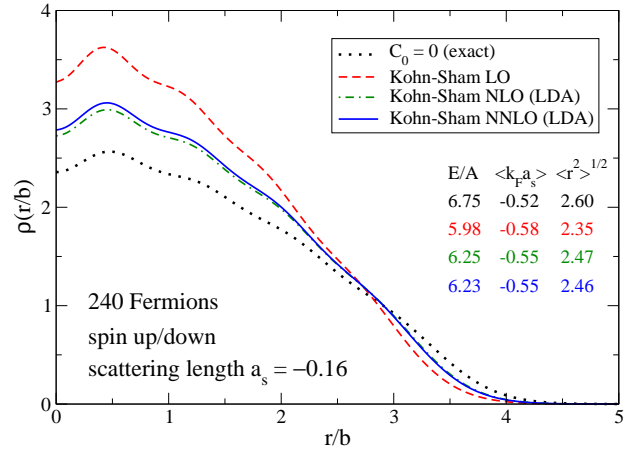}
 \caption{Density profile of a dilute system of fermions in a 
  trap~\cite{Puglia:2002vk}.}
 \label{fig:13}       
\end{figure}

Given (\ref{eq:Gammalda}) and (\ref{eq:J0lda}),
the iteration procedure is:
    \be
    
      \item Guess an initial density profile $\rho(r)$
         (e.g., the Thomas-Fermi density).
         
      \item Evaluate the local single-particle potential 
         ${ V_{\rm KS}(r) \equiv v_{\rm ext}(r) - J_0(r)}$.
         
      \item Find the wave functions and energies
      $\{\psi_\alpha,\epsilon_\alpha\}$
      of the lowest $A$ states (including degeneracies) 
       by solving:    
         \beq
          \bigl[ -\frac{{\nabla}^2}{2M}  +  V_{\rm KS}(r)
          \bigr]\, \psi_\alpha(\bfx) = \epsilon_\alpha \psi_\alpha(\bfx)
	  \; .
         \eeq
         
      \item Compute a new density
               $\rho(r) = \sum_{\alpha=1}^{A} |\psi_\alpha(\bfx)|^2$.
        Other observables are simple functionals of 
             $\{\psi_\alpha,\epsilon_\alpha\}$.
        
      \item Repeat 2.--4. until changes are small (``self-consistent'')
      
    \ee 
This sounds like a simple Hartree calculation!
Results at different EFT orders 
for a dilute Fermi gas in a harmonic oscillator
trap is given in Fig.~\ref{fig:13}.
Note the systematic progression from order to order.

\begin{figure}[t]
\centering
      \includegraphics*[width=2.1in]{error_plot2}
      \includegraphics*[width=2.3in]{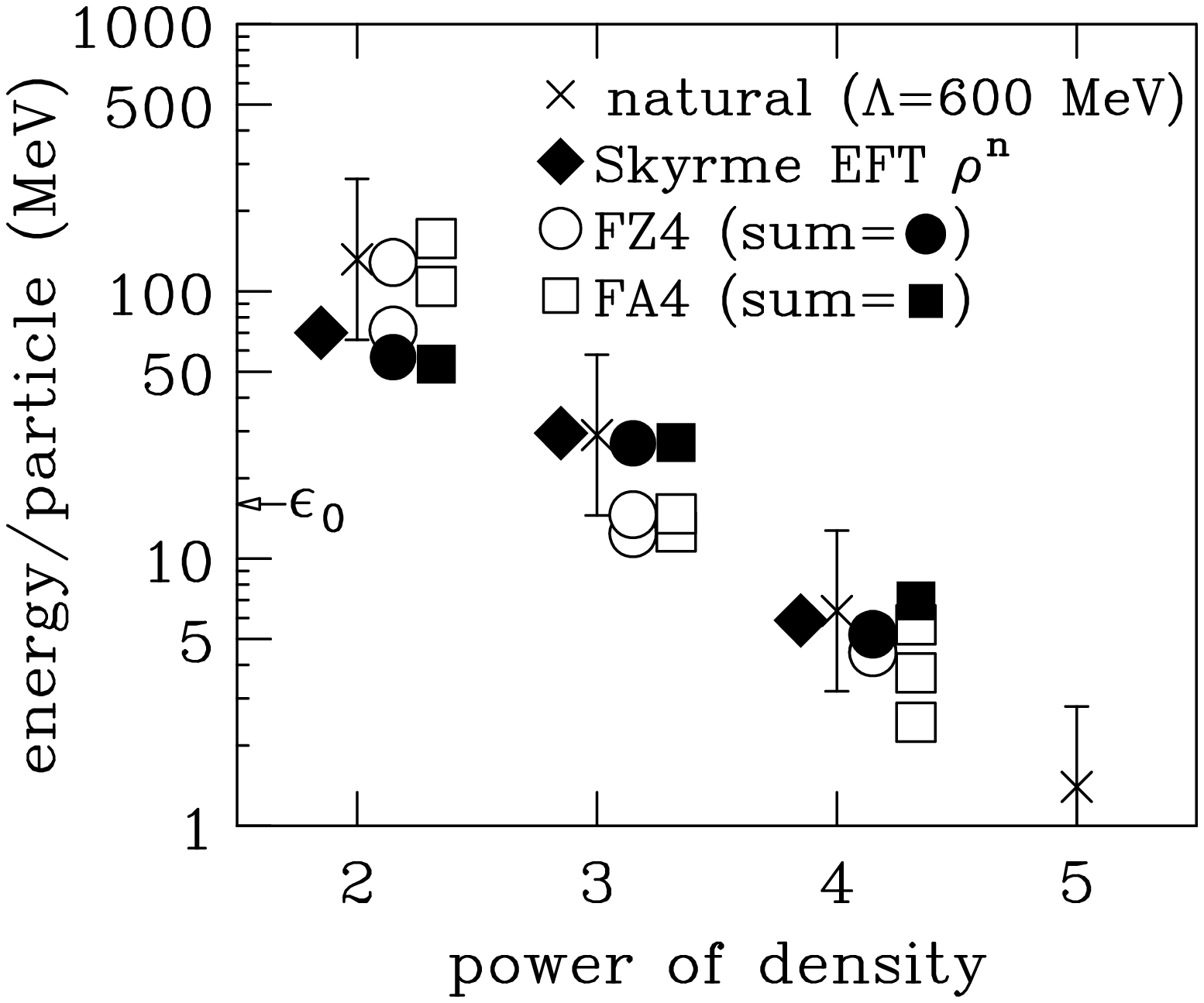}
 \caption{Estimates for energy functionals for a dilute
 fermions in a harmonic trap (left) and for phenomenological
 energy functionals for nuclei (right).}
 \label{fig:14}       
\end{figure}

An important consequence of the systematic EFT approach
is that we can also estimate individual terms in energy functionals.
If we scale contributions to the energy per particle
according to the average density or $\langle\kf\rangle$,
we can make estimates \cite{Puglia:2002vk,Bhattacharyya:2004qm}.
This is shown in Fig.~\ref{fig:14} for both the dilute trapped
fermions, which is under complete control, and for phenomenological
energy functionals for nuclei, to which a postulated QCD power counting is
applied.
In both cases, the estimates agree well with the actual numbers
(sometimes overestimating the contribution because of 
accidental cancellations), which means that
truncation errors are understood.

Conventional DFT is one example of using effective actions,
which feature sources coupled to composite operators.
It's possible that for some applications
a different type of effective action may be better.
There are many outstanding questions from the present discussion,
particularly as we try to adapt it to real nuclei.
We'll address some of them in the next lecture.
   

\section{Refinements: Toward EFT/DFT for Nuclei}
\label{sec:3}

Let's enumerate some questions about DFT and nuclear structure.
  \bi
    \item How is Kohn-Sham DFT more than mean field?
     That is, where are the approximations and
       how do we truncate?
     How do we include long-range effects (correlations)?
     
    \item  What can you calculate in a DFT approach?  
     Can we calculate single-particle properties? Or excited states?

    \item How does pairing work in DFT?  
      Can we (should we) decouple $pp$ and $ph$?
        Are higher-order contributions important?

  \item The Skyrme functional depends on multiple densities:
     $\rho(\xvec)$, $\tau(\xvec)$,
         and ${\bf J}(\xvec)$; how does that work?
      
    \item What about broken symmetries that arise
    with self-bound systems? (translation, rotation,  \ldots)  
      
    \item How do we connect to the free, microscopic 
    NN$\cdots$N interactions?      
     Can we use chiral EFT or low-momentum interactions/RG?           
  \ei
We'll explore some answers to these questions (and note which
ones are open) in this lecture.

\begin{figure}[t]
\centering
      \includegraphics*[width=2.2in]{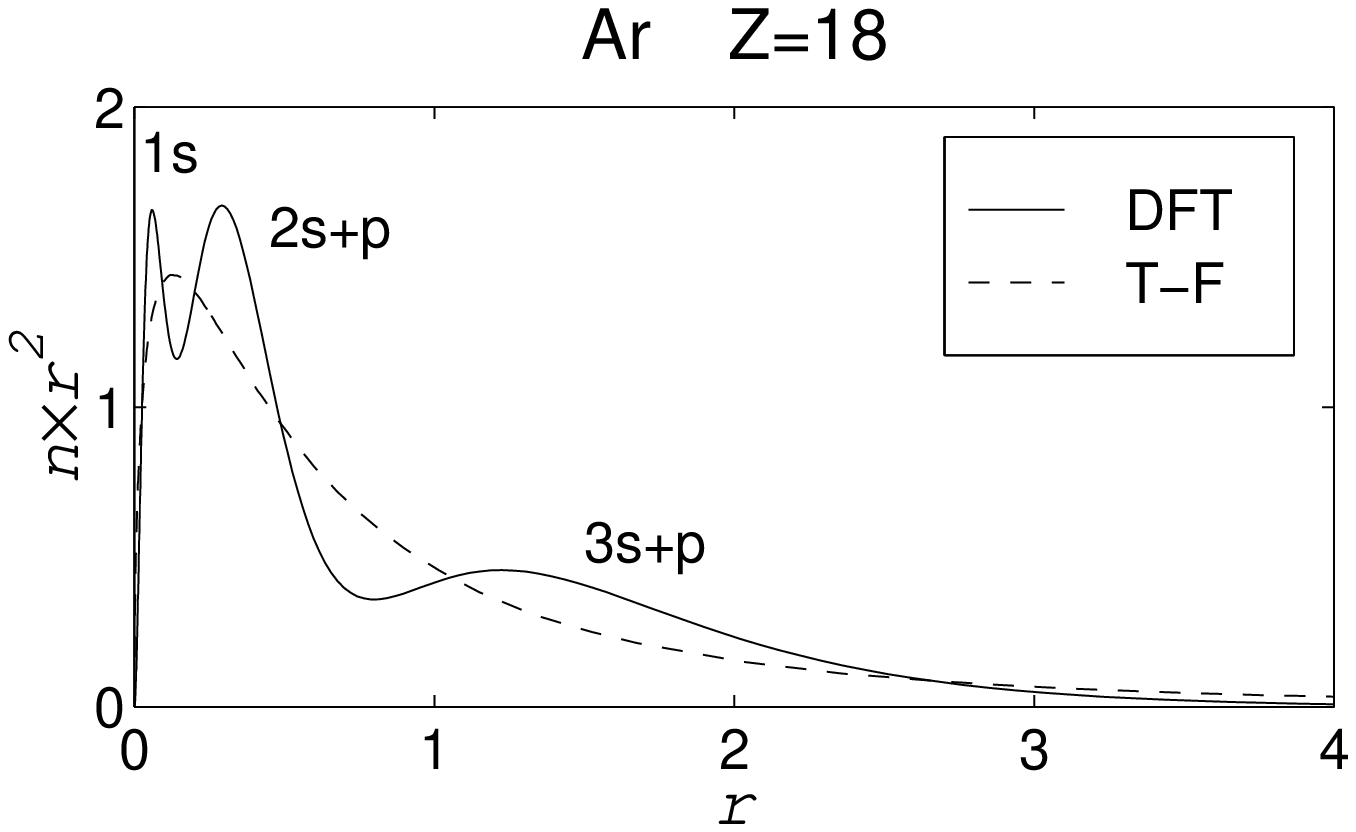}
      \includegraphics*[width=2.4in]{g2NF2asm16tf_bw} 
   \caption{Thomas-Fermi vs.\ DFT for atoms \cite{ARGAMAN00}
   (left) and
     trapped fermions (right).}
 \label{fig:15}       
\end{figure}

Consider Kohn-Sham DFT compared to the Thomas-Fermi energy functional,
for which the entire functional is treated in the local density
approximation (LDA).
In Kohn-Sham DFT, treating kinetic energy non-locally
leads to the shell structure of electrons in atoms and
in trapped atoms, as seen in Fig.~\ref{fig:15}.
This motivates going even further beyond the LDA.

As a simple step beyond Kohn-Sham LDA, we consider functionals
of the kinetic energy density in addition to the usual fermion density.
The phenomenological
Skyrme $E$ is a functional of $\rho$ \emph{and} 
$ \tau \equiv 
       \langle \bm{\nabla}\psi^\dagger\cdot\bm{\nabla}\psi \rangle$
(and ${\bf J}$):
  \beqa
    E[\rho,\tau,{\bf J}] 
      &\!=\!& \int\!d^3x\,
      \biggl\{ {1\over 2M}\tau + {3\over 8} t_0 \rho^2
    + {1\over 16} t_3 \rho^{2+\alpha}
   + {1\over 16}(3 t_1 + 5 t_2) \rho \tau  \nonumber
    \\ & & \hspace*{-.1in}\null
    + {1\over 64} (9t_1 - 5t_2) (\bm{\nabla} \rho)^2  
    - {3\over 4} W_0 \rho \bm{\nabla}\cdot{\bf J}
    + {1\over 32}(t_1-t_2) {\bf J}^2 \biggr\}
    \; . 
    \label{eq:skyrmeE} 
  \eeqa
To do this in DFT/EFT, add to the Lagrangian 
$\eta({\bf x})\,\bm{\nabla}\psi^\dagger\bm{\nabla}\psi$
and generalize our Legendre transformation and inversion
to $\Gamma[\rho,\tau]$,
  \beq
    \Gamma[\rho,\tau] = W[J,\eta]
     - \int\! J(x)\rho(x) - \int\! \eta(x)\tau(x)
     \; . 
  \eeq
Now there are two Kohn-Sham potentials:
     \beq
          J_0({\bf x}) = \frac{\delta \Gamma_{\rm int}[\rho,\tau]}{\delta
          \rho({\bf x})} \quad \mbox{and} \quad 
          \eta_0({\bf x}) = \frac{\delta \Gamma_{\rm int}[\rho,\tau]}{\delta
          \tau({\bf x})}
	  \; .      
     \eeq
The Kohn-Sham equation defines $1/\Mstar({\bf x}) \equiv 
      1/M - 2\eta_0({\bf x})$:
  \beq
   \Bigl( - \mathbf{\nabla}\cdot { \frac{1}{2M^*({\bf x})}} \mathbf{\nabla} 
      + \Vext(\xvec)-J_0(\xvec) 
   \Bigr)\,  
     \phi_{\alpha}({\bf x}) =
   \epsilon_{\alpha}\,\phi_{\alpha}({\bf x}) 
   \; .
  \eeq
  
\begin{figure}[t]
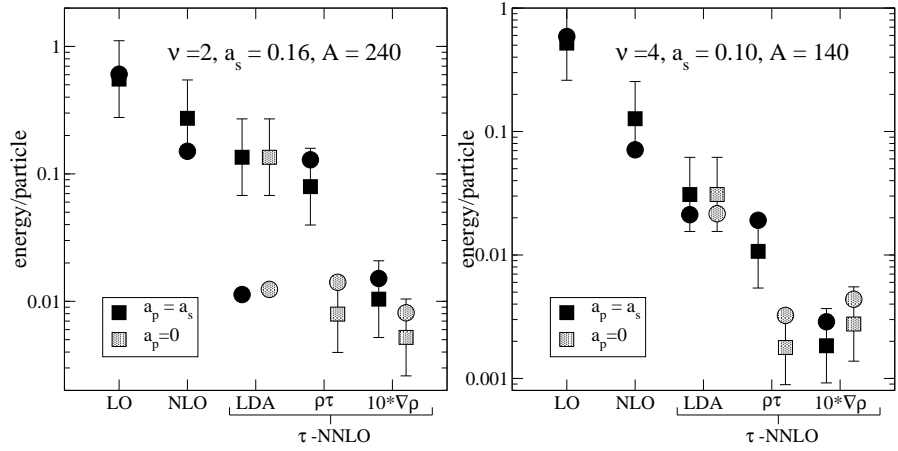

\centering
      \includegraphics*[width=2.3in]{error_plot_tau_breakup240}
      \includegraphics*[width=2.3in]{error_plot_tau_breakup140} 
   \caption{Estimates of terms in the energy functional, including
   those with gradients, compared to actual 
   values~\cite{Bhattacharyya:2004qm}.}
 \label{fig:16}       
\end{figure}

A simple first application is to 
evaluate Hartree-Fock diagrams including the quadratic gradient 
terms~\cite{Bhattacharyya:2004qm}.
Consider the HF ``bowtie diagrams'' 
 \begin{center}
    \includegraphics*[width=2.3in,angle=0]{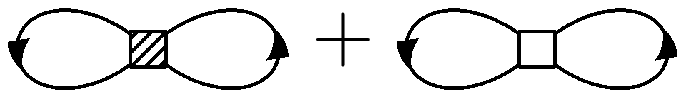}       
 \end{center}
\noindent
that have vertices with derivatives: 
 \beq
   {\cal L}_{\rm eft} = \ldots
        + \frac{{ C_2}}{16}\bigl[ (\psi\psi)^\dagger 
                                   (\psi\!\galnab\!{}^2\psi)+\mbox{ h.c.} 
                              \bigr]   
     +
          \frac{{ C_2'}}{8}(\psi\! \galnab\! \psi)^\dagger \cdot
               (\psi\!\galnab\!\psi) + \ldots
 \eeq
The energy density in Kohn-Sham LDA is
 \beq
   {\cal E}_{\rm int}[\rho] = \ldots + 
     \frac{C_2}{8}
     \Bigl[
       \frac35\left(\frac{6\pi^2}{\nu}\right)^{2/3}\rho^{8/3}
     \Bigr]
    +
     \frac{3C_2'}{8}
     \Bigl[
       \frac35\left(\frac{6\pi^2}{\nu}\right)^{2/3}\rho^{8/3}
     \Bigr]
     + \ldots
 \eeq
while the
energy density in Kohn-Sham with $\tau$ ($\nu=2$) is
  \beq
    {\cal E}_{\rm int}[\rho,\tau] = \ldots +
     \frac{C_2}{8}
     \bigl[
       \rho\tau + \frac34(\bm{\nabla}\rho)^2
     \bigr]
    +
     \frac{3C_2'}{8}
     \bigl[
       \rho\tau - \frac14(\bm{\nabla}\rho)^2
     \bigr]
    + \ldots
  \eeq
We find that power counting estimates for terms in the energy
functional also work with gradient terms
(see Fig.~\ref{fig:16}).

\begin{figure}[t]
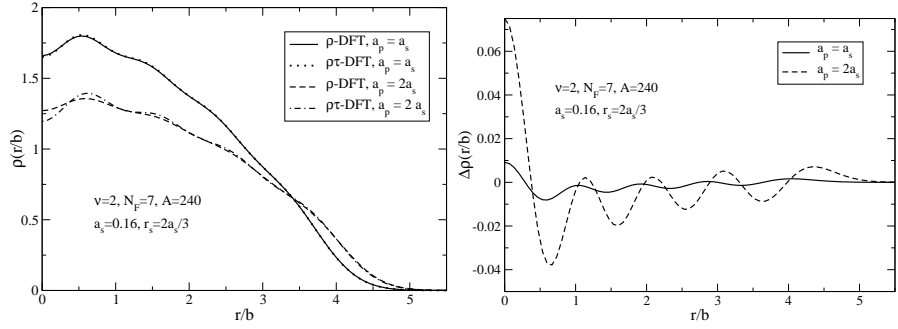

\centering
      \includegraphics*[width=2.3in]{DensityDistribution240}
     \includegraphics*[width=2.3in]{DensityDistributionDiff240}
   \caption{Comparing densities from energy functionals of $\rho$ only
     and $\rho$, $\tau$~\cite{Bhattacharyya:2004qm}.}
 \label{fig:17}       
\end{figure}
  
Now let's compare the dilute fermion functional to the phenomenological
Skyrme functional.
The Skyrme energy density functional (for $N = Z$) is
given in (\ref{eq:skyrmeE})
while the corresponding
dilute energy density functional for $\nu=4$ 
(and $V_{\rm external}=0$) is
  \beqa
    E[\rho,\tau,{\bf J}] 
      &=& \int\!d^3x\,
      \biggl\{ {{\tau\over 2M} + {3\over 8} C_0 \rho^2   
   + {1\over 16}(3 C_2 + 5 C'_2) \rho \tau}  
    {+ {1\over 64} (9C_2 - 5C'_2) (\bm{\nabla} \rho)^2}  
   \nonumber
    \\ & & \null
    - {3\over 4} C_2'' \rho \bm{\nabla}\cdot{\bf J}
    { \null + \frac{c_1}{2M} C_0^2  \rho^{7/3}
    + \frac{c_2}{2M} C_0^3  \rho^{8/3}
    + \frac{1}{16} D_0 \rho^3}
    + \cdots \biggr\} \;.
  \eeqa
They have the same terms after the association $t_i \leftrightarrow C_i$,
except that the Skyrme functional is missing non-analytic terms, the
three-body contribution, and other features.
We note after matching to empirical Skyrme coefficients that the 
``effective'' scattering length from $C_0$ is
$a_0 \approx -2\mbox{--}3\,\mbox{fm}$, but that
          $|\kf a_p|,\, |\kf r_0| < 1$ (with $a_p < 0$).
However, we want the Skyrme functional to account for the finite-ranged
pion, so it \emph{should not} be equivalent to a short-distance expansion.
Thus, the close correspondence suggests that the Skyrme functional is
lacking and should be generalized. 	  

\begin{figure}[t]
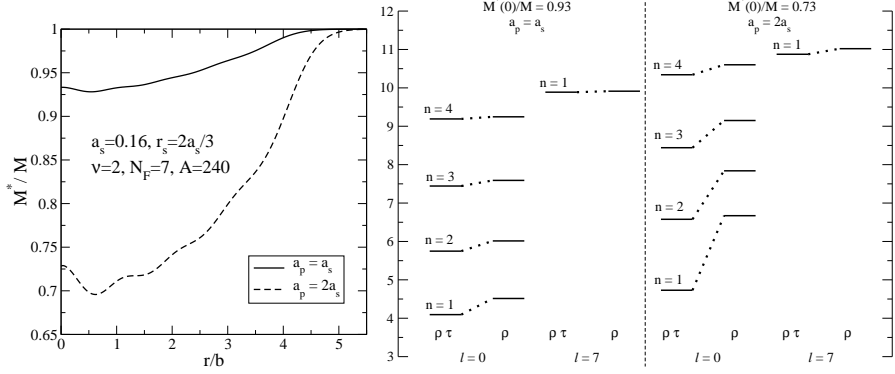

\centering
      \includegraphics*[width=1.9in]{EffectiveMass240n}
      \hfill
      \raisebox{0in}{%
     \includegraphics*[width=2.7in]{EnergySpectrum240b}}
   \caption{Comparing effective masses and single-particle spectra 
   from energy functionals of $\rho$ only
     and $\rho$, $\tau$~\cite{Bhattacharyya:2004qm}.}
 \label{fig:18}       
\end{figure}

It is useful to compare results from the 
$\rho$ only functional compared to the
$\rho$ and $\tau$ functional, as in Fig.~\ref{fig:17}
and these tables:
\begin{center}
 \renewcommand{\tabcolsep}{10pt}
 \begin{tabular}{ccc}
    { $a_p = a_s$} & $E/A$ & $\sqrt{\langle r^2\rangle}$ \\ \hline
    $\rho$ & 7.66 & 2.87 \\
    $\rho\tau$ & 7.65 & 2.87 \\ \hline
 \end{tabular}
\hspace*{.4in}
 \begin{tabular}{ccc}
    { $a_p = 2 a_s$} & $E/A$ & $\sqrt{\langle r^2\rangle}$ \\ \hline
    $\rho$ & 8.33 & 3.10 \\
    $\rho\tau$ & 8.30 & 3.09 \\ \hline
 \end{tabular}
\end{center} 

\noindent
We see very little difference in the Kohn-Sham observables, which are
the binding
energy and the density distribution.  However, the single-particle
Kohn-Sham spectrum, which is \emph{not} an observable, shows significant
differences, as evidenced by Fig.~\ref{fig:18}. 
(Note: The effective mass $M^*$ is closely
related to single-particle levels.)
We can show for a uniform system that the HF single-particle levels satisfy
\beq
 {\varepsilon_{\bf k}^{\rho} - \varepsilon_{\bf
         k}^{\rho\tau}}
          = \frac{\pi}{\nu}[(\nu-1) a_s^2 r_s + 2(\nu+1) a_p^3]\, 
          \frac{k_{F}^2 - {\bf k}^2}{2M}\rho
	  \; ,
	  \label{eq:vareps}
\eeq
and the $\rho\tau$ result is the one corresponding to the spectrum
from the full Hartree-Fock propagator. 
So the issue becomes how the full $G$ (as shown below
with the self-energy) is related to the Kohn-Sham 
$G_{\rm ks}$~\cite{Bhattacharyya:2004aw};
the closer they are, the better approximation $G_{\rm ks}$ will
provide for single-particle properties.
 
 \begin{center}
  \includegraphics*[angle=0.0,width=2.5in]{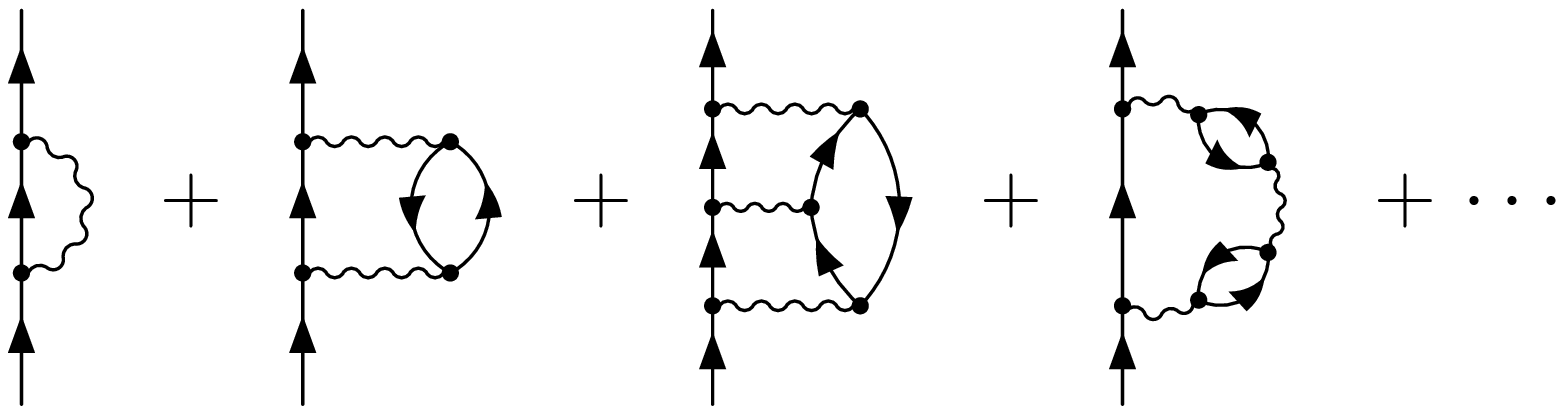}
      \raisebox{.31in}{ \Lra}
  \includegraphics*[angle=0.0,width=1.7in]{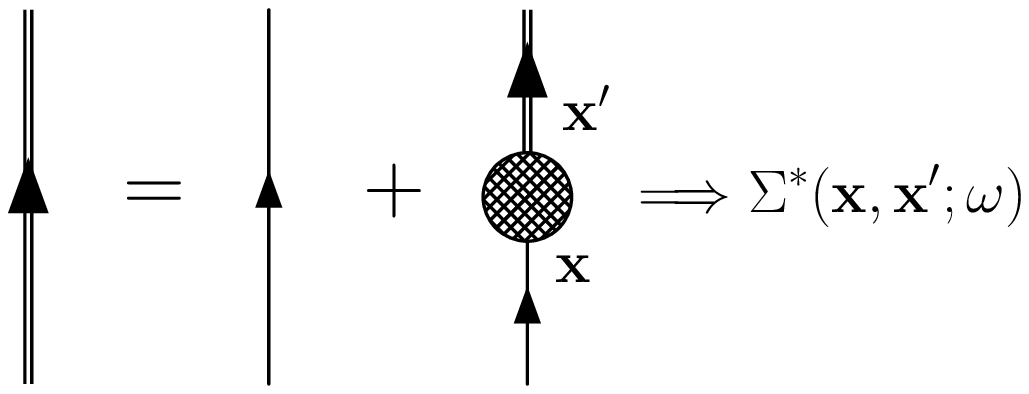}
  \end{center}

To explore this connection, we add a non-local source $\xi(x',x)$ coupled to
$\psi(x)\psi^\dag(x')$ [we're back in Minkowski space here for
no particular reason!]:
 \beq
    Z[J,\xi] = e^{iW[J,\xi]}
      = \int\! D\psi D\psi^\dag \ e^{i\int\! d^4x\ [{\cal L}\,+\,   J(x)
    \psi^\dag(x) \psi(x)\, { +\,
    \int\! d^4x'\, \psi(x)\xi(x,x')\psi^\dag(x')}]}
    \; .
\eeq
Writing $\Gamma[\rho,\xi] = \Gamma_0[\rho,\xi] 
     + \Gamma_{\rm int}[\rho,\xi]$,
 \beq
   G(x,x') = \left. \frac{\delta W}{\delta \xi}\right|_J
     = \left. \frac{\delta \Gamma}{\delta \xi}\right|_\rho
     = G_{\rm ks}(x,x') + G_{\rm ks}\Bigl[ 
     \frac{1}{i}\frac{\delta\Gamma_{\rm int}}{\delta G_{\rm ks}}
     + \frac{\delta\Gamma_{\rm int}}{\delta \rho}
     \Bigr] G_{\rm ks} \; ,
 \eeq
which is represented diagrammatically as:   

\begin{figure}[h]
  \begin{center} 
   \includegraphics*[width=2.7in]{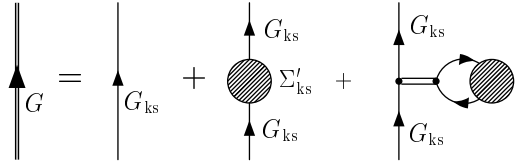}
 \end{center}
 \vspace*{-.1in}
 \caption{Full Green's function $G$ in terms of the
  Kohn-Sham Green's function $G_{\rm ks}$.}
  \label{fig:20a}
\end{figure} 

\noindent 
Now $G$ and $G_{\rm ks}$ yield the same density by 
\emph{construction}; that is,
$\rho_{\rm ks}(\bfx) = -i\nu G_{\rm KS}^0(x,x^+)$
    equals $\rho(\bfx) = -i\nu G(x,x^+)$.
Here is a simple diagrammatic demonstration
(the double line is minus the inverse of a single ph ring):
\begin{center}
   \includegraphics*[width=4.0in]{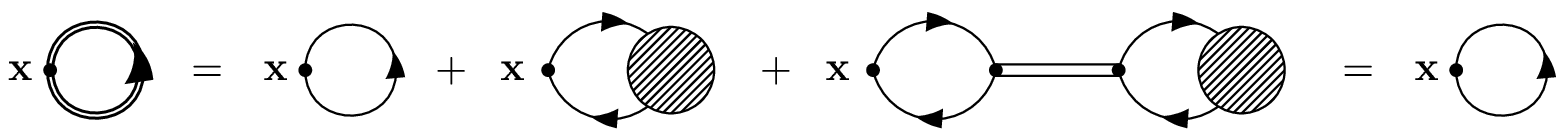}
\end{center}
But other single-particle properties (e.g., the spectrum)
are generally different,
since the last two terms in Fig.~\ref{fig:20a} will not cancel.

We can ask whether  
the Kohn-Sham basis is a useful one for $G$.
Or, more simply,
ask how close is $G_{\rm KS}$ to $G$.
We find that it depends on what sources are used,
as shown by the comparison of single-particle spectra
in (\ref{eq:vareps}), with more sources implying less
difference.
This is a topic that merits further investigation.   

\begin{figure}[t] 
  \begin{center}
    \includegraphics*[width=2.5in]{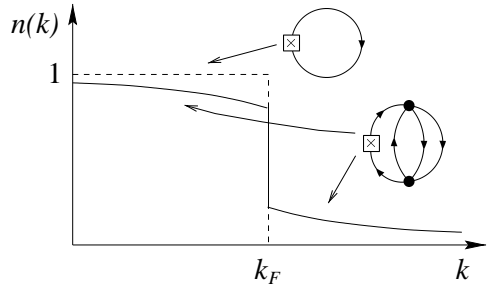}
  \end{center}
  \vspace*{-.1in}
  \caption{Schematic momentum occupation number $n(k)$
  for mean-field (Hartree-Fock) and with correlations.}
  \label{fig:occupf2}
\end{figure}
   
The comparison of   
Kohn-Sham DFT and ``mean-field'' models often leads
to misunderstandings, as when considering ``occupation numbers'',
because of a confusion between $G$ and $G_{\rm KS}$.
Figure~\ref{fig:occupf2} suggests that occupation numbers are
equal to 0 or 1 if and only if correlations are \emph{not} included.  
The Kohn-Sham propagator \emph{always} has a ``mean-field'' structure,
which means that (in the absence of pairing) the Kohn-Sham occupation
numbers are always 0 or 1.
But correlations are certainly included
in $\Gamma[\rho]$!
(In principle, all correlations can be included; in practice, certain
types like long-range particle-hole correlations may be largely omitted.)
Further, 
$n({\bf k}) = \langle a^\dagger_{\bf k} a^{\phantom{\dagger}}_{\bf k}
            \rangle$ is resolution dependent (not an observable!);
the operator related to experiment is more complicated.
Additional discussion on these issues can be found in
\cite{Furnstahl:2001xq}.

\subsection{Pairing in Kohn-Sham DFT} 

\begin{figure}[t]
\centering
      \includegraphics*[width=2.1in]{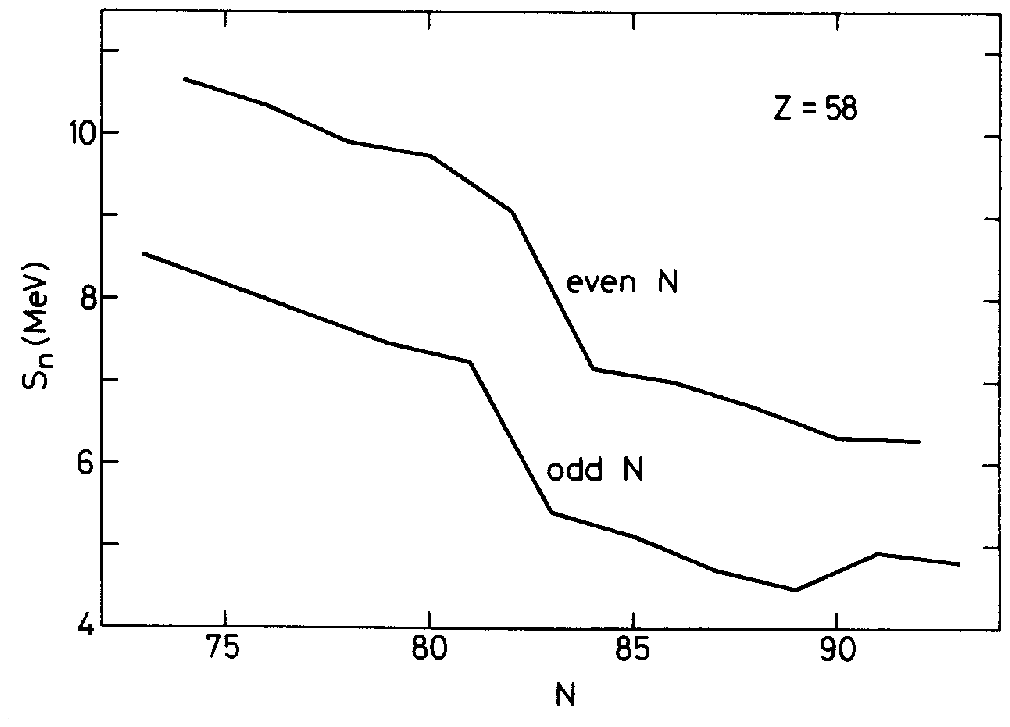}
      \hfill
      \includegraphics*[width=2.5in]{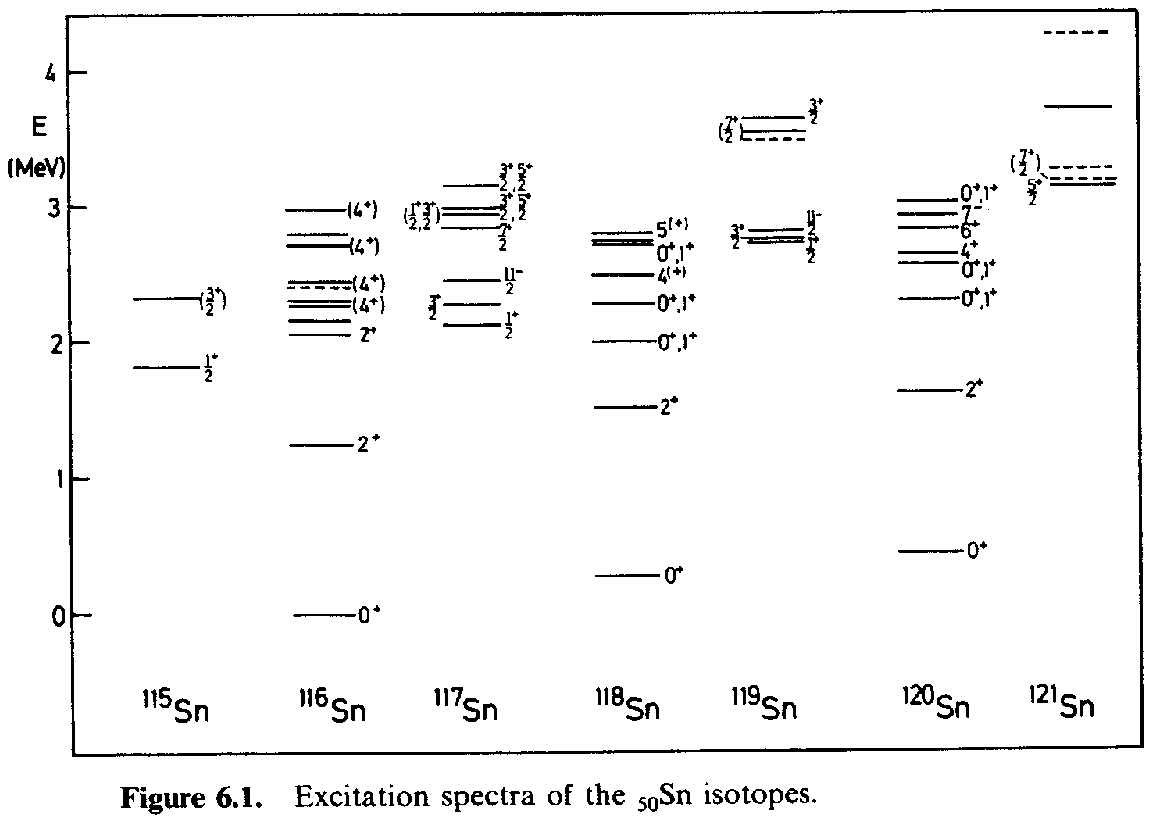}
   \caption{Evidence for pairing in nuclei (see text and
   \cite{RINGSCHUCK}).}
 \label{fig:19}       
\end{figure}

There is abundant evidence for pairing in nuclei.
The semi-empirical mass formula reproduces nuclear masses only
with a pairing term (the last one):
 \beqa
     B(N,Z)  &=& 
     (15.6\,\mbox{MeV})
      \left[1-1.5\left(\frac{N-Z}{A}\right)^2 \right]A 
     - (17.2\,\mbox{MeV})A^{2/3}
    \nonumber \\ && \null
     - (0.70\,\mbox{MeV})\frac{Z^2}{A^{1/3}}
     + (6\,\mbox{MeV})[(-1)^N + (-1)^Z]/A^{1/2} \; , 
  \eeqa
which implies an odd-even staggering of binding energies
(left panel of Fig.~\ref{fig:19}).
Other evidence is
the energy gap in the spectra of deformed nuclei,
low-lying $2^+$ states in even nuclei (right panel of Fig.~\ref{fig:19}),
and deformations and moments of inertia (the theory requires pairing
to reproduce data).  

\begin{figure}[t]
\centering
       \includegraphics*[angle=0.0,width=1.2in]{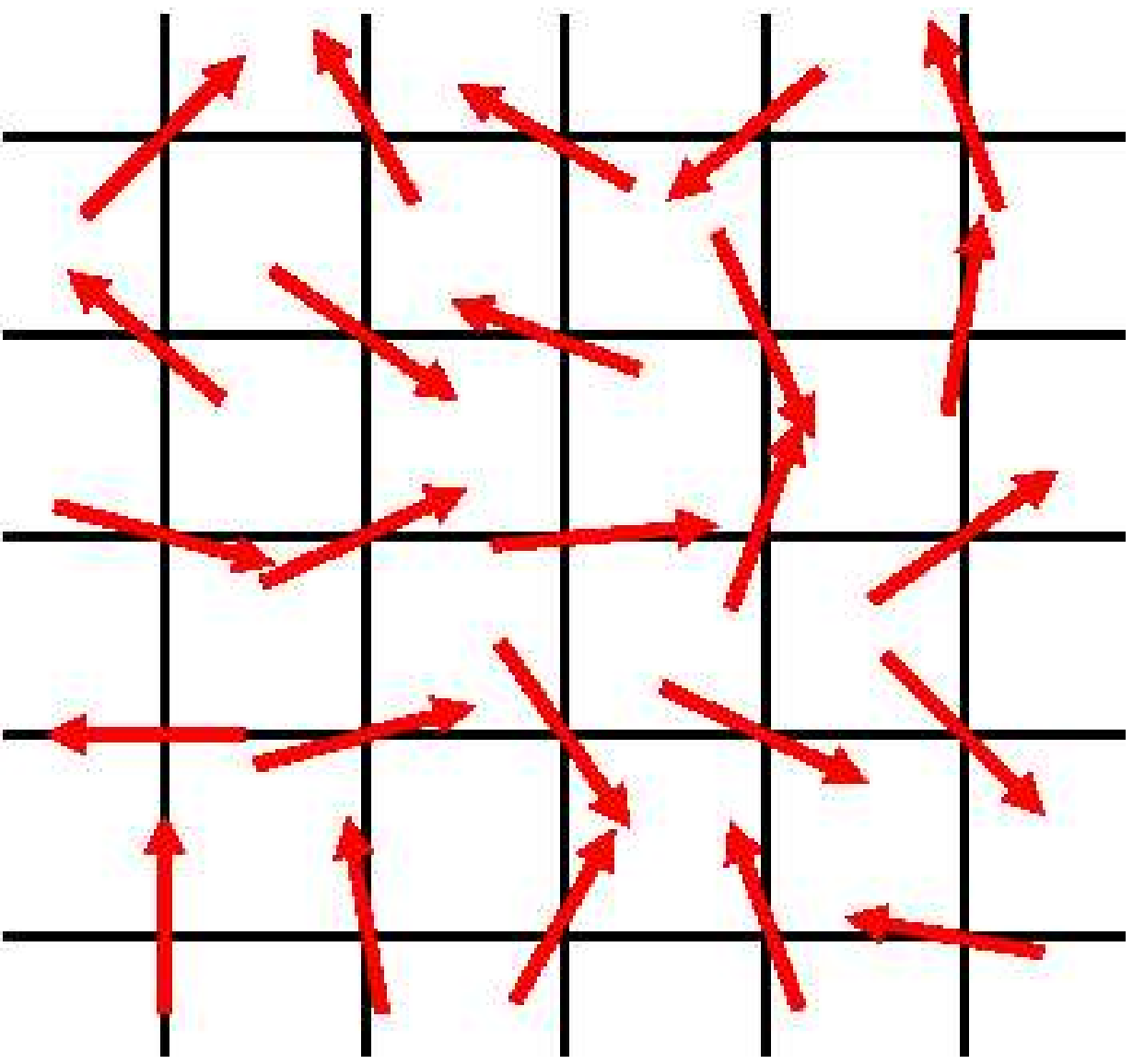}
       \hspace*{.2in}
       \includegraphics*[angle=0.0,width=1.2in]{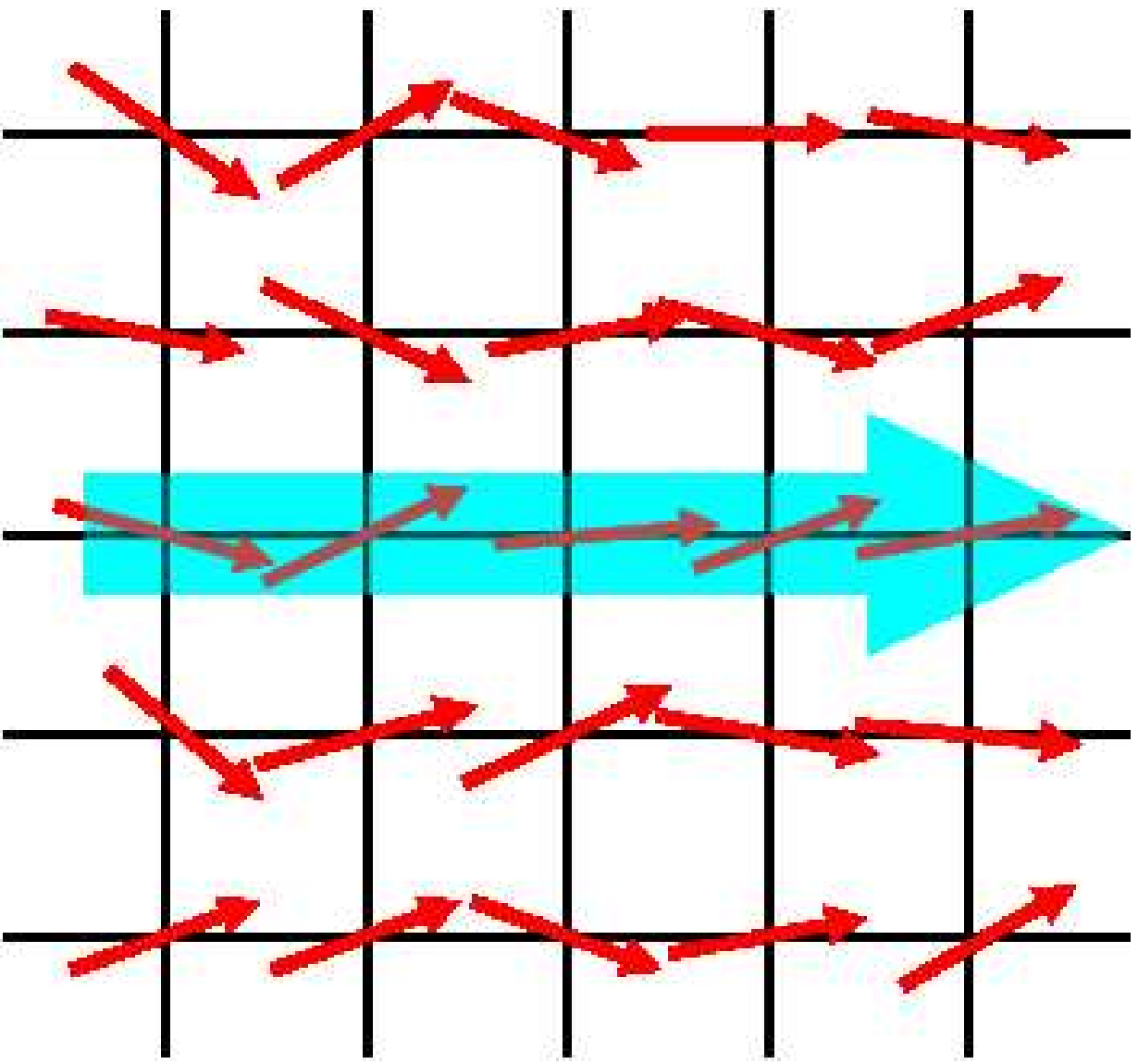}
       \hspace*{.2in}
       \includegraphics*[angle=0.0,width=1.2in]{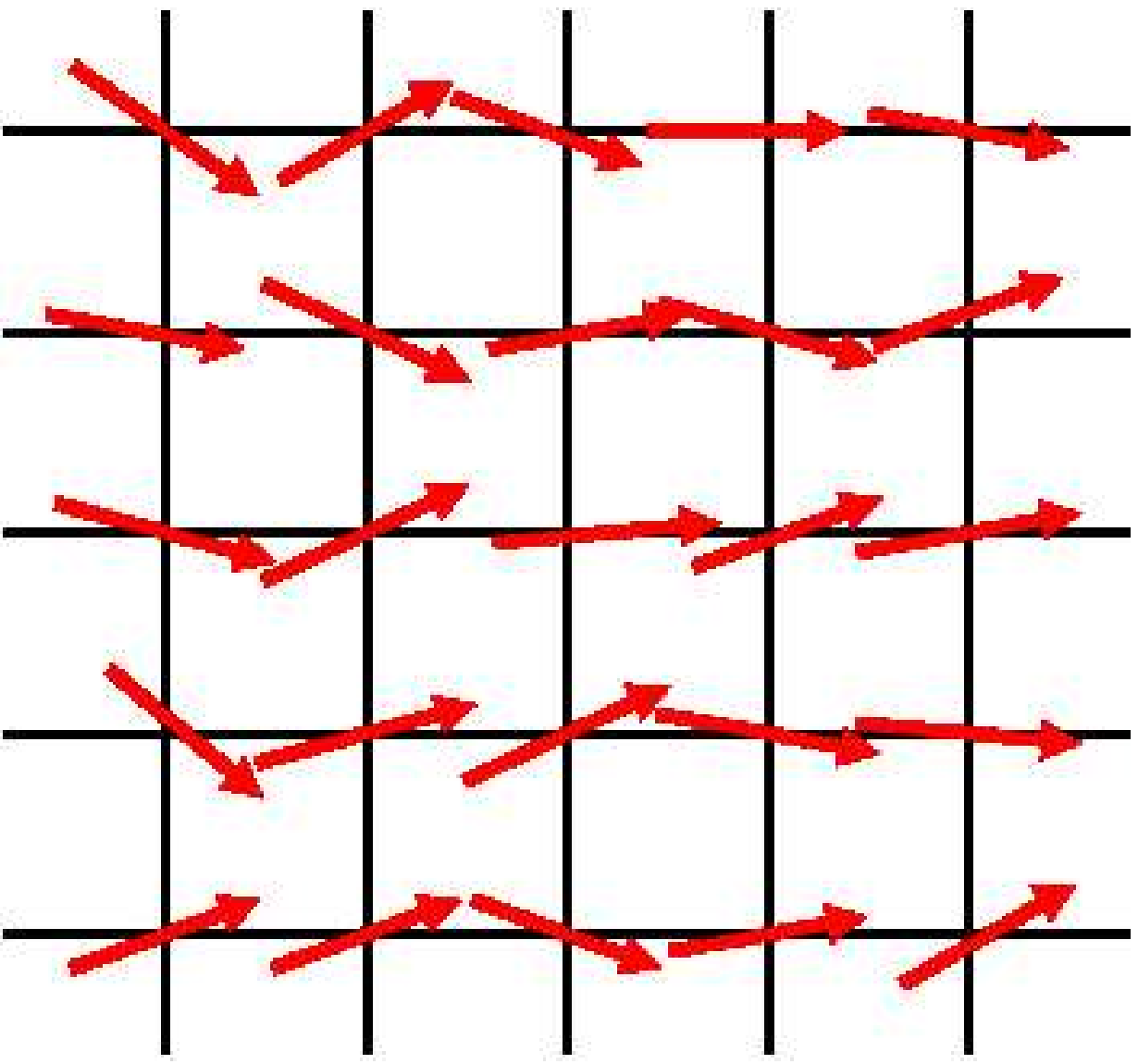}
   \caption{Spontaneous symmetry breaking analogy with spins.}
 \label{fig:21}       
\end{figure}
   
Pairing is an example of spontaneous symmetry breaking, which is
naturally accommodated in an effective action framework.   
For example, consider testing for zero-field magnetization $M$ in a spin
system by 
introducing an external field $H$ to break the rotational symmetry.
Legendre transform the Helmholtz free energy $F(H)$:
     \beq
      \mbox{\normalsize invert\ \ } M = -\partial F(H)/\partial H \quad
        \Longrightarrow \quad  \Gamma[M] = F[H(M)] + M H(M)
	\; .
     \eeq
Since $H = \partial \Gamma/\partial M \longrightarrow 0$, we look
for the stationary points of 
$\Gamma$ to identify possible ground states, including whether
the symmetry broken state is lowest.
           
For pairing, the broken symmetry is a $U(1)$ [phase] symmetry.
The textbook effective action treatment 
in condensed matter 
is to introduce a contact interaction~\cite{NAGAOSA,STONE}:
$g\,\psi^\dagger\psi^\dagger\psi\psi$,
and perform a  Hubbard-Stratonovich transformation with an auxiliary
pairing field $\hat\Delta(x)$ 
coupled to $\psi^\dagger\psi^\dagger$,
which eliminates the contact interaction.
Then one constructs the 1PI effective action $\Gamma[\Delta]$ with
$\Delta = \langle \hat\Delta \rangle$,
and looks for values for which 
$\delta\Gamma/\delta\Delta = 0$.
To leading order in the loop expansion (mean field), this yields the 
BCS weak-coupling gap equation with gap $\Delta$.
              
The natural alternative here
is to combine an expansion (e.g., EFT power counting) and the 
\emph{inversion} method for effective actions~\cite{Fukuda:im,VALIEV97,VALIEV}.
Thus we introduce another external current $j(x)$, which is coupled
to the fermion pair density in order to explicitly breaks the 
phase symmetry.
This is a natural generalization of Kohn-Sham
DFT~\cite{Bulgac:2001ai,Bulgac:2001ei,Yu:2002kc}.
 cf.\ DFT with nonlocal source~\cite{OLIVEIRA88,KURTH99}.
           
So we consider a local composite effective action with 
pairing~\cite{Furnstahl:2006pa}.
The generating functional has sources $J,\pairj$
coupled to the corresponding densities:
 \beq
     Z[J,\/\pairj] = e^{-W[J,\/\pairj]}
       = \int\! D(\psi^\dag\psi) 
       \, e^{-\!\int\! d^4x\ [{\cal L} +   J(x)
     \,\psi_\alpha^\dag \psi_\alpha
     + {\pairj(x)(\psidagup\psidagdown + \psidown\psiup)}]}
     \; .
 \eeq
Densities are found by functional derivatives with respect to $J$
and $\pairj$:
 \beq
   \rho(x)\equiv \langle \psi^\dag(x)\psi (x)\rangle_{J,\pairj}
   = \left.\frac{\delta W[J,\pairj]}{\delta J(x)}\right|_{\pairj} 
   \; ,
 \eeq
and
 \beq
  {\phi(x) \equiv \langle 
     \psidagup(x)\psidagdown(x)+ \psidown(x)\psiup(x)
   \rangle_{J,\pairj}
   = \left.\frac{\delta W[J,\pairj]}{\delta \pairj(x)}\right|_{J} }
   \; .
 \eeq
The effective action $\Gamma[\rho,\phi]$
follows as before by functional Legendre transformation: 
    \beq
    \Gamma[\rho,\phi]  = 
      W[J,\pairj] - \int\! d^4x\, J(x)\rho(x)
      - \int\! d^4x\, \pairj(x)\phi(x)
      \; ,
    \eeq
and is proportional to the (free) energy functional $E[\rho,\phi]$; 
at finite temperature, the proportionality constant is $\beta$.
The sources are given by functional derivatives
wrt $\rho$ and $\phi$:
     \beq
         \frac{\delta E[\rho,\phi]}{\delta\rho({\bf x})}
       = J({\bf x})
       \qquad\mbox{and}\qquad
         \frac{\delta E[\rho,\phi]}{\delta\phi({\bf x})}
       = \pairj({\bf x})
       \; .
     \eeq
But the sources are zero in the ground state,
so we determine the ground-state $\rho({\bf x})$
     and $\phi({\bf x})$ by stationarity:
      \beq
         \left.
           \frac{\delta E[\rho,\phi]}{\delta \rho({\bf x})}
          \right|_{{\rho=\rho_\grounds,\phi=\phi_\grounds}}
          =
         \left.
           \frac{\delta E[\rho,\phi]}{\delta \phi({\bf x})}
          \right|_{{\rho=\rho_\grounds,\phi=\phi_\grounds}}
          = 0
	  \; .
      \eeq
This is Hohenberg-Kohn DFT extended to pairing!

We need a method to carry out the Legendre transforms
to get Kohn-Sham DFT; an obvious choice is to apply 
the Kohn-Sham inversion method again,
with order-by-order matching in the counting parameter $\lambda$. 
Once again,
  \beqau
    \mbox{ diagrams}&\Longrightarrow&
    {W}[J,\pairj,\lambda] = 
    {{W}_0[J,\pairj]} + \lambda{W}_1[J,\pairj] +
    \lambda^2{W}_2[J,\pairj] 
    + \cdots 
    \\
   \mbox{ assume}&\Longrightarrow&
    J[\rho,\phi,\lambda] = 
    {J_0[\rho,\phi]} + \lambda J_1[\rho,\phi] +
    \lambda^2J_2[\rho,\phi] + \cdots 
     \\
   \mbox{ assume}&\Longrightarrow&
    \pairj[\rho,\phi,\lambda] = {\pairj_0[\rho,\phi]} + 
    \lambda\pairj_1[\rho,\phi] +
    \lambda^2\pairj_2[\rho,\phi] + \cdots 
     \\
   \mbox{ derive}&\Longrightarrow&
    {\Gamma}[\rho,\phi,\lambda] = 
    {{\Gamma}_0[\rho,\phi]} 
    + \lambda{\Gamma}_1[\rho,\phi] + \lambda^2{\Gamma}_2[\rho,\phi] +
    \cdots 
  \eeqau
 Start with the exact expressions for $\Gamma$ and $\rho$ 
 \beq
   \Gamma[\rho,\phi] = W[J,j] - \int\! J\,\rho 
      - \int\! j\,\phi \;,
 \eeq
 and
 \beq     
    \rho(x)
      = \frac{\delta W[J,j]}{\delta J(x)}\,, \quad
      \phi(x) = \frac{\delta W[J,j]}{\delta j(x)} \; ,   
 \eeq     
and plug in the expansions, with $\rho,\phi$ treated as order unity.
Zeroth order is the Kohn-Sham system with potentials 
$J_0({\bf x})$ and $\pairj_0({\bf x})$,
 \beq
   \Gamma_0[\rho,\phi] = W_0[J_0,j_0] - \int\! J_0\,\rho 
      - \int\! j_0\,\phi \; ,
 \eeq     
so the \emph{exact} densities $\rho({\bf x})$ and $\phi({\bf x})$
are  by \emph{construction}
 \beq
   \rho(x)
      = \frac{\delta W_0[J_0,j_0]}{\delta J_0(x)}\,, \quad
      \phi(x) = \frac{\delta W_0[J_0,j_0]}{\delta j_0(x)} \; .   
 \eeq     
Now introduce single-particle orbitals and solve
 \beq
  \left(
    \begin{array}{cc}
     h_0(\xvec) - \mu_0 & \pairj_0(\xvec) \\
     \pairj_0(\xvec)         & -h_0(\xvec) + \mu_0 
    \end{array}
  \right)
  \left(
    \begin{array}{c}
    u_i(\xvec) \\ v_i(\xvec)
    \end{array}
  \right)
  = E_i
  \left(
    \begin{array}{c}
    u_i(\xvec) \\ v_i(\xvec)
    \end{array}
  \right)
 \eeq
where
  \beq
   h_0 (\xvec) \equiv -\frac{\bm{\nabla}^2}{2M} 
      + V_{\rm trap}(\xvec)
      -J_0(\xvec) \; .
  \eeq
This is just like Skyrme Hartree-Fock Bogliubov approach~\cite{RINGSCHUCK}.  

The diagrammatic expansion of the $W_i$'s is the same as without pairing, 
except now
lines in diagrams are KS Nambu-Gor'kov Green's functions, 
\begin{center}
    \includegraphics*[angle=0.0,width=4.5in]{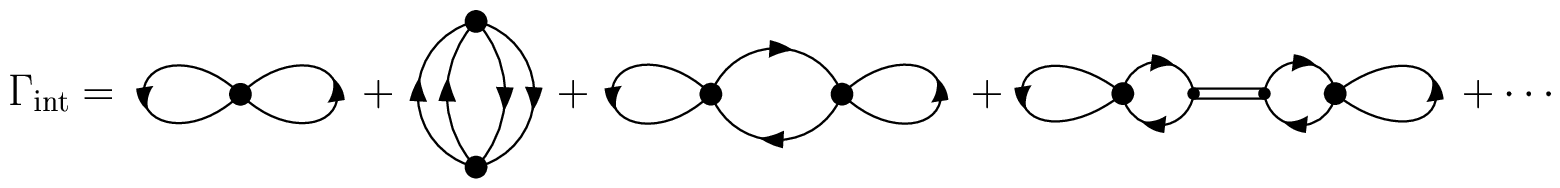}
\end{center}  
 \beq
   \renewcommand{\arraycolsep}{2pt}
   \hspace*{-.1in}
   {\bf G} = 
     \left(
     \begin{array}{cc}
     \langle T_\tau\psiup(x)\psidagup(x') \rangle_0 &
     \langle T_\tau\psiup(x)\psidown(x') \rangle_0 \\
     \langle T_\tau\psidagdown(x)\psidagup(x') \rangle_0 &
     \langle T_\tau\psidagdown(x)\psidown(x') \rangle_0 
     \end{array}
     \right)
    \equiv
    \left(
     \begin{array}{cc}
     \GKS^0  &  \FKS^0 \\
     {\FKS^0}^\dagger & -\wt G_{\ks}^0
     \end{array}
    \right)
    \; . 
 \eeq
The extra diagram shown follows from the inversion (here it removes anomalous
diagrams).
In frequency space, the Kohn-Sham Green's functions are
  \beq
     \GKS^0 (\xvec, \xvec'; \omega) = \sum_j\, \left[
      \frac{u_j (\xvec)\, u_j^* (\xvec')}{i\omega - E_j} 
      + \frac{v_j (\xvec')\, v_j^* (\xvec)}{i\omega + E_j} 
      \right] \; ,
  \eeq
  \beq
     \FKS^0 (\xvec, \xvec'; \omega) = -\sum_j\, \left[
      \frac{u_j (\xvec)\, v_j^* (\xvec')}{i\omega - E_j} 
      - \frac{u_j (\xvec')\, v_j^* (\xvec)}{i\omega + E_j} 
      \right] \; .
  \eeq

The Kohn-Sham self-consistency procedure
involves the same iterations 
as in phenomenological Skyrme HF (or relativistic mean-field) 
when pairing is included.
In terms of the orbitals, the fermion density is
 \beq
   \rho(\xvec) =
   2\sum_i\, |v_i(\xvec)|^2 \; ,
 \eeq
 and the pair density is (warning: this is unrenormalized!)
 \beq
   \phi(\xvec) =
   \sum_i\, [ u_i^*(\xvec) v_i(\xvec) + u_i(\xvec) v_i^*(\xvec) ]
   \; .
 \eeq
The chemical potential $\mu_0$
is fixed by $\int\!\rho(\xvec) = A$.
Diagrams for
$\Gamma[\rho,\phi] \propto E_0[\rho,\phi] + E_{\rm int}[\rho,\phi]$ 
yield the Kohn-sham potentials
  \beq
    J_0(\xvec)\Bigr|_{\rho=\rho_{\rm gs}} = 
      \left.\frac{\delta E_{\rm int}[\rho,\phi]}{\delta \rho(\xvec)}
    \right|_{\rho=\rho_{\rm gs}}
  \  \mbox{and} \quad
    \pairj_0(\xvec)\Bigr|_{\phi=\phi_{\rm gs}} = 
      \left.\frac{\delta E_{\rm int}[\rho,\phi]}{\delta \phi(\xvec)}
    \right|_{\phi=\phi_{\rm gs}} \; .
  \eeq

\subsection{Renormalization of Pairing}    
  
When we carry out the DFT pairing
calculation for a uniform dilute Fermi system,
we find divergences almost immediately.
The generating functional with constant sources ${\mu}$ 
and ${\pairj}$ is:
 \beqa
    e^{-W[\mu,j]}
      &=& \int\! D(\psi^\dag\psi) 
      \exp\Bigl\{ -\!\int\! d^4x\  \bigl[\psi_\alpha^\dagger
        (\frac{\partial}{\partial\tau}  
      - \frac{\bm{\nabla}^{\,2}}{2M} - {\mu} )
             \psi_\alpha\,
    \\
    & & \hspace*{0.2in}
  \null  + \frac{C_0}{2}\psidagup\psidagdown\psidown\psiup
    + \, {\pairj}(\psiup\psidown + \psidagdown\psidagup )
     {\bigr]\Bigr\}}
	         {+ \frac12\zeta\, j^2}\,\bigr]\Bigr\}
\eeqa
(cf.\ adding an integration over an auxiliary field 
$\int\! D(\Delta^\ast,\Delta)\ e^{-\frac{1}{|C_0|} \int\!
|\Delta|^2}$,
then shifting variables to eliminate
$\psidagup\psidagdown\psidown\psiup$
for $\Delta^\ast \psiup\psidown$).     
There are new divergences because of $\pairj$,
e.g., expand $W$ to ${\cal O}(\pairj^2)$:
   \begin{center}
    \includegraphics*[angle=0.0,width=3.in]{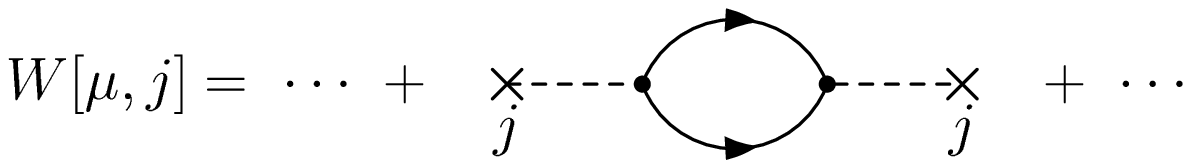}
   \end{center}
which has the same linear divergence as in 2-to-2 scattering.
To renormalize, we add the counterterm 
$\frac12\zeta |\pairj|^2$ to ${\cal L}$
(see~\cite{ZINNJUSTIN}), which is
additive to $W$ (cf. $|\Delta|^2$), so there is no effect on scattering.
 
We'll use dimensional regularization again, but
generalize from DR/MS (as used by Papenbrock and Bertsch~\cite{PaB99}) 
to DR/PDS, which generates explicit $\Lambda$ dependence to
``check'' renormalization (by verifying that $\Lambda$ dependence cancels).
The basic free-space integral in $D$ spatial dimensions is  
 \beq 
   \left(\frac{\Lambda}{2}\right)^{3-D} \int\! \frac{d^D k}{(2\pi)^D}
     \, \frac{1}{p^2-k^2+i\epsilon}
     \stackrel{{\rm PDS}}{\longrightarrow} -\frac{1}{4\pi} (\Lambda + ip)
     \; ,
 \eeq
where  $\int\! \frac{1}{\epsilon_k^0}
\rightarrow \frac{M\Lambda}{2\pi}$.
Renormalizing and matching free-space scattering yields for $C_0(\Lambda)$:
  \beq
   C_0(\Lambda) 
     =  \frac{4\pi a_s}{M}\frac{1}{1-a_s\Lambda}
     = {\frac{4\pi a_s}{M}} + {\frac{4\pi a_s^2}{M} \Lambda} + 
     {\cal O}(\Lambda^2) = {C_0^{(1)}} + {C_0^{(2)}} + \cdots
  \eeq
Note: we recover DR/MS by taking $\Lambda= 0$.
As an exercise, you
can verify that NLO renormalization 
in free space (left):
\begin{center}
   \includegraphics*[angle=0.0,width=1.3in]{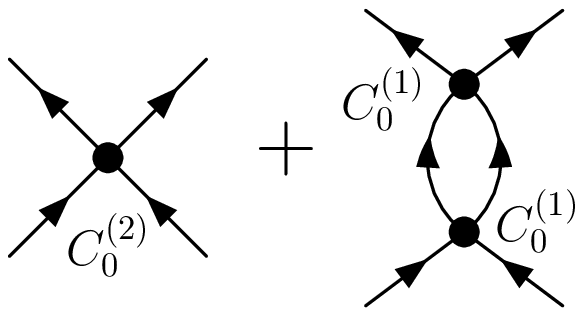}
        \quad \raisebox{.3in}{\Lra} \quad
    \raisebox{-.0in}{%
   \includegraphics*[angle=0.0,width=1.5in]{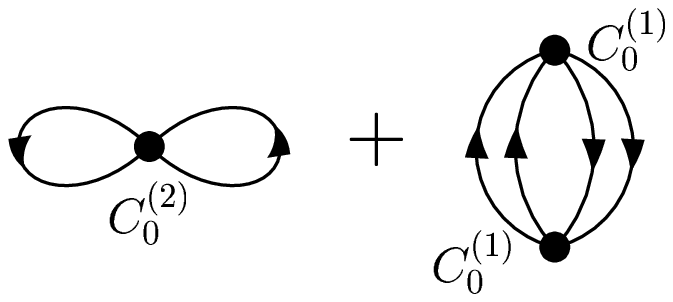}}
\end{center}
implies that the corresponding sum of diagrams at finite density (right)
is independent of $\Lambda$.

\begin{figure}[t]
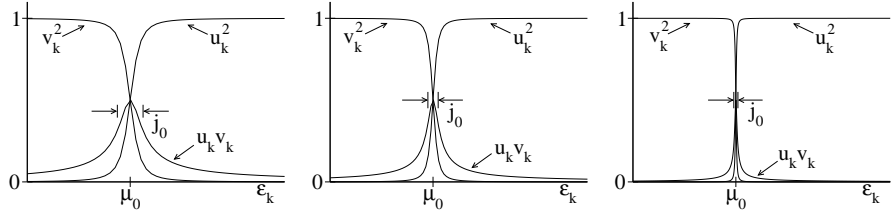

\centering
       \includegraphics*[angle=0.0,width=1.45in]{occupation2_bw}
       \hspace*{.05in}
       \includegraphics*[angle=0.0,width=1.45in]{occupation3_bw}
       \hspace*{.05in}
       \includegraphics*[angle=0.0,width=1.45in]{occupation4_bw}
   \caption{Quasiparticle wave functions for a uniform system
   for several values of $j_0/\mu_0$.
   As $j_0/\mu_0$ decreases, $u_k v_k$ becomes sharply
   peaked at $\mu_0$.}
 \label{fig:22}       
\end{figure}

\begin{figure}[t]
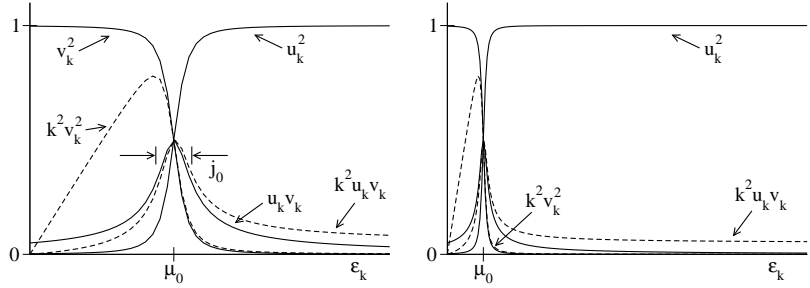

\centering
       \includegraphics*[angle=0.0,width=2.0in]{occupation2_ksq_bw}
       \hspace*{.1in}
       \includegraphics*[angle=0.0,width=2.0in]{occupation2b_ksq_bw}
   \caption{Illustration of the divergence
   in $\sum_k u_k v_k$.}
 \label{fig:23}       
\end{figure}

Now consider the Kohn-Sham noninteracting system for a uniform system,
where we have constant chemical potential $\mu_0$ and pairing source
$j_0$ (rather than spatially dependent sources). 
The bare density $\rho$ is: 
    \beq
     \rho =
       -\frac{1}{\beta V}\frac{\partial W_0[\mu_0,j_0]}{\partial \mu_0} = 
       \frac{2}{V}\sum_{\bf k} v_k^2
       =
        \int\! \frac{d^3k}{(2\pi)^3}\,
            \biggl(1 - \frac{\epsilon_k^0-\mu_0}{E_k} \biggr) 
	    \; ,          
    \eeq
and the    
bare pair density $\phi_B$ is: 
    \beq
     \phi_B =
       \frac{1}{\beta V}\frac{\partial W_0[\mu_0,j_0]}{\partial j_0} = 
       \frac{2}{V}\sum_{\bf k} u_k v_k 
        =
       - \int\! \frac{d^3k}{(2\pi)^3}\,
            \frac{j_0}{E_k}
	    \; .  
    \eeq
In these expressions, $j_0$ plays role of constant gap; e.g., the
spectrum is
     \beq
       {E_k = \sqrt{(\epsilon_k^0-\mu_0)^2 + j_0^2}}
       \; ,
       \qquad
       \epsilon_k^0 = \frac{k^2}{2M} \; .
     \eeq
(See also Fig.~\ref{fig:22}.) 
The divergence in $\phi_B$ is illustrated in Fig.~\ref{fig:23}.    
   
The basic DR/PDS integral in $D$ dimensions, 
    with $x\equiv j_0/\mu_0$, is
 \beqa
   {I(\beta)} &\equiv& \Bigl( \frac{\Lambda}{2}\Bigr)^{3-D} 
     \int\!  \frac{d^Dk}{(2\pi)^D}
    \, \frac{(\epsilon_k^0)^\beta}
            {\sqrt{(\epsilon_k^0-\mu_0)^2 + \pairj_0^2}}
	    \nonumber \\
    &=&
    {\frac{M\Lambda}{2\pi}
    \, \mu_0^\beta \,  \Bigl(1- \delta_{\beta,2} \frac{x^2}{2}
    \Bigr)}
    \nonumber
    \\ & & \quad
   \null + (-)^{\beta+1}\, \frac{M^{3/2}}{\sqrt{2}\pi} \,
      [\mu_0^2 (1+x^2)]^{(\beta+1/2)/2} \,
      P^0_{\beta+1/2}\Bigl(\frac{-1}{\sqrt{1+x^2}}\Bigr)
      \; .
 \eeqa 
We can
check that in the KS density equation the $\Lambda$ dependence cancels:
 \beq
  {\rho} =
    -\frac{1}{\beta V}\frac{\partial W_0[\,]}{\partial \mu_0} = 
     \int\! \frac{d^3k}{(2\pi)^3}\,
         \biggl(1 - \frac{\epsilon_k^0}{E_k} + \frac{\mu_0}{E_k}\biggr)
         \longrightarrow {0 - I(1) + \mu_0\, I(0)}
	 \; .
 \eeq
The KS equation for the pair density $\phi$ fixes $\zeta^{(0)}$: 
 \beq
  {\phi} = 
    \frac{1}{\beta V}\frac{\partial W_0[\,]}{\partial j_0} = 
    - \int\! \frac{d^3k}{(2\pi)^3}\,
         \frac{j_0}{E_k} + \zeta^{(0)}j_0 
    \longrightarrow {- j_0\,I(0) + \zeta^{(0)}j_0}
 \eeq    
so that 
 \beq	
      \zeta^{(0)} = \frac{M\Lambda}{2\pi} \; .
  \eeq
 
Calculating to $n^{\rm th}$ order,
we find $\Gamma_{1 \leq i \leq n}[\rho,\phi]$ by first
constructing all of the 
$W_{1 \leq i \leq n}[\mu_0(\rho,\phi),j_0(\rho,\phi)]$,
including additional Feynman rules~\cite{VALIEV},
\begin{center}
    \includegraphics*[width=4in]{fig_Gamma_int2}
\end{center}  
So the procedure is to
calculate $\mu_i$, $j_i$ from $\Gamma_i$, then
use $\sum_{i=0}^n j_i = j \rightarrow 0$ to find $j_0$.
The renormalization conditions mean that there is
no freedom in choosing $C_0(\Lambda)$, so the
           $\Lambda$'s must cancel!
In leading order, diagrams for 
$\Gamma_1[\rho,\phi] = W_1[\mu_0(\rho,\phi),j_0(\rho,\phi)]$ are
\begin{center}
 \includegraphics*[width=4.6in]{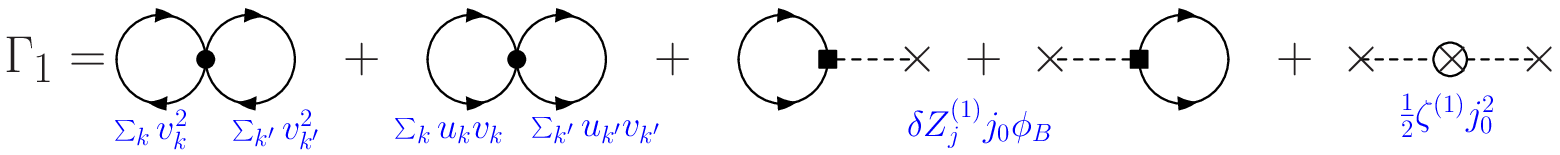}
\end{center} 
and we choose $\delta Z_j^{(n)}$ and $\zeta^{(n)}$
to convert $\phi_B$ to the renormalized $\phi$, yielding
 \beq
   {\frac{1}{\beta V} \Gamma_1[\rho,\phi]
     = \frac14 C_0^{(1)} \rho^2 + \frac14 C_0^{(1)} \phi^2}
     \quad \mbox{with\ } C_0^{(1)} = \frac{4\pi a_s}{M} \; .
 \eeq

The $\Gamma_1$ dependence on $\rho$ and $\phi$ is explicit,
so it is easy to find $\mu_1$ and $j_1$:
 \beq
   \mu_1 =  \frac{1}{\beta V} \frac{\partial \Gamma_1}{\partial \rho}
     = \frac12 C_0^{(1)} \rho 
   \qquad \mbox{and} \qquad
   j_1 = - \frac{1}{\beta V} \frac{\partial \Gamma_1}{\partial \phi}
     = -\frac12 C_0^{(1)} \phi  \; .
 \eeq 
The ``gap'' equation then follows from $\pairj = \pairj_0 + \pairj_1 = 0$:
 \beq
    \pairj_0 = -\pairj_1 = -\frac{1}{2} |C_0^{(1)}| \phi 
     =  
      \frac{1}{2} |C_0^{(1)}|\, \pairj_0 
      \left[
       \int\!\frac{d^3k}{(2\pi)^3}\,
        \frac{1}{\sqrt{(\epsilon_k^0-\mu_0)^2 + \pairj_0^2}}
       -  \zeta^{(0)} 
      \right] \; .
 \eeq
DR/PDS reproduces the Papenbrock/Bertsch result~\cite{PaB99} 
(with $x\equiv |j_0/\mu_0|$) 
 \beqa
  1 &=& -\sqrt{2M\mu_0} |a_s| (1+x^2)^{1/4} P^0_{1/2}
    \left(\frac{-1}{\sqrt{1+x^2}}\right)
    \nonumber \\
   & & \stackrel{x \rightarrow 0}{\longrightarrow}
     \kf a_s \Bigl[\frac{4 - 6\log{2}}{\pi} + \frac{2}{\pi}\log{x} \Bigr]
     \; ,
\eeqa 
and if $\kf a_s < 1$, then $j_0/\mu_0 = (8 / e^2) e^{-\pi/2\kf|a_s|}$ holds
to very good approximation.

The renormalized effective action $\Gamma = \Gamma_0 + \Gamma_1$ is
 \beq
   \frac{1}{\beta V} \Gamma
   = \int\! (\epsilon_k^0 - \mu_0 - E_k) 
     + \frac12 \zeta^{(0)}j_0^2 + \mu_0\rho - j_0\phi
     + \frac14 C_0^{(1)}\rho^2 + \frac14 C_0^{(1)} \phi^2 \; .
 \eeq
We check for $\Lambda$'s again,
\beq
 \frac{1}{\beta V} \Gamma
   = 0 - I(2) + 2\mu_0 I(1) - (\mu_0^2+j_0^2) I(0)
     + \frac12\frac{M\Lambda}{2\pi}j_0^2 + \cdots 
\eeq
and find they do cancel:
\beq
  \frac{M\Lambda}{2\pi}\Bigl(
    -\mu_0^2(1-j_0^2/2\mu_0^2) + 2\mu_0^2 - \mu_0^2 - j_0^2 + \frac12j_0^2
  \Bigr) = 0
  \; .
\eeq
 To find the energy density, evaluate $\Gamma$ at the stationary
    point  $j_0 = -\frac12 |C_0^{(1)}|\phi$ with $\mu_0$ 
    fixed
    by the equation for $\rho$, yielding the
same results as Papenbrock/Bertsch (plus an HF term)~\cite{Furnstahl:2006pa}.

\begin{figure}[t]
\centering
       \includegraphics*[angle=0.0,width=0.7in]{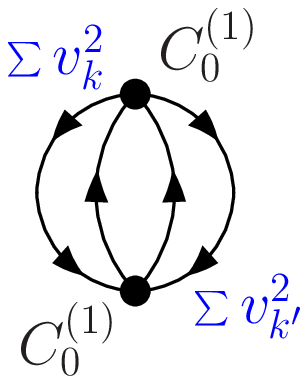}
       \hspace*{.2in}
       \includegraphics*[angle=0.0,width=1.9in]{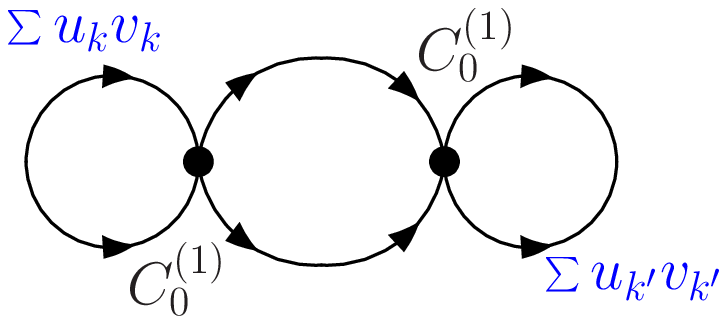}
   \caption{Contributions to the NLO energy density.}
 \label{fig:23b}       
\end{figure}

Life gets more complicated at Next-to-Leading Order (NLO), where
dependence of $\Gamma_2$ on $\rho$ and $\phi$
is no longer explicit
and analytic formulas for DR integrals not available.
$\Gamma_2$ at NLO is~\cite{Furnstahl:2006pa} 
\beqa
  & &
    - \bigl(C_0^{(1)}\bigr)^2
     \int\!\! \frac{d^3p}{(2\pi)^3}
     \int\!\! \frac{d^3k}{(2\pi)^3}
     \int\!\! \frac{d^3q}{(2\pi)^3}
     \frac{1}{E_p + E_k + E_{p-q} + E_{k+q}}
   \nonumber \\ & & \qquad\qquad \null\times
    \bigl[\,
      {u_p^2\, u_k^2\, v_{p-q}^2\, v_{k+q}^2}
      -2 u_p^2\, v_k^2\, (uv)_{p-q}\, (uv)_{k+q}
   \nonumber  \\ & & \qquad\qquad\qquad \null
      + (uv)_p\, (uv)_k\, (uv)_{p-q}\, (uv)_{k+q}
    \,\bigr]
\eeqa
and  
\beqa
 & & 
   - \bigl(C_0^{(1)}\bigr)^2
     \int\!\! \frac{d^3k}{(2\pi)^3}
   \,  \frac{1}{2 E_k}
    \bigl[
      \rho (u_k v_k)^2 + \frac12 {\phi_B} ({u_k^2} - v_k^2)
    \bigr]^2 \; .
\eeqa  
The  
UV divergences can be identified from 
   \beq
  v_k^2 = \frac12 \left( 1 - \frac{\xi_k}{E_k} \right)
     \stackrel{k\rightarrow\infty}{\longrightarrow}
        \frac{j_0^2 M^2}{k^4} \; ,
   \qquad
  u_k^2 = \frac12 \left( 1 + \frac{\xi_k}{E_k} \right)
     \stackrel{k\rightarrow\infty}{\longrightarrow}
       1 - \frac{j_0^2 M^2}{k^4} \; ,
   \eeq
and   
   \beq
  u_k v_k = - \frac{j_0}{2E_k}
     \stackrel{k\rightarrow\infty}{\longrightarrow}
    - \frac{j_0 M}{k^2} \; ,
   \qquad \qquad
  \frac{1}{E_k} 
     \stackrel{k\rightarrow\infty}{\longrightarrow}
  \frac{2M}{k^2} \; .
   \eeq
For the renormalization at NLO to work,
the bowtie diagram with the $C_0^{(2)} = 
(4\pi a_s^2/M)\Lambda$
vertex must precisely cancel the $\Lambda$ dependence from 
the beachball with $C_0^{(1)} = 4\pi a_s/M$ vertices:
 \begin{center}
   \includegraphics*[width=3.6in]{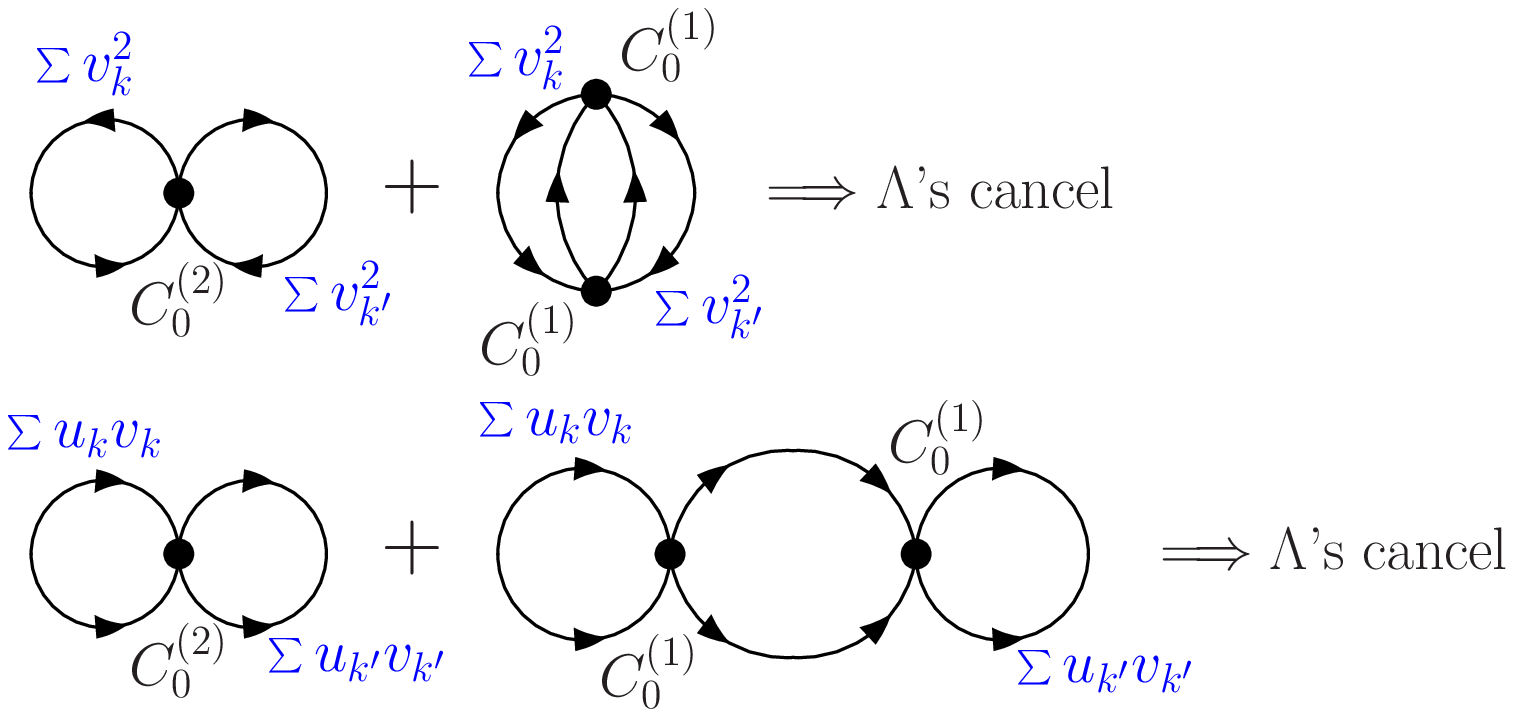}
 \end{center}
(Note that the $\delta Z_j^{(1)}$ vertex takes $\phi_B \rightarrow \phi$.) 
How do we see cancellation of $\Lambda$'s
and evaluate renormalized results without analytic formulas? 
 
Before addressing that issue, we first see how the       
standard induced interaction result~\cite{GORKOV61,HEISELBERG00} is recovered here.
As $j_0 \rightarrow 0$, $u_k v_k$ peaks at $\mu_0$ (see Fig.~\ref{fig:22}).
At leading order (for $T = 0$),
\beq 
  \Delta_{LO}/\mu_0 = \frac{\ts 8}{\ts e^2}\,e^{-1/N(0)|C_0|}
    = \frac{\ts 8}{\ts e^2}\,e^{-\pi/2\kf |a_s|} \; ,
\eeq
where we make the association $j_0 \rightarrow \Delta_{LO}$.    
At NLO the exponent is modified, which changes the prefactor,
$\Delta_{NLO} \approx \Delta_{LO}/(4 e)^{1/3}$, using 
   \begin{center}
    \includegraphics*[width=4.6in]{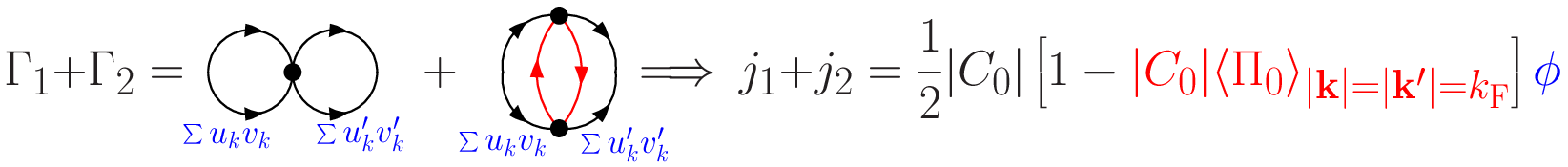}
   \end{center}
Further details can be found in \cite{Furnstahl:2006pa}.
An unexplored question is   
how does the Kohn-Sham gap compare to ``real'' gap?

Now we return to the question of renormalizing in practice; an alternative
approach is to use subtractions.
The NLO integrals with $E_k = \sqrt{(\epsilon_k-\mu_0)^2 + j_0^2}$
are intractable, but we directly obtain a renormalized
result  with the substitution 
\beq
  \int\!  \frac{1}{E_1 + E_2 + E_3 + E_4}  
\longrightarrow
  \int \biggl[ {\frac{1}{E_1 + E_2 + E_3 + E_4} 
   -  
    \frac{{\cal P}}{\epsilon^0_1 + \epsilon^0_2
        - \epsilon^0_3 - \epsilon^0_4}} \biggr] 
\eeq
plus a DR/PDS integral that is proportional to $\Lambda$.
When applied at LO,
  \beq
    \int\! \frac{1}{E_k}
      = \int\! \Bigl[ {\frac{1}{E_k} - \frac{\cal P}{\epsilon^0_k}}
	       \Bigr]
     + \frac{M\Lambda}{2\pi} \; .
  \eeq
This is the same sort of subtraction used
to eliminate $C_0$ in the gap equation,
 \beq
   \frac{M}{4 \pi a_s} + \frac{1}{|C_0|} = 
    \frac{1}{2} \int\!\frac{d^3k}{(2\pi)^3}\,
      \frac{1}{\epsilon_k^0}
     \Longrightarrow 
  \frac{M}{4\pi a_s} = -\frac{1}{2} 
   \int\!\frac{d^3k}{(2\pi)^3}\,
   \biggl[
      \frac{1}{E_k} -
      \frac{1}{\epsilon_k^0}
   \biggr]
   \; .     
 \eeq  
Any equivalent subtraction also works, e.g.,
 \beq
    \int\!\frac{d^3k}{(2\pi)^3}\, \frac{\cal P}{\epsilon^0_k - \mu_0}
      = \int\!\frac{d^3k}{(2\pi)^3}\, \frac{1}{\epsilon^0_k}
      \; .
 \eeq   

So how do we renormalize the divergent pair (anomalous) density,
 \beq
   \phi(\xvec) =
   \sum_i\, [ u_i^*(\xvec) v_i(\xvec) + u_i(\xvec) v_i^*(\xvec) ]
   \longrightarrow \infty \;,
 \eeq
 in a finite system? 
(Cf.\ the scalar density $\rho_s = \sum_i
\overline\psi(\xvec)\psi(\xvec)$ for relativistic Hartree theory).
Answer: use the subtracted expression for $\phi$ in the uniform system, 
 \beq
   \phi = \int^{k_c}\! \frac{d^3k}{(2\pi)^3} \, j_0 
      \left(
        \frac{1}{\sqrt{(\epsilon_k^0-\mu_0)^2 + j_0^2}}
        - \frac{1}{\epsilon_k^0}
      \right)
  \stackrel{k_c\rightarrow \infty}{\longrightarrow} \mbox{finite,}
  \label{eq:phi1}
 \eeq
and apply this in a local density approximation (Thomas-Fermi):
     \beq
       \phi(\xvec) = 2\sum_i^{E_c} u_i(\xvec)v_i(\xvec)
          - j_0(\xvec) \frac{M\, {k_c(\xvec)}}{2\pi^2}
       \quad \mbox{with} \quad
      {E_c = \frac{k_c^2(\xvec)}{2M} + J(\xvec) - \mu_0 }
      \; .
     \eeq
This procedure was worked out by Bulgac and
collaborators~\cite{Bulgac:2001ei,Bulgac:2001ai,Yu:2002kc}.
Convergence is very slow as the energy cutoff is increased, so
Bulgac/Yu devised a different subtraction,
 \beq
   \phi = \int^{k_c}\! \frac{d^3k}{(2\pi)^3} \, j_0 
      \left(
        \frac{1}{\sqrt{(\epsilon_k^0-\mu_0)^2 + j_0^2}}
      { - \frac{{\cal P}}{\epsilon_k^0-\mu_0} }
      \right)
  \stackrel{k_c\rightarrow \infty}{\longrightarrow} \mbox{finite}
  \; .
  \label{eq:phi2}
 \eeq
A comparison of convergence in uniform system for the two
subtraction schemes (\ref{eq:phi1}) and (\ref{eq:phi2}):  
  \begin{center}
     \includegraphics*[angle=0.0,width=2.4in]{convergence_plot_bw}
  \end{center}
shows dramatic improvement for the Bulgac/Yu subtraction.
Bulgac et al.\ have demonstrated that this works in finite systems, and there
have been recent practical applications.  

We finish this lecture with a brief mention of alternatives to a local 
Kohn-Sham formalism for pairing.
One alternative
is to couple a source to the \emph{non-local} pair field~\cite{OLIVEIRA88,KURTH99}: 
 \beq 
   \widehat H \longrightarrow
     \widehat H -  
       \int\! dx\,dx'\, [D^*(x,x')\psiup(x)\psidown(x') + \mbox{H.c.}]
       \; , 
 \eeq
which yields essentially a two-particle-irreducible (2PI)
effective action $\Gamma[\rho,\Delta]$
with $\Delta(x,x') = \langle \psiup(x)\psidown(x') \rangle$.
Or one could use auxiliary fields:
introduce $\widehat\Delta^*(x) \psi(x)\psi(x) + \mbox{H.c.}$
via a Hubbard-Stratonovich transformation to obtain a
1PI effective action in
$\Delta(x) = \langle \widehat\Delta(x)\rangle$.
By adopting a 
special saddle point evaluation, one can obtain Kohn-Sham DFT.
Finally there is the possibility of deriving a density functional
(without Kohn-Sham orbits) by direct renormalization group
evolution~\cite{Schwenk:2004hm}.


\section{Loose Ends and Challenges plus Cold Atoms}
\label{sec:4}


In this final lecture, we touch on some loose ends raised in previous
lectures, and outline some of the plans and challenges for moving
forward toward a microscopic DFT for nuclei based on effective field
theory and renormalization group ideas and
methods~\cite{Furnstahl:2004xn}.
We'll also briefly consider cold atom physics and some recent work
on density functionals for that problem. 

\subsection{Toward a Microscopic Nuclear DFT}

We have outlined a framework for generating density functional theory
based on effective actions.
A key ingredient is a tractable hierarchy of many-body approximations
to which we can apply the inversion method.
A scenario for carrying this out has emerged, which combines chiral
effective field theory (EFT) interactions with renormalization group
techniques.  While many challenges remain, it is a plausible and 
systematically improvable
path to a microscopic nuclear DFT.  

This scenario goes like this:
 \be
   \item Construct a chiral EFT to a given order, including all
    many-body forces.  At present, the NN chiral EFT has been worked
    out to N$^3$LO~\cite{N3LO,N3LOEGM}, 
    while three-body forces at the N$^2$LO level
    are used.  The latter will soon be extended to N$^3$LO and already
    the leading four-body force (which appears at N$^3$LO) has
    been tested.
    To minimize the truncation error following Lepage's prescription,
    one should increase the 
    cutoff regulator $\Lambda$ until the truncation error is minimized.
    (Note: it is still a matter of investigation where the breakdown scale
    actually lies.)

   \item Evolve the Hamiltonian to lower $\Lambda$ with 
   renormalization group (RG) methods.
   There are choices here, including the $V_{{\rm low}\,k}$ approach,
   the Similarity Renormalization Group
   (SRG)~\cite{Bogner:2006pc,Bogner:2007}, 
   and possibly simply
   a direct construction of the chiral EFT at a lower 
   cutoff~\cite{Coraggio:2007mc}.
   Cutoffs in the range of
   $\Lambda \approx 2\,\mbox{fm}^{-1}$ appear to be appropriate for
   ordinary nuclei.
   One needs the consistent evolution of 
   \emph{all} interactions \emph{and} other operators.
   As discussed in the first lecture,
   by decoupling high and low momentum the nuclear many-body problem
   becomes perturbative in the particle-particle channel, in stark
   contrast to the situation with 
   conventional interactions~\cite{Bogner_nucmatt,Bogner:2007}.

   \item Generate the density functional in effective action form.
   A by-product of evolving to low momentum is that the convergence
   of the many-body diagrams no longer is critically dependent on the
   choice of single-particle potential.  This opens the door to choosing
   it to maintain the density as in a Kohn-Sham approach.  In the short
   term, a direct construction of the functional in the Skyrme form
   is possible via an adaption (and extension) of Negele and Vautherin's 
   density
   matrix expansion (DME)~\cite{NEGELE72,NEGELE75}.  
   In the long term, chain-rule constructions
   will allow non-local effects to be included
   [see after (\ref{eq:J0d})].
 \ee
This program is well underway and is part of a larger project to
construct and constrain
a universal nuclear energy density functional (UNEDF).   
A detailed overview and an explanation
of the DME will be available in a forthcoming publication \cite{Platter}.

\newcommand{\rvec}{{\bf r}}

\newcommand{\rone}{{\bf r}_1}
\newcommand{\rtwo}{{\bf r}_2}
\newcommand{\svec}{{\bf s}}
\newcommand{\Rvec}{{\bf R}}

We'll briefly describe the idea of the DME~\cite{NEGELE72,NEGELE75}, 
which starts by expressing
the Hartree-Fock energy using the density matrix.
Recall that we take the best single Slater determinant in a variational sense
 \beq
   | \Psi_{\rm HF} \rangle = \det\{\psi_i(\xvec), i=1\cdots A  \}
   \,, \quad \xvec = ({\bf r}, \sigma, \tau) \; ,
 \eeq
to find the Hartree-Fock energy (suppressing $\sigma,\tau$):
  \beqa
   \hspace*{-.5in} 
   \langle  \Psi_{\rm HF} | \widehat H | \Psi_{\rm HF} \rangle
     &\!\!\!\!=\!\!\!\!&
    \cdots
   {+ \frac12\sum_{i,j=1}^A
     \int\!\! d\rone \! \int\!\! d\rtwo \, 
      |\psi_i(\rone)|^2 v(\rone,\rtwo) |\psi_j(\rtwo)|^2} 
    \\ & & \hspace*{-.7in}
   \null {- \frac12\sum_{i,j=1}^A
     \int\!\! d\rone \! \int\!\! d\rtwo \, 
      \psi^\dagger_i(\rone) \psi_i(\rtwo) v(\rone,\rtwo) 
      \psi^\dagger_j(\rtwo)\psi_j(\rone) } 
      \; .
  \eeqa
We can trivially express this in
terms of the single-particle density matrix:
     \beq 
        \rho(\rone,\rtwo) = \nu \sum_{\epsilon_\alpha \le \epsilon_{\rm F}}
	   \psi_\alpha^\dagger(\rone)\psi_\alpha(\rtwo) \; .
     \eeq
The idea is to write this in
the \emph{Kohn-Sham basis} (i.e., the $\psi_\alpha$'s are Kohn-Sham
orbitals), so that it is compatible with the DFT diagrammatic
expansion.
If we
change to $\Rvec = \frac12(\rone+\rtwo)$
and $\svec = \rone - \rtwo$, we can expand in $\svec$
 \beq
  \rho(\Rvec+\svec/2,\Rvec-\svec/2)
    = e^{\svec\bm{\cdot}(\bm{\nabla}_1-\bm{\nabla}_2)/2}
    \left.\rho(\rone,\rtwo)\right|_{\svec=0} \; .
 \eeq
Negele and Vautherin obtained an expansion in terms of the
fermion, kinetic energy, and other densities:
 \beq
   \rho(\rone,\rtwo) = \frac{3j_1(sk)}{sk}\rho(\Rvec)
     + \frac{35j_3(sk)}{2sk^3} 
     \left(
       \frac14 \nabla^2\rho(\Rvec) - \tau(\Rvec) + \frac35k^2\rho(\Rvec)
       + \cdots
     \right) \; ,	 
 \eeq
which leads to functionals of these densities, for which we can  
take the $\delta/\delta\rho(\Rvec)$, $\delta/\delta\tau(\Rvec)$,
etc.\ derivatives directly.
(See also DME applied to 
ChPT in nuclear medium by Kaiser and
collaborators~\cite{Kaiser:2002jz,Kaiser:2003uh,Kaiser:2005}.)
This is clear at the Hartree-Fock level, but generalizations
are needed for higher-order diagrams.  These are also in
progress~\cite{Vincent:2007}.

  \begin{figure}[t]
  \centering
	\includegraphics*[angle=0.0,width=2.8in]{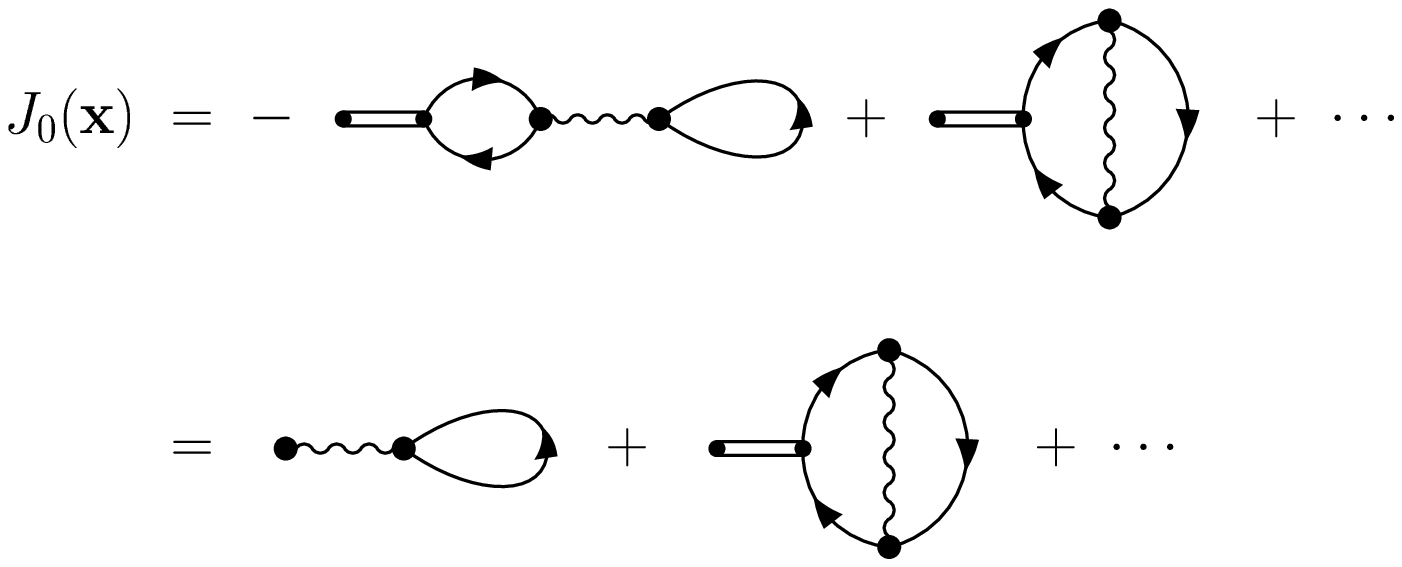}
	\hspace*{.1in}
	\includegraphics*[angle=0.0,width=2.8in]{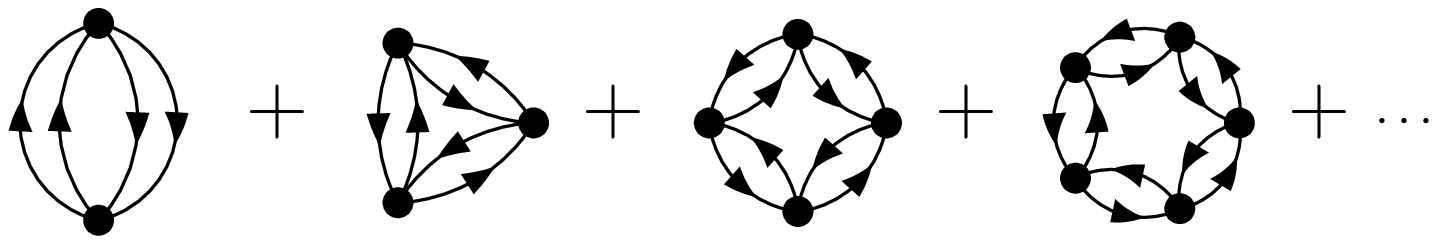}
    \caption{Long-range effects contributing to energy density
    functionals.}
  \label{fig:23c}       
  \end{figure}

There are some important open questions for this approach or 
any DFT treatment of finite nuclei.  These include:
\begin{itemize}
  \item
  For pairing, the energy interpretation, number projection,
  renormalization in finite systems, and efficient numerical
  implementation.  Also, a unified microscopic treatment of
  particle-particle and particle-hole physics.

  \item
  DFT for self-bound systems.
  Self-bound systems have no external potential, which implies
  that the true ground-state density is uniform!
  More generally, how
  do we deal with symmetry breaking
  (translational, rotational invariance, 
  particle number) and restoration.
  There has been little or no guidance from Coulomb DFT.
  There are analogous issues and methods for effective actions,
  namely soliton zero modes and projection methods.
  Work on an 
  energy functional for the intrinsic density is
  in its infancy~\cite{Engel:2006qu}. 

  \item
  Long-range effects, as illustrated schematically in Fig.~\ref{fig:23c}.
  This includes long-range forces (i.e., pion and Coulomb exchange)
  but also long-range correlations.
  The latter can be understood as
  non-localities from near-on-shell particle-hole excitations,
  as in the lower diagrams pictured in Fig.~\ref{fig:23c}.

\end{itemize}

\subsection{Covariant DFT}

Thus far, our discussion has included only nonrelativistic EFT and DFT.
However, there is a successful phenomenology based on relativistic
mean field energy functionals~\cite{SEROT86,RING96,SEROT97}.  
Can we make a connection?

In principle we could proceed by deriving a covariant EFT.
We start by observing that \emph{all} low-energy effective theories
have incorrect UV behavior.
Sensitivity to short-distance physics is signalled by
divergences but finiteness (e.g., with cutoff)
doesn't mean there is not sensitivity! 
One must absorb (and correct) sensitivity by renormalization.
Instances of UV divergences for low-energy nuclear physics are
    \begin{center}
    \renewcommand{\tabcolsep}{10pt}
    \begin{tabular}{cc}
      nonrelativistic     &  covariant \\ \hline
      scattering  & scattering \\
      pairing     & pairing    \\
                  & {anti-nucleons}
    \end{tabular}
    \end{center}

  \begin{center}
     \includegraphics*[width=2.4in,angle=0]{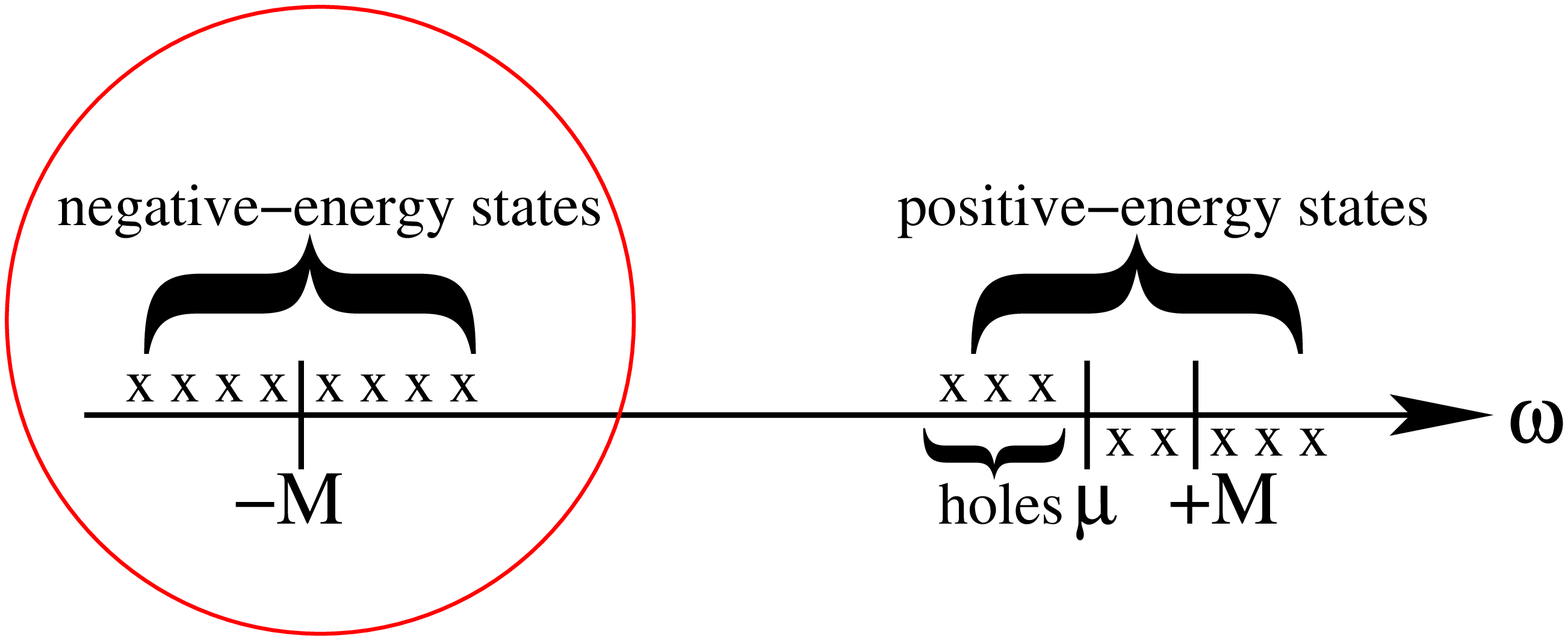}
  \end{center}
Thus, there is an additional source of divergences in the covariant
case from the ``Dirac sea''.

Gasser, Sainio, Svarc~\cite{GASSER88} derived 
chiral perturbation theory (ChPT) for ${\pi}N$ physics using
relativistic nucleon degrees of freedom.
But they found that the
loop and momentum expansions no longer agree (as they do in nonrelativistic
ChPT), which means that
systematic power counting was lost.
The heavy-baryon EFT restores power counting by a ${1/M}$ expansion,
and has been the basis for nonrelativistic NN EFT
treatments~\cite{JENKINS91}.

However,
Hua-Bin Tang~\cite{TANG96} (and with Paul Ellis~\cite{ELLIS98}) observed:
  \begin{quote} 
      ``\ldots EFT's permit useful low-energy expansions only if we
      absorb {\emph{all}} of the hard-momentum effects into the parameters
      of the Lagrangian.'' 
     ``When we include the nucleons relativistically, the anti-nucleon
     contributions are also hard-momentum effects.''
  \end{quote}
They advocated moving the ``Dirac sea'' physics into the
coefficients,  
thereby absorbing the ``hard'' part of a diagram into parameters,
while the remaining ``soft'' part satisfies chiral power counting.
The original ${\pi}N$ prescription by Tang and Ellis
(expand, integrate term-by-term, and resum propagators)
was systematized for ${\pi}N$
by Becher and Leutwyler under the name ``infrared regularization'' or 
IR~\cite{BECHER99}. 
It is not unique; e.g., Fuchs et al.\ have used
additional finite subtractions in DR~\cite{FUCHS03}.  
The extension of IR to multiple heavy particles 
by Lehmann and Pr\'ezeau~\cite{Lehmann:2001xm},
with a 
convenient reformulation by Schindler, Gegelia, and 
Scherer~\cite{Schindler:2003xv},
offers the possibility of a working covariant EFT.

If we restrict our attention to purely short-ranged, natural interactions,
there are tremendous simplifications.
In particular,
tadpoles and $N\overline N$ loops in free space vanish!
For example, leading order (LO) has scalar, vector, etc.\ vertices,
\beq
  {\cal L}_{\rm eft} = 
      \cdots 
           - \frac{{C_s}}{2} (\psibar\psi)(\psibar\psi) 
           - \frac{{C_v}}{2}
           (\psibar\gamma^\mu\psi)(\psibar\gamma_\mu\psi)
      + \cdots 
\eeq       
which we designate as
  \begin{center}
    \includegraphics*[width=0.6in,angle=0]{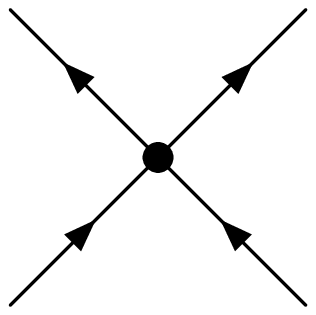}
  \end{center}
and consider all possible diagrams at NLO:
    \begin{center}
    \includegraphics*[width=2.1in,angle=0]{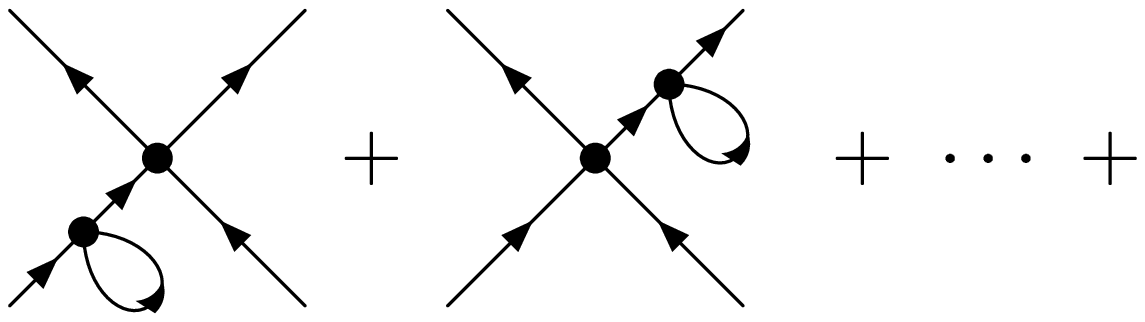}
    \includegraphics*[width=2.1in,angle=0]{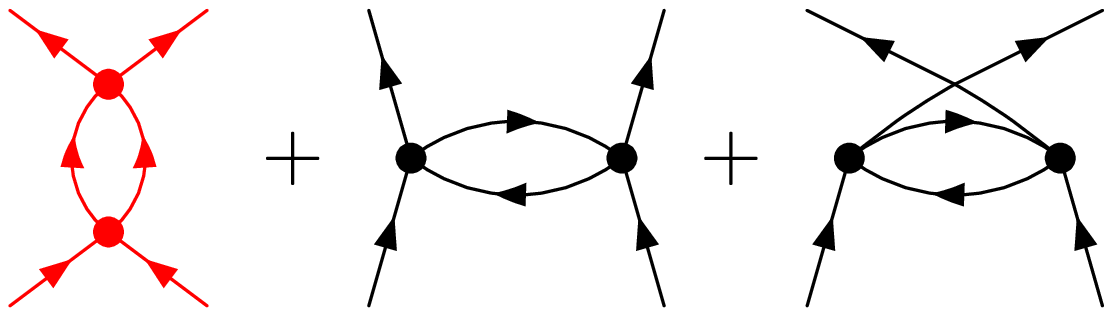}
    \end{center}
Only the particle-particle loop diagram survives IR
and all of the others pictured here vanish.    
Since only forward-going nucleons
contribute in the end, one obtains the same scattering 
amplitude as in nonrelativistic DR/MS for small $k$.	 

Unlike QED DFT, ``no sea'' for nuclear structure is a misnomer;
one should include $\overline N N$ ``vacuum physics'' in coefficients via
renormalization.
But note that
requiring renormalizability at the hadronic level corresponds to making 
a model for the short-distance behavior, which has proven to be
a poor model phenomenologically.
Fixing short-distance behavior is not the same thing as 
throwing away negative-energy states.
For a long time, people searched for \emph{unique} ``relativistic effects''; 
these were largely misguided efforts.

The further investigation of covariant EFT and its extension to DFT
is motivated by the successes of relativistic
mean-field phenomenology and other arguments about low-energy
QCD.  But there is much to be done for it to be competitive
with the nonrelativistic EFT.

\subsection{DFT for Cold Atoms with Large Scattering Length}

Finally we return to the large scattering length problem,
which is realizable with cold atoms.
The total cross section for scattering is expressed in
term of partial-wave phase shifts as
 \beq 
   \sigma_{\rm total} = \frac{4\pi}{k^2} \sum_{l=0}^{\infty} 
              (2l+1)\sin^2\delta_l(k) \; .
 \eeq
Recall that 
an attractive potential pulls the asymptotic wave function
(outside the potential) in, by an amount at each energy
or $k$ called the phase shift.
At low energy ($\lambda = 2\pi/k \gg 1/R$), the S-wave phase
shift $\delta_0(k)$ satisfies:
   \beq k \cot \delta_0(k) \stackrel{k\rightarrow 0}{\longrightarrow}
      -\frac{1}{a_0} + \frac12 r_0 k^2 + \ldots
   \eeq 
where
${a_0}$ is the ``scattering length'' and $r_0$ is the ``effective range''.
The effective range expansion for low-energy scattering goes back to
Schwinger and Bethe and others.
 The effective range expansion typifies the general principles
we have stated for EFT's:  If a complicated potential produces scattering with a given
$a_0$ and $r_0$, we can replace it by a simpler potential with the same
values and everything agrees at low energies.
In general, the effective-range expansion is reproduced and extended by EFT.

Having a bound-state or near-bound state at zero energy
means large scattering lengths ($a_0 \rightarrow \pm \infty$).
For $kR \rightarrow 0$,
the total cross section is
 \beq
  \sigma_{\rm total} = \sigma_{l=0} = 
    \frac{4\pi a_0^2}{1 + (k a_0)^2}
    = \left\{
    \begin{array}{cl}
      4\pi a_0^2 & \mbox{for\ } ka_0 \ll 1 \;, \\
      \frac{\textstyle 4\pi}{\textstyle k^2} & \mbox{for\ } ka_0 \gg 1
      \mbox{\ (unitarity limit).}
    \end{array}
    \right.
 \eeq 
We are particularly interested in cases where there is a bound-state 
near zero energy or there just misses being a bound state.
These pictures: 
  \begin{center}
     \includegraphics*[width=2.2in,angle=0.0]{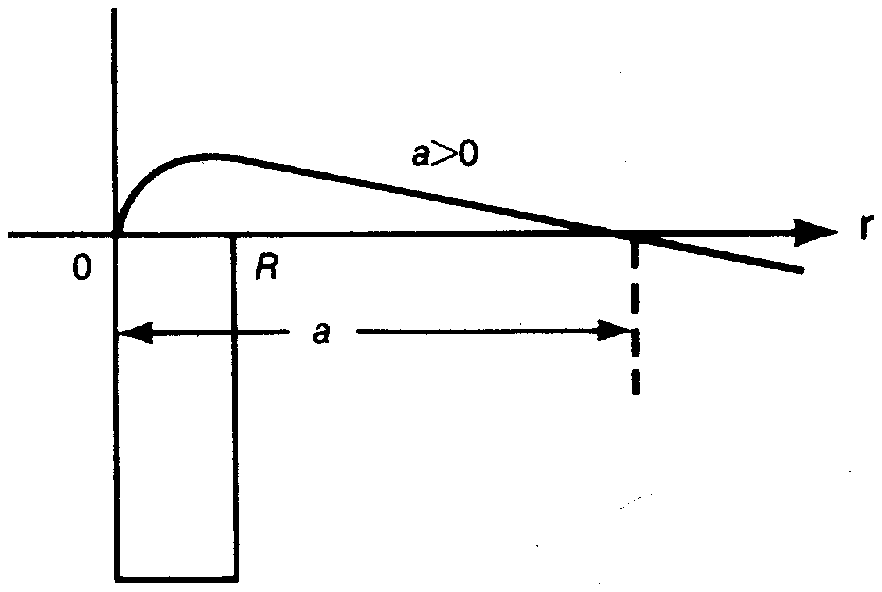}
     \includegraphics*[width=1.7in,angle=0.0]{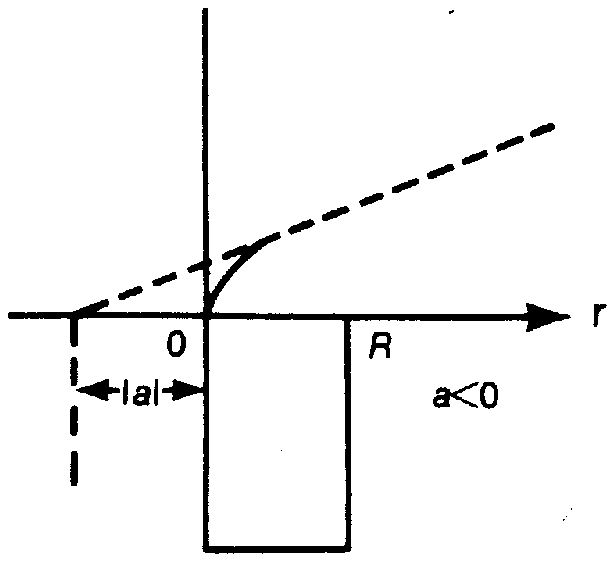}
  \end{center}
are a reminder of the interpretation of the scattering 
length in terms of the intercept of the zero energy scattering wave function,
which is a straight line outside the potential.
For potentials that just have a bound state, the wave function just
turns over and $a_0$ is large and positive.
If the potential just fails to have a bound state, it doesn't quite
make it to horizontal and $a_0$ is large and negative.

At low energies, depending on the size of $k$ times $a_0$,
the cross section first goes like the
square of $a_0$ and then saturates at the unitarity limit.  
So if we could adjust the depth of the bound state,
we can control the ``strength'' of the interaction in a sense. 
This is possible for atoms by changing an external
magnetic field to produce resonant scattering.
For QCD is it possible to
adjust the quark mass \emph{theoretically} so that $m_\pi$ changes
and the nuclear $a_0$ can be tuned to $\pm\infty$!

\newcommand\dlr{\raisebox{0.1em}{$\stackrel{\scriptstyle\leftrightarrow}\partial$}}
\newcommand\+{\dagger}
\renewcommand\d{\partial}

Let's consider the large scattering length many-body problem, which means 
we have an
attractive two-body potential with $a_0 \rightarrow \infty$.
If $R \ll 1/\kf \ll a_0$, as in this figure,
  \begin{center}
    \includegraphics*[width=1.3in,angle=0.0]{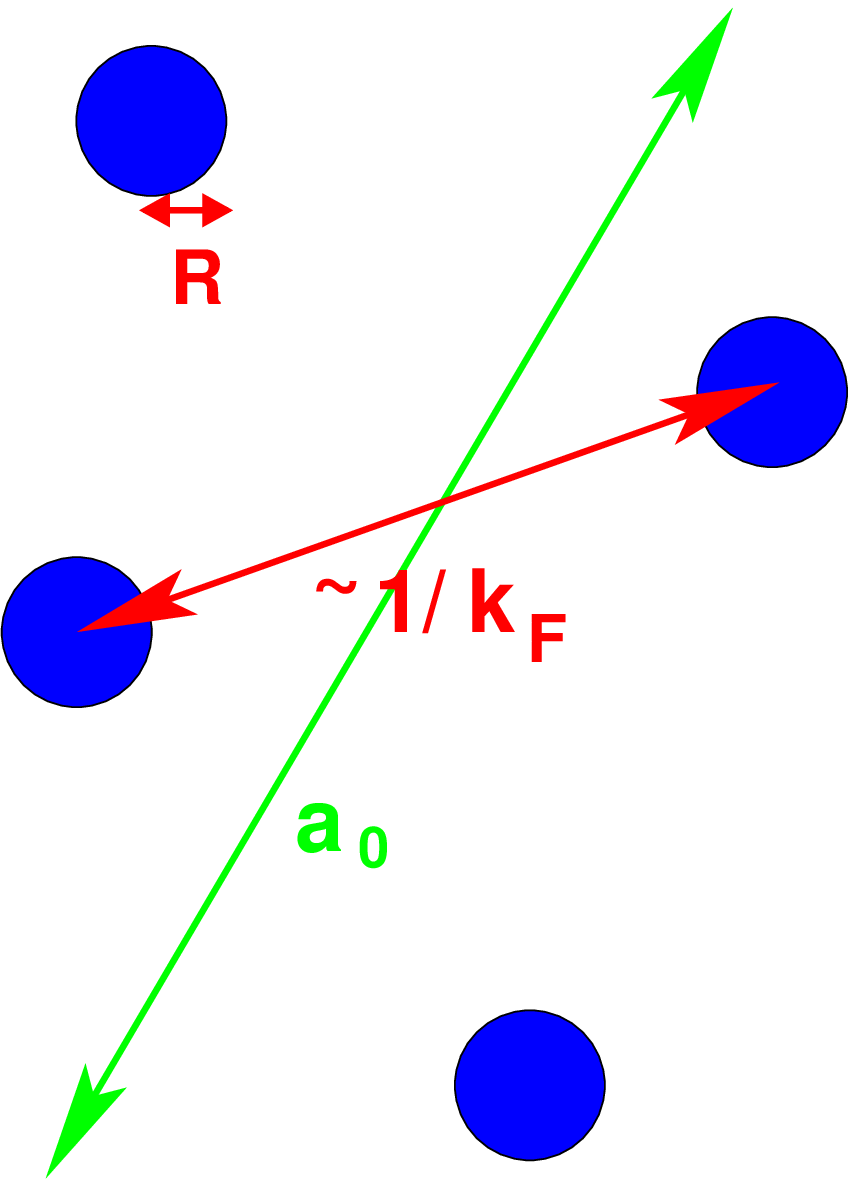}
  \end{center}
then we expect scale invariance (since we lose both $R$ and $a_0$ as
possible scales).
This means that the energy and superfluid gap should
be pure numbers times
           $E_{\rm FG} = \frac35\frac{\kf^2}{2M}$.

Recall that for the natural scattering length case,    
EFT power counting led to an organized perturbative expansion:
 \begin{center}
   \includegraphics*[width=3.5in,angle=0]{fig_ere}
 \end{center}  
\noindent
with $C_0 = \frac{\textstyle 4\pi}{\textstyle M}a_0$
and $k a_0 \ll 1$.  
But in the large scattering length limit, 
$k a_0 \gg 1$ so the bubble series diverges.
This is not a difficulty in free space,
because the geometric sum of bubbles is easily performed.
This sum yields the $f_0(k)$ expansion by keeping $k a_0$
to all orders and expanding the rest:
 \beq
   f_0(k) \propto 
   \frac{1}{-1/a_0 + r_0k^2/2 - ik}
   \longrightarrow
   \frac{-1}{1/a_0+ik} \left[ 1 + \frac{r_0/2}{1/a_0+ik}k^2 + \cdots \right]
 \eeq

With a natural $a_0$ and a perturbative expansion, we found the
DR/MS (minimal subtraction) scheme particularly convenient.
With large $a_0$, we need a new renormalization scheme.
DR/PDS was proposed by Kaplan, Savage, Wise~\cite{KSW} and counts $\mu \sim k$:
 \beq
    \raisebox{-.2in}{\includegraphics*[width=.4in,angle=0]{fig_C0}}
   \ \Longrightarrow\ C_0{(\mu) = \frac{4\pi}{M} 
       \left( \frac{1}{-\mu + 1/a_0} \right)
       \stackrel{a_0\rightarrow\infty}{\longrightarrow}
       -\frac{4\pi}{M\mu}} \; ,
  \eeq

     \beq
      \raisebox{-.2in}{\includegraphics*[width=0.7in,angle=0]{fig_C0sq}}
       \ \Longrightarrow\ 
       \int\! \frac{d^Dq}{(2\pi)^3} \frac{1}{k^2-q^2 + i\epsilon}
         \stackrel{D\rightarrow 3}{\longrightarrow} 
           - \frac{{\mu + } ik}{4\pi} \; .
     \eeq
In medium, each additional $C_0$ vertex gives a factor
      \beq
         C_0(\kf) \left(\frac{M}{\kf^2}\right)^2
	  \left( \frac{\kf^5}{M} \right) \sim \kf^0\; ,
      \eeq
which means that all $C_0$ diagrams are leading order!
Thus, we are told to sum \emph{all} many-body diagrams with
$C_0$ vertices.
This is only possible numerically (or possibly
with an additional expansion).
   
\begin{figure}[t]
\centering
       \includegraphics*[angle=0.0,width=3.2in]{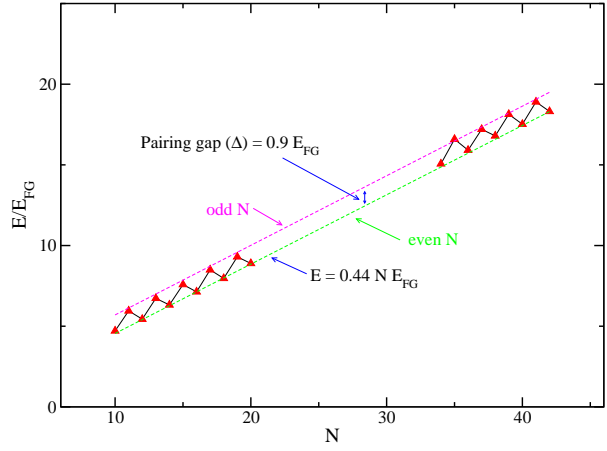}
   \caption{GFMC results from Chang \textit{et al.} \cite{Chang:2004sj}
    for the unitary fermion system.}
 \label{fig:27}       
\end{figure}

So we turn to numerical calculations.
GFMC results from Chang \textit{et al.} \cite{Chang:2004sj}, in which  
one extrapolates to large numbers of fermions,
are shown in Fig.~\ref{fig:27}.
They find the energy per particle is $E/N = 0.44(1) 
       E_{\rm FG}$.
Diffusion Monte Carlo (DMC) results~\cite{Astrak}, with
a square-well potential tuned to $a_0 \rightarrow \infty$
and an extrapolation to large numbers of fermions, find
a similar result, namely an energy per particle of 
$E/N = 0.42(1) E_{\rm FG}$.  

Thomas Papenbrock has considered
DFT for the unitary regime~\cite{Papenbrock:2005bd}.
He assumes a simple, constrained form of the density functional,
  \beq
    {\cal E}[\rho] = \frac{\hbar^2}{m}
    \left[
    \frac{m}{2m_{\rm eff}} \sum_{j=1}^{N} |\nabla\phi_j(r)|^2
      + \Bigl(\xi - \frac{m}{m_{\rm eff}}  \Bigr) c\rho^{5/3}
    \right]
    + \frac12 m\omega^2 r^2\rho \; ,
  \eeq
with non-localities and gradient terms via the effective mass 
$m_{\rm eff}$.
The parameters in $E[\rho] = \int\! d\xvec \, {\cal E}[\rho(\xvec)]$
can be fit for $N=2$ to exact results for two fermions in 
harmonic trap [Busch et al (1998)]:
\beq
  \psi_{\rm rel}(r)
   = \frac{1}{\sqrt{2^{3/2}\pi l^3}} \frac{l}{r} e^{-r^2/4l^2} \; ,
   \qquad l = \sqrt{\hbar/m\omega}\; , \quad E = 2\hbar\omega \; ,
\eeq
and a gaussian center-of-mass wave function. 
The result is 
 \beq 
   \rho_{\rm exact}(r) = \frac{4}{\pi^{3/2}l^3}\frac{l}{r}
     e^{-2(r/l)^2}  \int_0^r\! dx\, e^{x^2}
     \; .
 \eeq

\begin{figure}[t]
\centering
       \includegraphics*[angle=0.0,width=3.2in]{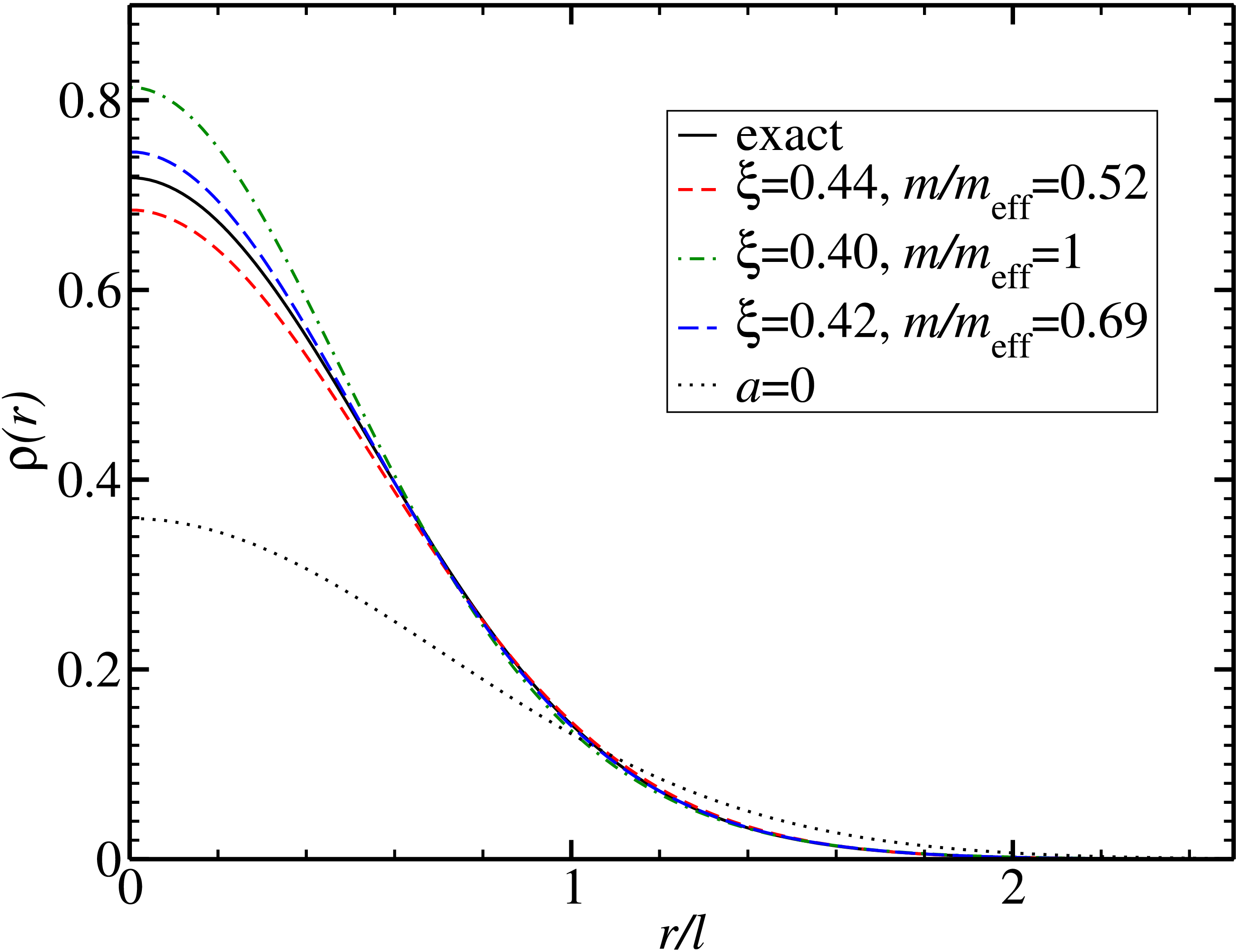}
   \caption{Papenbrock results from a density functional
   for the unitary regime.}
 \label{fig:29}       
\end{figure}

\begin{figure}[t]
\centering
       \includegraphics*[angle=-90.0,width=3.2in]{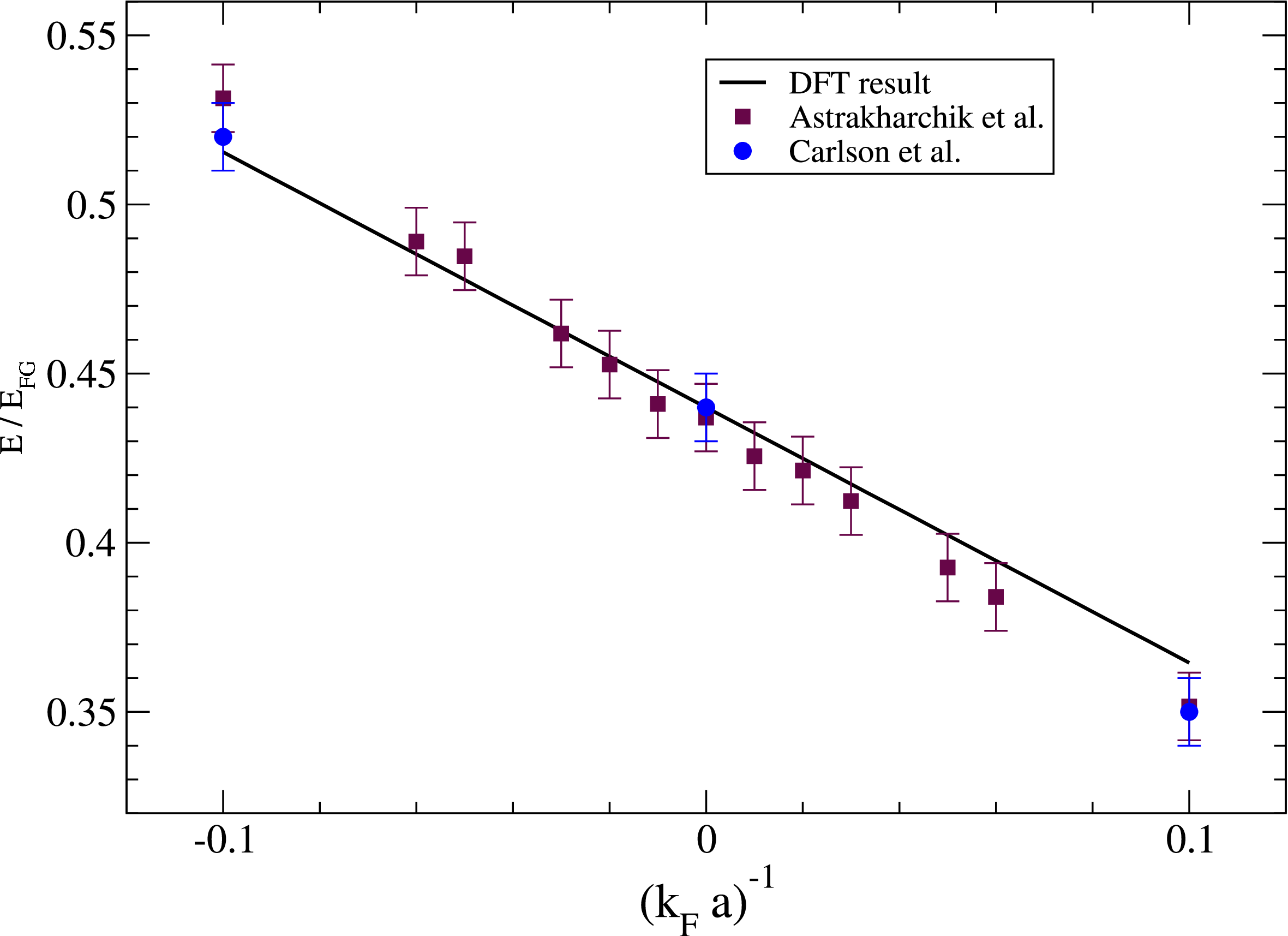}
   \caption{Papenbrock/Bhattacharyya DFT results for finite
   scattering length $a$ compared to Monte Carlo calculations.}
 \label{fig:30}       
\end{figure}

Results of such fits are shown in Fig.~\ref{fig:29}.
They predict $\xi = 0.42$ and $m/m_{\rm eff} = 0.69$ from the best fit.
The value for $\xi$ is amazingly close to the Monte Carlo result.
Papenbrock and
Bhattacharyya~\cite{Bhattacharyya:2006fg,Papenbrock:2006hg} 
consider corrections for an 
LDA density functional close to the unitary limit
  \beq
  {\cal E}[\rho] = {\cal E}_{\rm FG}
  \left(
    \xi + \frac{c_1}{a\rho^{1/3}}
    + c_2 r_0 \rho^{1/3}
  \right) \; .
  \eeq
Again, this is fit to the harmonically trapped two-fermion system.
Results are given in Fig.~\ref{fig:30} and show impressive agreement
with Monte Carlo calculations.

Finally, there are interesting investigations of the
constraints of general coordinate and conformal invariance by
Dam Son and collaborators \cite{Son:2005rv,Son:2005tj}.
They ask:
Is there more than scale invariance for the unitary Fermi gas?
The symmetries can be exposed  by adding a background gauge field $A_\mu$ 
and curved space with metric $g_{ij}(t,\xvec)$:
\beqa
        && S \longrightarrow
  \int\!dt\,d\xvec\,\sqrt g\biggl[ \frac i2 \psi^\+ \dlr_t \psi
    - \frac{g^{ij}}{2m}(\d_i+iA_i)\psi^\+(\d_j-iA_j)\psi \\
  &&  
  \hspace*{.5in}
  \null + (q_0\sigma-A_0)\psi^\+\psi - \frac{g^{ij}}2\d_i\sigma\d_j\sigma
    - \frac{\sigma^2}{2r_0^2} \biggr] \; .
\eeqa 
This is more than scale and Galilean invariance!
Direct consequences include extra constraints on ${\cal L}_{\rm eff}$ at NLO,
which naively involve five arbitrary functions but these symmetries
show there are only three.
For the unitary Fermi gas, three
constants from scale invariance are reduced to two constants from
conformal invariance.
This leads us to ask: 
What additional constraints can we find for the energy functional?
This and the other open questions are ripe for investigation!

Acknowledgments:
I thank the organizers of the Trento school, Janos Polonyi and Achim
Schwenk, for the opportunity to participate in an excellent lecture
series,
and all the student participants, who made giving the lectures a
pleasure.
This work was supported in part by the National Science 
Foundation under Grant No.~PHY--0354916 and the Department of 
Energy under Grant No. DE-FC02-07ER41457.


%
%

%

\begin{thebibliography}{99.}
\bibitem{Fermions}
Recall that the wave function of a system of identical fermions
must be antisymmetric under exchange of any two particles.  This has
important consequences for many-body systems, such as the development
of a Fermi sea and ``Pauli blocking''.
\bibitem{Achim1}This figure was adapted by Achim Schwenk from a
  picture by A. Richter.
\bibitem{JoeC}This discussion is adapted from a talk by Joe Carlson.
\bibitem{caveat1}The term ``ab initio'' is often used in this context,
but the calculations do not in practice
proceed only from the Coulomb interaction.
\bibitem{Table2}
 The goal of devising a potential that fits the scattering data
as best as possible (that is, $\chi^2/\mbox{dof} \approx 1$) 
below where inelastic effects become important has been realized by
several potentials, including  CD-Bonn, Nijmegen I and II, and Argonne
$v_{18}$.
\bibitem{Table3}
Calculations of the energy of few-body nuclei using accurate
two-body potentials have demonstrated the need for a three-body force.
\bibitem{Table4}
 All NN potentials have a cut-off on high-momentum
(short-distance) contributions.  Putting in a cut-off doesn't mean the
high-energy physics is thrown out, however!  We need to renormalize
(this is one of the main points of our discussion).
\bibitem{Table5}
 One way of specifying a many-body approximation is to say which
Feynman diagrams are included.  (If all are included, we're solving
the problem exactly!)  Usually one needs to select an (infinite)
subset (for example, the ladder or ring diagrams).  In some case there
are rigorous justifications for which to sum, but not always and there
are seldom estimates available for what quantitative error is made by the
truncation.
\bibitem{Extra} To derive the second equality for $f_l(k)$, first put the
$e^{i\delta_l}$ in the denominator, then replace it by $\cos + i\sin$
and divide top and bottom by $\sin$.
\end{thebibliography}
%


\printindex
\end{document}